\documentclass{mclass}

\begin{document}

\markboth{Benatti, Floreanini}
{Open Quantum Dynamics and Entanglement}

\catchline{}{}{}{}{}

\title{Open Quantum Dynamics: Complete Positivity and Entanglement}

\author{Fabio Benatti}
\address{Dipartimento di Fisica Teorica, Universit\`a di Trieste,
Strada Costiera 11 and\\ 
Istituto Nazionale di Fisica Nucleare, Sezione di Trieste,
34014 Trieste, Italy; benatti@ts.infn.it}

\author{Roberto Floreanini}
\address{Istituto Nazionale di Fisica Nucleare, Sezione di Trieste, 
Dipartimento di Fisica Teorica, Universit\`a di Trieste,
Strada Costiera 11, 34014 Trieste, Italy; florean@ts.infn.it}


\maketitle


\begin{abstract}
We review the standard treatment of open quantum systems in relation
to quantum entanglement, analyzing, in particular, the behaviour
of bipartite systems immersed in a same environment. 
We first focus upon the notion 
of complete positivity, a physically motivated algebraic constraint 
on the quantum dynamics, in relation to quantum entanglement,
{\it i.e.} the existence of statistical correlations which can not be accounted
for by classical probability.
We then study the entanglement power of
heat baths versus their decohering properties, a topic of increasing
importance in the framework of the fast developing fields of quantum 
information, communication and computation.
The presentation is self contained and, through several
examples, it offers a detailed survey of the physics and 
of the most relevant and used techniques
relative to both quantum open system dynamics and quantum entanglement.
\end{abstract}



\section{Introduction}
\label{sec1}

Standard quantum mechanics mainly deals with \textit{closed}
physical systems that can be considered isolated from any external environment,
the latter being generically a larger system consisting of
(infinitely) many degrees of freedom. 
The time-evolution of closed systems is described by one-parameter
groups of unitary 
operators embodying the reversible character of the dynamics.
On the contrary, when a system $S$ interacts with an environment $E$
in a non-negligible way, it must be treated as an 
\textit{open quantum system}, namely as a subsystem embedded within $E$, 
exchanging with it energy and entropy, and whose time-evolution
is irreversible.%
\footnote{The literature on the theory of open quantum systems and
their phenomenological applications is vast; in the  
References, only papers that are strictly relevant to the
present exposition are therefore included. We start by providing
a list of general reviews and monographs on the topic,
Refs.[\refcite{alicki-lendi}-\refcite{weiss}], where
additional references can be found.}

In general, the time-evolution of $S$ is inextricably linked to 
that of $E$. The compound system $S+E$ is closed and develops
reversibly in time; however, the global time-evolution rarely permits
the extraction of a meaningful dynamics for the system $S$ alone.
This can be done if the coupling among subsystem and environment is
sufficiently weak, in which case physically plausible approximations
lead to \textit{reduced dynamics} that involve only the $S$ degrees of
freedom and are generated by \textit{master equations}. 
Such reduced dynamics provide effective descriptions of how
$E$ affects the time-evolution of $S$ which, on time-scales that are
specific of the given physical contexts, typically
incorporates dissipative and noisy effects.

In order to better appreciate the physical motivations underlying the
concept of reduced dynamics, 
classical Brownian motion is a helpful guide.\cite{gardiner2}
On the fast microscopic time-scale, the scattering of the environment
particles off the mesoscopic Brownian particle $S$ are described by  
a reversible dynamics.
On the slower mesoscopic time-scale, the memory effects related 
to microscopic interactions are averaged out and the effective
dynamics of $S$ is diffusion, that is an irreversible Markov process.
Physically speaking, on the slow time-scale the environment degrees of 
freedom act as a sink where $S$ dissipates energy, but also
as a source of (white) noise with long-run stabilizing effects.  
We shall see that, under certain conditions, irreversible Markovian
time-evolutions, leading to dissipation and noise, find a concrete description
in quantum mechanics by means of the so-called \textit{quantum
dynamical semigroups}. 

Classical Brownian motion indicates that, when the typical
time-scale of $S$ is much larger than the time-scale 
governing the decay of time-correlations of the environment, then 
$E$ can be described as an effective source of damping and noise. 
In the framework of open quantum systems, this possibility is
technically implemented either by letting the typical variation time of $S$, 
$\tau_S$, go to infinity, while the environment correlation time $\tau_E$
stays finite, or by letting $\tau_E$ go to zero, 
while $\tau_S$ stays finite.\cite{spohn}
As we shall see, these two regimes give rise to two different
procedures to arrive at a reduced dynamics described by
Markovian semigroups: the so-called
\textit{weak coupling} and \textit{singular coupling} limits.

Since their first appearance, open quantum systems have been providing 
models of non-equilibrium quantum systems in diverse fields as
chemical-physics, quantum optics and magnetic resonance.
Recently, the rapid development of the theory of quantum
information, communication and computation%
\footnote{For general reviews on these topics, see
Refs.[\refcite{nielsen}-\refcite{zeilinger}].}
has revived the interest in
open quantum systems in relations to their decohering properties, but
also in their capacity of creating entanglement in multi-partite
systems immersed in certain environments.
The typical open quantum systems in these contexts  
are $n$-level systems, like atoms, photons or neutrons embedded in
optical cavities or heat baths consisting of bosonic or fermionic
degrees of freedom.
These will be the cases studied in this review. 
 
It is worth mentioning that, more in general, one can expect
dissipative and noisy effects to affect also
the behaviour of elementary particle systems, 
like neutral mesons, neutrinos and photons:
the ensuing irreversibility is the emerging result of
the dynamics of fundamental degrees of freedom at very short
distances, typically the Planck-scale.\cite{bf1}
These effects produce distinctive signatures in the behaviour
of selected physical observables, allowing direct experimental
testing (for specific treatments and further details, 
see Refs.[\refcite{bf2}-\refcite{bf12}]).

Providing different concrete experimental contexts to study dissipative 
quantum dynamics is important not only from the point of view of the
physics of open quantum systems, but also for 
the investigations of certain still debated aspects
of these generalized time-evolutions, specifically regarding
the property of positivity {\it v.s.} that of complete
positivity, notions on which we now briefly elaborate.

A primary consequence of a noisy environment $E$ is that pure states
of $S$, that is projections onto Hilbert space vectors, are usually mapped 
into mixtures of projections, that is into generic density matrices.
These are operators with discrete spectrum consisting of positive eigenvalues
summing up to one; within the statistical interpretation of quantum
mechanics, they represent probabilities and are used to calculate
the mean values of all physical observables pertaining to $S$.
In order to be physically consistent, any reduced dynamics must thus
correspond to maps on the space of states of $S$ that preserve for all times 
the positivity of the spectrum.
In other words, any physically consistent reduced dynamics must
transform density matrices into density matrices at all (positive) times.

The simplest way to arrive at dissipative time-evolutions
is to construct them as solutions of suitable master equations of
\textit{Liouville} type: $\partial_t\rho_t=\mathbb{L}[\rho_t]$, 
where $\mathbb{L}$
operates linearly on the density matrices $\rho_t$ describing
the state of $S$ at time $t$.
These are usually obtained by first tracing away the
environment degrees of freedom, an operation that leads to a dynamical 
equation plagued in general by non-linearities, secular terms and
memory effects.
Non-linearities and secular terms can be eliminated by
requiring the initial states of $S$ to be statistically decoupled
from an initial reference state of $E$. 
On the other hand, memory effects are expected to
be relevant on short time-scales and to become negligible if one looks at
phenomena associated to longer time-scales when those effects have
already died out.

In technical terms, studying the time-evolution on a slow time-scale,
thus neglecting memory effects,
means operating a so-called
{\it Markovian approximation}; this gives an efficient
description of the reduced dynamics of $S$ in terms of time-evolutions 
consisting of one-parameter semigroups of linear maps 
$\gamma_t=\exp(t\mathbb{L})$, $t\geq 0$.
These are maps that, acting on all possible initial density matrices
$\rho$, provide their time-evolved partners
$\rho_t\equiv\gamma_t[\rho]$ at time $t$.
The dynamical maps $\gamma_t$ are formally obtained by
exponentiating the \textit{generators} $\mathbb{L}$, so that
$\gamma_{t+s}=\gamma_t\circ\gamma_s=\gamma_s\circ\gamma_t$, $s,t\geq0$.

In most open systems, the memory effects are due to a transient regime
with a very small time-span that makes Markov approximations legitimate;
nevertheless, if not carefully performed, a likely outcome is
a physically inconsistent time-evolution that does not 
preserve the positivity of the spectrum of all initial density matrices.
As we shall explicitly see in the following, 
brute force Markov approximations may provide semigroups of linear
maps $\gamma_t$ that at some time transform an initial state $\rho$ into an 
operator $\rho_t$ with negative eigenvalues, which can not be used as 
probabilities.
In other words, without due care in the derivation, one is very
likely to obtain $\gamma_t$ that do not transform the space of states 
into itself.
In order to avoid the appearance of negative probabilities,
one has to perform Markov approximations that lead to semigroups
of linear maps that preserve positivity: maps with such a property are called
\textit{positive}. 

Remarkably enough, both the mathematical property of linear maps of being positive
and the derivation of positive dynamical maps through Markov
approximations are elusive and still not fully understood.
However, positivity is not the end of the story and even a semigroup
of positive linear maps would not be fully physical consistent. 
Indeed, suppose $S$ is coupled to an inert system $A$, a so-called 
\textit{ancilla}, with which it does not interact; then, the time-evolution of
the compound system $S+A$ is $\gamma_t\otimes {\rm id}_A$.
The identity map ${\rm id}_A$ means that the coupling to the
ancilla does not possess a dynamical character: it 
manifests itself purely in terms of statistical correlations,
{\it i.e.} in the possibility of initial correlations
between $S$ \hbox{and $A$.}

It turns out that the positivity of $\gamma_t$ does not guarantee
the positivity of $\gamma_t\otimes{\rm id}_A$; namely, even if
$\gamma_t$ transforms any initial state of $S$ into a state, 
$\gamma_t\otimes{\rm id}_A$ can map an initial state of $S+A$ into
an operator which exhibits negative eigenvalues and can not thus be
interpreted as a physical state of $S+A$.
In order to be sure that, for whatever ancilla $A$, the evolution
$\gamma_t\otimes{\rm id}_A$ consistently maps any state of $S+A$ into
a state, $\gamma_t$ must be a so-called \textit{completely positive}
map for all $t\geq0$, a stronger property than positivity.

We shall show that the ultimate physical meaning of
complete positivity lies in its relation to the existence of
\textit{entangled states},
the foremost instance of them being a vector state with
a singlet-like structure that can not be written as a tensor product
of vector states.
Indeed, the only states that could be mapped out of the state-space by
$\gamma_t\otimes{\rm id}_A$ and thus made not anymore
acceptable as physical states, are only the entangled states of $S+A$.

Unitary time-evolutions are automatically completely positive, while 
this is not in general true of many Markovian approximations
encountered in the literature on open quantum systems. 
Complete positivity or its absence depend on the way Markovian
approximations are performed.
Since the coupling of the open system $S$ with a generic ancilla
can not be excluded, the request of complete positivity is hardly
dismissible when one deals with dynamical semigroups.
Such a request has strong mathematical and physical consequences
in that the structure of the generator $\mathbb{L}$ is fixed; from this  
a typical hierarchy among the characteristic relaxation times of $S$ follows. 
The uncontrollability of the ancilla and the abstractness
of the whole justification of why complete positivity should be
physically compelling is scarcely appealing from a concrete point of
view, all the more so since complete positivity strongly constrains the
reduced dynamics.

However, we shall see that the same physical inconsistencies typical
of the abstract contexts, $S+A$, also occur when one considers
not a generic inert ancilla $A$, but another system $S$ also immersed
in $E$, that is an open quantum system $S+S$.
If the two parties weakly interact with $E$ and do not interact
between themselves, the dissipative time-evolution of the compound
system $S+S$ is of the form $\gamma_t\otimes\gamma_t$ instead of
the abstract $\gamma_t\otimes{\rm id}_A$ and the setting becomes
more physical. 

Consider, for instance, two optically active non-interacting 
atoms (or molecules) 
initially prepared in an entangled polarization state and
evolving in contact with a same heat bath, weakly coupled to them.
Then, if the dynamical maps $\gamma_t$ describing the dissipative evolution
of each single atom is not completely
positive, there are physically admissible initial entangled states of
the two atoms that, after some time, lose their interpretation as 
physical states because of the emergence of negative probabilities
generated by $\gamma_t\otimes\gamma_t$.

The rising of quantum entanglement as a physical resource
enabling the performance of quantum information and
computation tasks otherwise impossible as teleportation, superdense
coding, quantum cryptography and quantum computation, has spurred the
investigation of how entanglement can be generated, detected and manipulated.
Contrary to standard expectations, the presence of an
environment, and thus of noise, need not necessarily spoil the
entanglement properties of states of systems immersed in it, but in
some cases can even have entangling effects.
Using the previous examples, the two optically active atoms could
be prepared in an initial separable state without either classical or
quantum correlations and put into a heat bath.
The covariance structure of the bath correlation functions may be such
that not only the two atoms become entangled, but also such that a 
certain amount of entanglement survives over longer and longer times.
The reason is not hard to see: even if not directly coupled, two
systems immersed in a same heat bath can interact through the bath itself;
it does then depend on how strong this indirect interaction is with
respect to the decoherence whether entanglement can be created and
maintained.
 
The plan of this review is as follows. 
In Section 2, we shall set the mathematical framework appropriate to
the description of open quantum systems and to the derivation of their 
reduced dynamics, with particular emphasis on the notions of positive
and completely positive maps on the space of states. 

In Section 3, we shall review standard derivations of dissipative
semigroups, discussing the role of the Markov approximations in
achieving completely positive dynamical maps or not and illustrating 
the physical inconsistencies arising from lack of complete positivity.

In Section 4, we shall address the entangling effects that the presence
of an environment may have on non-directly interacting bipartite systems.
All topics will be illustrated by means of concrete examples.

\section{Open Quantum Systems: Mathematical Setting}
\label{sec2}

In the first part of this Section we review some basic kinematics
and dynamics of open quantum systems, fixing, in passing, the necessary
notation.
In the second part we shall discuss positivity and complete positivity
of linear maps with particular emphasis on their relations with
quantum entanglement.

We shall consider finite ($n$-level)
open quantum systems $S$; they are described by means of 
$n$-dimensional Hilbert spaces $\mathbb{C}^n$,
where $\mathbb{C}$ is the set of complex numbers, and by the 
algebras $M_n(\mathbb{C})$ of $n\times n$ complex matrices $X$.%
\footnote{General monographs on quantum physics relevant
for our exposition are 
Refs.[\refcite{alicki-fannes},\refcite{peres},\refcite{thirring}].}

The hermitian matrices $X=X^\dagger$ correspond to the system observables
and their mean values depend on the physical states of $S$.
The latter can be divided into two classes, the
\textit{pure states} described by projectors
$P_\psi=\vert\psi\rangle\langle\psi\vert$ 
onto normalized vectors $\vert\psi\rangle\in\mathbb{C}^n$ and the
\textit{statistical mixtures} 
described by \textit{density matrices}, {\it i.e.} by linear convex
combinations of (not necessarily orthogonal) projectors 
\begin{equation}
\label{mixst}
\rho=\sum_{j}\lambda_j\,P_{\psi_j}\ ,\ \lambda_j\geq0\ ,\
\sum_j\lambda_j=1\ .
\end{equation}
In the following we shall refer to them simply as states, specifying whether
they are pure or mixtures when necessary; further, 
\begin{equation}
\label{normal}
{\rm Tr}(\rho)=\sum_j\lambda_j=1\ ,
\end{equation}
where $\rm Tr$ represents the trace-operation ${\rm
Tr}(X)=\sum_i\langle i\vert\,X\,\vert i\rangle$, with $\vert
i\rangle$, $i=1,2,\cdots,n$ any orthonormal basis in $\mathbb{C}^n$.

Projectors and density matrices are thus normalized hermitian elements of 
$M_n(\mathbb{C})$; further, they are positive semi-definite matrices.
\bigskip

\noindent
{\bf Definition 2.1}\quad
{\it
A matrix $X^\dagger=X\in M_n(\mathbb{C})$ is positive semi-definite if
\begin{equation}
\label{posop}
\langle\psi\vert\,X\,\vert\psi\rangle\geq 0\ ,\quad
\forall\ \psi\in\mathbb{C}^n\ .
\end{equation}
}
\bigskip

\noindent
Positive semi-definite matrices will be called positive for sake of
simplicity and denoted by $X\geq0$; their spectrum
necessarily consists of positive (non-negative) eigenvalues.

Quantum states are thus positive, normalized operators; the
eigenvalues of the pure ones
$P_\psi$ are $1$ (non-degenerate) and $0$ ($n-1$ times
degenerate),  while those of density matrices are generic $0\leq r_j\leq1$,
$j=1,2,\ldots,n$, such that $\sum_j r_j=1$.
It follows that pure states and mixtures are distinguished by the fact that 
$\rho^2=\rho$ if and
only if $\rho=P_\psi$ for some $|\psi\rangle\in \mathbb{C}^n$.
Also, the eigenvalues of a density matrix $\rho$ correspond to the weights 
$\lambda_j$ in~(\ref{mixst}) if and only if~(\ref{mixst}) is the spectral
representation of $\rho$, the $|\psi_j\rangle$ are its eigenstates and
the $P_{\psi_j}$ the corresponding orthogonal
eigenprojectors,
$P_{\psi_j}P_{\psi_k}=\delta_{jk}\, P_{\psi_j}$.

Given a state $\rho$, the mean value $\langle X\rangle_\rho$ of any
observable $X=X^\dagger\in M_n(\mathbb{C})$ is calculated as follows:
\begin{equation}
\label{meanv}
\langle X\rangle_\rho\equiv{\rm Tr}\bigl(\rho\,X\bigr)=
\sum_j\lambda_j\langle\psi_j\vert\,X\,\vert\psi_j\rangle\ .
\end{equation}

The previous considerations constitute the bulk of the
statistical interpretation of quantum mechanics:
\medskip

\centerline{
\begin{minipage}[t]{12cm}
\textsf{\noindent
Given the spectral 
decomposition of a state,
\begin{equation}
\label{statint}
\rho=\sum_{i=1}^n \, r_i\,\vert r_i\rangle\langle r_i\vert\ ,\quad 
\sum_{i=1}^nr_i=1\ ,\quad \langle r_i\vert r_j\rangle=\delta_{ij}\ ,
\end{equation}
the eigenvalues $r_i$ constitute a probability distribution which
completely defines the statistical properties of the system.
}
\end{minipage}
}
\bigskip

Accordingly, one associates to quantum states the von Neumann entropy,
\begin{equation}
\label{vNe}
S(\rho)=\,-\,{\rm Tr}(\rho\log\rho)=\,-\,\sum_{i=1}^n\,r_i\,\log r_i\ ,
\end{equation}
which measures the amount
of uncertainty about the actual state of $S$.
It turns out that $S(\rho)=0$ if and only if $\rho^2=\rho$, otherwise
$0<S(\rho)\leq \log n$.
\bigskip

\noindent
{\bf Remark 2.1}\quad
{
Density matrices form a convex subset 
${\cal S}(S)\subset M_n(\mathbb{C})$ which we shall refer to as the
state-space of $S$.
Namely, combining different mixtures
$\sigma_j\in\mathcal{S}(S)$ 
with weights 
$\lambda_j\geq 0$, $\sum_j\lambda_j=1$, into the convex combination
$\sum_j\lambda_j\sigma_j$, the latter also belongs to
$\mathcal{S}(S)$.
Pure states are extremal elements of $\mathcal{S}(S)$, that is they
can not be convexly decomposed, while with them, by linear convex
combinations, one generates the whole of the state-space. $\Box$
}
\bigskip

\subsection{Reversible and Irreversible Dynamics}
\label{subs2.1}

The state-space and the algebra of observables fix the kinematics of $S$. 
Its dynamics as a closed system is determined by 
a Hamiltonian operator $H\in M_n(\mathbb{C})$ through the Schr\"odinger
equation (we shall set $\hbar=1$):
\begin{equation}
\label{Scheq}
\partial_t\vert\psi_t\rangle=-i\,H\,\vert\psi_t\rangle\ .
\end{equation}
By direct inspection, first on projectors $P_\psi$
and then on mixtures, this gives rise to the so-called 
Liouville-von Neumann equation on the state-space $\mathcal{S}(S)$:
\begin{equation}
\label{Liouv}
\partial_t\rho_t=-i\,\bigl[H,\,\rho_t\bigr]\ ,
\end{equation}
whose solution, with initial condition $\rho_{t=0}=\rho$, is 
\begin{equation}
\label{Liouv1}
\rho_t=U_t\,\rho\, U_{-t}\ ,\quad U_t={\rm e}^{-iH t}\ .
\end{equation}

Denoting by $\rho\mapsto\mathbb{U}_t[\rho]\equiv\rho_t$ the
dynamical map~(\ref{Liouv1}) and by 
\begin{equation}
\label{Liouv11}
\rho\mapsto \mathbb{L}_H[\rho]\equiv-i\,\bigl[H,\,\rho\bigr]\ ,
\end{equation}
the linear action of the generator on the left hand side of~(\ref{Liouv}),
the Schr\"odinger unitary dynamics amounts to exponentiation of $\mathbb{L}$:
\begin{equation}
\label{Liouv2}
\rho_t=\mathbb{U}_t[X]={\rm e}^{t\mathbb{L}_H}[\rho]=\sum_k\frac{t^k}{k!}
\underbrace{\mathbb{L}_H\circ \mathbb{L}_H\cdots\circ
\mathbb{L}_H}_{\mathbb{L}_H^k}[\rho]\ ,
\end{equation}
where $\circ$ means compositions of maps.
Therefore, the dynamical maps $\mathbb{U}_t$ form a one-parameter group of
linear maps on ${\cal S}(S)$:
$\mathbb{U}_t\circ\mathbb{U}_s=\mathbb{U}_{t+s}$ for all
$t,s\in\mathbb{R}$.
This fact mathematically describes the reversible character of the
unitary Schr\"odinger dynamics; in particular,
the dynamical maps $\mathbb{U}_t$ can be inverted, preserve the 
spectrum of all states $\rho$, leave  the von Neumann entropy
invariant and transform pure states into pure states:
\begin{equation}
\label{pure}
\rho^2=\rho\Longrightarrow\bigl(\mathbb{U}_t[\rho]\bigr)^2
=\mathbb{U}_t[\rho]\ .
\end{equation}
\bigskip

\noindent
{\bf Remark 2.2}\quad
{
One can pass from the Schr\"odinger to the Heisenberg time-evolution
through the definition of mean values~(\ref{meanv}) and the so-called
\textit{duality relation}
\begin{equation}
\label{Heis}
{\rm Tr}\Bigl[\mathbb{U}_t[\rho]\,X\Bigr]={\rm Tr}\Bigl[
\rho\,\mathbb{U}_t^\ast[X]\Bigr]\ ,
\end{equation}
which holds for all $\rho\in{\cal S}(S)$,  $X\in M_n(\mathbb{C})$ and defines
the \textit{dual linear map} $\mathbb{U}_t^\ast$.
The latter acts on $M_n(\mathbb{C})$ as
\begin{equation}
\label{Heis1}
X\mapsto\mathbb{U}_t^\ast[X]=U_{-t}\,X\,U_t={\rm e}^{-t\mathbb{L}_H}[X]\ .
\end{equation}
\hfill $\Box$
}
\bigskip

The unitary dynamics is not the only dynamical transformation affecting 
quantum states.
According to the postulates of quantum 
mechanics, if the state
of $S$ is $\vert\psi\rangle\langle\psi\vert$, upon measuring the 
(spectralized) observable 
$X=\sum_{i=1}^n\, x_i \vert x_i\rangle\langle x_i\vert$, then
\begin{itemize}
\item
the eigenvalues $x_i$ are
obtained with probabilities $\vert\langle x_i\vert\psi\rangle\vert^2$,
\item
if the measure gives $x_i$, then the post-measurement state
of $S$ is $\vert x_i\rangle\langle x_i\vert$.
\end{itemize}

It follows that, by repeating the measurement many times on copies of $S$ 
equally prepared in the pure state $\vert\psi\rangle\langle\psi\vert$
and collecting all the resulting post-measurement states, the outcome
is a physical mixture described by the density matrix
\begin{equation}
\label{red}
\sum_{i=1}^n\,\vert\langle x_i\vert\psi\rangle\vert^2\,
\vert x_i\rangle\langle x_i\vert=
\sum_{i=1}^n\,\vert x_i\rangle\langle x_i\vert\,\Bigl(
\vert\psi\rangle\langle\psi\vert\Bigr)\, \vert x_i\rangle\langle x_i\vert\ .
\end{equation}
Setting $P_i=\vert x_i\rangle\langle x_i\vert$ and
extending~(\ref{red}) linearly to any density matrix 
$\rho\in{\cal S}(S)$, one gets the following linear 
map on $\mathcal{S}(S)$:
\begin{equation}
\label{red1}
\rho\mapsto\mathbb{P}[\rho]=\sum_{i=1}^n\,P_i\rho\,P_i\ .
\end{equation}
This map is a mathematical description of the 
so-called \textit{wave-packet reduction}.

Contrary to the unitary dynamics $\mathbb{U}_t$, 
$\mathbb{P}$ transforms pure states into mixtures thus increasing 
their von Neumann entropy; the process it describes is sometimes identified
with a randomizing \textit{quantum noise}, with decohering properties.
By \textit{decoherence} it is meant the result of any dynamical
transformation that suppresses the phase-interferences present in a
linear superposition of vector states.
\bigskip

\noindent
{\bf Remark 2.3}\quad
{
The wave-packet reduction mechanism $\mathbb{P}$
effectively describes what happens to
$S$ when it is not closed but in interaction with an external specific
environment, in this case an apparatus measuring the observable $X$. 
Specifically, the wave packet reduction is the final
effect on the open system $S$ of its interaction with the macroscopic
environment. $\Box$
} 
\bigskip

In the following, we shall be concerned with open quantum systems $S$
immersed in an environment $E$.
In principle, the environment should consist of infinitely many
degrees of freedom and thus be addressed by means of the more
proper algebraic approach to quantum statistical mechanics;\cite{thirring,bratteli} 
however, 
for sake of clarity, we shall describe it by means of an infinite
dimensional Hilbert space $\mathfrak{H}$ and represent its states 
by density matrices $\rho_E$.%
\footnote{At a certain stage, one nevertheless has to perform the so-called
thermodynamic limit, in which volume and number of degrees of freedom
are let to infinity, while keeping the density finite.
In practical terms, this amounts to substituting
integrals for discrete summations, {\it e.g.} see \hbox{Example 3.3.}}

Subsystem $S$ and environment $E$ make a closed compound system $S+E$,
its Hilbert space being the tensor product 
$\mathbb{C}^n\otimes\mathfrak{H}$; the total system evolves reversibly 
according to a group of dynamical maps 
$\mathbb{U}_t^{S+E}=\exp(t\mathbb{L}_{S+E})$ that act on the
state-space $\mathcal{S}(S+E)$. 
The group is generated by formally exponentiating the commutator 
$\mathbb{L}_{S+E}[\,\rho_{S+E}\,]\equiv
-i\bigl[H_{S+E}\,,\,\rho_{S+E}\,\bigr]$ with respect to a total
Hamiltonian 
\begin{equation}
\label{global1}
H_{S+E}=H_S\otimes{\bf 1}_E\,+\,{\bf 1}_S\otimes H_E\,+\,\lambda\,H'\ ,
\end{equation}
where $\lambda$ is an adimensional coupling constant, $H'$ describes
the $S-E$ interaction, while $H_S$ and $H_E$
are Hamiltonian operators pertaining to $S$, respectively $E$ and
${\bf 1}_{S,E}$ are identity operators.
It follows that the generator decomposes as a sum of commutators,
\begin{equation}
\label{global2}
\mathbb{L}\equiv\mathbb{L}_{S+E}=\mathbb{L}_S\,+\,
\mathbb{L}_E\,+\,\lambda\,\mathbb{L}'\ ,
\end{equation}
where the subscripts $S,E$ identify which degrees of freedom are involved.

Given a state $\rho_{S+E}\in{\cal S}(S+E)$ of the compound closed
system, the statistical properties of the embedded subsystem $S$ are
described by a state $\rho_S\in\mathcal{S}(S)$ which
is obtained by the
\textit{partial trace} over the degrees of freedom of $E$:
\begin{equation}
\label{partrace}
\mathcal{S}(S+E)\ni\rho_{S+E}\mapsto
\rho_S\equiv{\rm Tr}_E(\rho_{S+E})=
\sum_j\,\langle\psi^E_j\vert\,\rho_{S+E}\,\vert\psi^E_j\rangle\ ,
\end{equation}
where $\{\vert\psi^E_j\rangle\}$ is any orthonormal
basis in $\mathfrak{H}$.
The right hand side of~(\ref{partrace}) belongs to $M_n(\mathbb{C})$
and can be easily checked to be positive  
and normalized according to~(\ref{posop}) and~(\ref{normal}), so that
$\rho_S\in\mathcal{S}(S)$.

Analogously, given any $\rho_S$ at time $t=0$, the state of $S$ at
any time $t$ is 
\begin{equation}
\label{partrace1}
\rho_S(t)={\rm Tr}_E\Bigl(\mathbb{U}_t^{S+E}[\rho_{S+E}]\Bigr)\ .
\end{equation}
This gives rise to a family of maps $\mathbb{G}_t$,
\begin{equation}
\label{partrace2}
\rho_S\mapsto\rho_S(t)=\mathbb{G}_t[\rho_S]\ ,
\end{equation}
which in general depend on $\rho_S$ and can not be extended to
the whole of the state-space.
If we ask that $\mathbb{G}_t$ preserve the convex structure of 
$\mathcal{S}(S)$, that is
\begin{equation}
\label{partrace3}
\mathbb{G}_t\Bigl[\sum_j\lambda_j\rho^j_S\Bigr]=
\sum_j\lambda_j\mathbb{G}_t\Bigl[\rho^j_S\Bigr]\ ,
\end{equation}
then the initial state of the
compound system must factorize:
$\rho_{S+E}=\rho_S\otimes\rho_E$, where $\rho_S\in{\cal S}(S)$
and $\rho_E$ is a fixed state of the environment.\cite{pechukas}

The factorized form of the initial state means in particular that open
system and environment are initially completely uncorrelated; though
not true in general, in
many interesting physical contexts such a condition is fully
consistent and gives rise to a family of dynamical
maps $\mathbb{G}_t$ depending on the environment reference state
$\rho_E$, but otherwise acting linearly on the state-space of $S$.%
\footnote{As already mentioned,
in presence of initial correlations between subsystem and
environment, a reduced dynamics that acts linearly on all
${\cal S}(S)$ can not in general be defined: 
its explicit form crucially
depends also on the initial reduced state ${\rm Tr}_E[\rho_{S+E}]$;
for further investigations on this issue, see 
Refs.[\refcite{royer1}, \refcite{royer2}, \refcite{petruccione}],
and references therein.}

The family of maps $\mathbb{G}_t$, $t\geq 0$, describes a
forward-in-time irreversible dynamics, for the partial trace breaks 
time-reversal; what the family lacks is a semigroup
composition law, since in general
$\mathbb{G}_t\circ\mathbb{G}_s\neq\mathbb{G}_{t+s}$ for $t,s\geq0$.
An equality would express the absence of cumulative memory effects, as such
it is expected to be a good approximation of the time-evolution of
open quantum systems only when their interaction with the environment is 
sufficiently weak or the environment time-correlations decay rapidly
with respect to time-variation of the system.

The technical procedures to eliminate memory effects and recover
semigroups of dynamical maps as reduced time-evolutions are known as
\textit{Markov approximations}.
These will be discussed in detail in the next Section; 
in the remaining part of the present Section we shall study 
the structure of $\mathbb{G}_t$ as a linear map, focusing on its
physical properties.

\subsection{Positivity and Complete Positivity}

As motivated before, we take an initial state of the compound system
$S+E$ in factorized form
$\rho_{S+E}=\rho_S\otimes \rho_E$.
By  performing the partial trace with respect to the orthonormal basis of
eigenvectors $\vert r^E_j\rangle$ of $\rho_E$ with corresponding
eigenvalues $r^E_j$, one obtains:\cite{alicki-lendi,kraus}
\begin{eqnarray}
\nonumber
\rho_S(t)={\rm Tr}_E\Bigl(\mathbb{U}_t^{S+E}[\rho_S\otimes\rho_E]\Bigr)&=&
\sum_{j,k}\,r^E_k\langle r^E_j\vert U^{S+E}_t\vert
r^E_k\,\rangle\,\rho_S\,\langle r^E_k\vert U^{S+E}_{-t}\vert r^E_j\rangle\\
\label{cp0}
&=&\sum_\alpha\, {\cal V}_\alpha(t)\,\rho_S\, {\cal V}_\alpha^\dagger(t)\ ,
\end{eqnarray}
where $\alpha$ is a double summation index and
\begin{equation}
{\cal V}_\alpha(t)\equiv \sqrt{r^E_k}\,\langle r^E_j\vert U^{S+E}_t\vert
r^E_k\rangle\ .
\end{equation}
The matrix elements 
$\langle r^E_j\vert U^{S+E}_t\vert r^E_k\rangle$ 
are elements of $M_n(\mathbb{C})$; further, from normalization
(${\rm Tr}(\rho_{S}(t))=1$), it follows that
\begin{equation}
\label{norm}
\sum_\alpha\,{\cal V}_\alpha^\dagger(t)\,{\cal V}_\alpha(t)={\bf 1}_n\ ,
\end{equation}
where ${\bf 1}_n\in M_n(\mathbb{C})$ denotes the identity matrix.

The duality relation~(\ref{Heis}) associates to
the linear map 
\begin{equation}
\label{cp1}
\rho_S\mapsto
\mathbb{G}_t[\rho_S]=\sum_\alpha\,{\cal V}_\alpha(t)\,\rho_S\,\,
{\cal V}^\dagger_\alpha(t)\ ,
\end{equation}
on the state-space ${\cal S}(S)$, the dual action 
\begin{equation}
\label{cp3}
X\mapsto
\mathbb{G}^\ast_t[X]=\sum_\alpha\,{\cal V}^\dagger_\alpha(t)\,X\,\,{\cal V}_\alpha(t)\  ,
\end{equation}
on the algebra of observables $M_n(\mathbb{C})$. 
The latter fulfils $\mathbb{G}^\ast_t[{\bf 1}_n]={\bf 1}_n$, a
property called \textit{unitality}.

The structure of both $\mathbb{G}_t$ and $\mathbb{G}^\ast_t$ is similar to 
the wave-packet reduction~(\ref{red1}), with the difference that
the ${\cal V}_\alpha(t)$ need not be orthogonal one-dimensional projectors as
the $P_i$.
By means of~(\ref{posop}), it is easy to see that these linear maps preserve
the positivity of operators: they belong to the class of
\textit{positive maps}.\cite{takesaki}
\bigskip

\noindent
{\bf Definition 2.2}\quad
{\it
A linear map 
$\Lambda:\,M_n(\mathbb{C})\mapsto M_n(\mathbb{C})$ is termed positive if 
it sends positive matrices into positive matrices, namely if 
\begin{equation}
\label{pos0}
0\leq X\,\mapsto\, \Lambda[X]\geq 0\ .
\end{equation}
}
\bigskip

\noindent
{\bf Remark 2.4}\quad
{
While positivity of hermitian matrices means positivity of their
eigenvalues, positivity of linear maps means their property of
transforming a matrix with positive eigenvalues into another such matrix.
Positive linear maps are sometimes more properly referred to as
positivity-preserving maps. $\Box$
}
\bigskip

It turns out that linear maps as $\mathbb{G}_t$, $\mathbb{G}^\ast_t$ and
$\mathbb{P}$ belong to a special
subclass of positive maps on $M_n(\mathbb{C})$: the so-called
\textit{completely positive} maps.\cite{alicki-fannes,takesaki}

\textit{Complete positivity} is a stronger property than positivity.
It concerns the possibility that the system $S$ be statistically
coupled to a so-called \textit{ancilla}, $A$, that is to a generic
remote and inert, finite-dimensional system.
If $A$ is $m$-dimensional, any linear map $\Lambda$ on $M_n(\mathbb{C})$
lifts to a map 
$\Lambda\otimes{\rm id}_A$ on the algebra $M_n(\mathbb{C})\otimes
M_m(\mathbb{C})$ of the compound system $S+A$, where ${\rm id}_A$
denotes the identical action on $M_m(\mathbb{C})$ leaving all
operators unaffected.
In the following the identical action will appear either as ${\rm
id}_A$ or as ${\rm id}_m$, explicitly indicating the dimensionality of the
ancillary system.

The physical interpretation of such a coupling is as follows:
the system $S$ of interest may have interacted in
the past with $A$ and become correlated (entangled) with it; afterwards and prior
to the start of the evolution of $S$, $S$ and $A$ has ceased to
interact so that what is left are the statistical correlations between them
incorporated in the state $\rho_{S+A}$.
When the evolution (or more in general any state change) of $S$ sets in, it has to be
considered as a transformation of the compound system $S+A$, which does
not affect the ancilla and is thus mathematically described by 
the \textit{lifting} $\Lambda\otimes{\rm id}_A$ of the map $\Lambda$.
\bigskip

\noindent
{\bf Definition 2.3}\quad
{\it
A linear map $\Lambda\, :\, M_n(\mathbb{C})\mapsto M_n(\mathbb{C})$ is
\textit{completely positive}, if and only if $\Lambda\otimes{\rm
id}_A$ is positive on $M_n(\mathbb{C})\otimes M_m(\mathbb{C})$ for all
$m\geq 1$, that is for all possible statistical couplings with
finite-dimensional ancillas.
}
\bigskip

When $m=1$, $S+A$ is $S$ itself as the
Hilbert space $\mathbb{C}^n\otimes \mathbb{C}$ is isomorphic to $\mathbb{C}^n$.
When $m\geq2$, all elements of $M_n(\mathbb{C})\otimes M_m(\mathbb{C})$ can
be written as $m\times m$ $M_n(\mathbb{C})$-valued matrices
$[X_{ij}]_{i,j=1}^m$, $X_{ij}\in M_n(\mathbb{C})$; therefore,
the definition concerns the linear actions 
$\Lambda\otimes{\rm id}_A[[X_{ij}]]=[\Lambda[X_{ij}]]$ and
selects among the positive linear maps $\Lambda$ 
those such that\cite{choi1}
$$
0\leq\pmatrix{X_{11}&\cdots&X_{1m}\cr
\cdot&\cdots&\cdot\cr
\cdot&\cdots&\cdot\cr
\cdot&\cdots&\cdot\cr
X_{m1}&\cdots&X_{mm}}\stackrel{\Lambda\otimes{\rm id}_A}{\longmapsto}
\pmatrix{\Lambda[X_{11}]&\cdots&\Lambda[X_{1m}]\cr
\cdot&\cdots&\cdot\cr
\cdot&\cdots&\cdot\cr
\cdot&\cdots&\cdot\cr
\Lambda[X_{m1}]&\cdots&\Lambda[X_{mm}]}\geq0\ .
$$

In the following Theorem, 
we recall two important results characterizing completely positive maps 
(see Refs.[\refcite{choi1}, \refcite{kraus}, \refcite{alicki-fannes}] 
for their proofs).
The first one states that, in order to
ascertain the complete positivity of 
$\Lambda: M_n(\mathbb{C})\mapsto M_n(\mathbb{C})$, 
it is not necessary to check all dimensions $m$, 
while the second one establishes that complete positive $\Lambda$ are
exactly those with the same structure as $\mathbb{U}_t$, $\mathbb{P}$, 
$\mathbb{G}_t$ and $\mathbb{G}^\ast_t$.
\bigskip

\noindent
{\bf Theorem 2.1}
\vspace{-.2cm}
{\it
\begin{itemize}
\item
A linear map $\Lambda\, :\, M_n(\mathbb{C})\mapsto M_n(\mathbb{C})$ is
\textit{completely positive}, if and only if $\Lambda\otimes{\rm
id}_n$ is positive on $M_n(\mathbb{C})\otimes M_n(\mathbb{C})$. 
\item
A linear map $\Lambda\, :\, M_n(\mathbb{C})\mapsto M_n(\mathbb{C})$ is
\textit{completely positive}, if and only if it is expressible in
\textit{Kraus-Stinespring form} 
\begin{equation}
\label{cp2}
\Lambda[X]=\sum_\alpha\,{\cal V}^\dagger_\alpha\,X\,{\cal V}_\alpha\ ,
\end{equation}
where ${\cal V}_\alpha\in M_n(\mathbb{C})$ are such that 
$\sum_\alpha {\cal V}_\alpha^\dagger {\cal V}_\alpha$ converges (to ${\bf 1}_n$ if
$\Lambda$ is unital).
\end{itemize}
}
\bigskip

\noindent
Suppose one has a linear map on $M_n(\mathbb{C})$ given in the form 
\begin{equation}
\label{cp4}
\Lambda[X]=\sum_{\alpha,\beta}\,C_{\alpha\beta}\,
{\cal W}_\alpha^\dagger\,X\,{\cal W}_\beta\ ,
\end{equation}
with ${\cal W}_\alpha\in M_n(\mathbb{C})$ and $C_{\alpha\beta}$ making a
hermitian matrix of coefficients such that
$\sum_{\alpha\beta}\,C_{\alpha\beta}\,{\cal W}^\dagger_\alpha {\cal W}_\beta$ converges.
By diagonalizing 
$[C_{\alpha\beta}]=U^\dagger\hbox{diag}(d_1,d_2,\ldots)U$, one
recovers~(\ref{cp2}) if and only if $[C_{\alpha\beta}]$
is positive definite and the eigenvalues $d_i$ positive; indeed, 
$C_{\alpha\beta}=\sum_j\,d_jU^*_{j\alpha}U_{j\beta}$ and this serves to
redefine ${\cal V}_j=\sum_\alpha\sqrt{d_j}U^*_{j\alpha}{\cal W}_\alpha$.
\bigskip

\noindent
{\bf Remark 2.5}\quad
{
If not all $d_j$ are positive, then the map can be separated into 
two sums, one over the positive and the other over the negative
eigenvalues; by extracting an overall minus sign, $\Lambda$ can then be
written as the difference of two completely positive maps $\Lambda_{1,2}$: 
$\Lambda= \Lambda_1 -\Lambda_2$.
The property of complete positivity fixes the structure
of the map by excluding the presence of a $\Lambda_2$, while that of
positivity is still far from being understood: no general
prescriptions on $\Lambda_{1,2}$ are known that ensure the positivity
of $\Lambda$. $\Box$
}
\bigskip

\noindent
Completely positive maps are a proper subset of all positive maps,
for not all positive maps $\Lambda$ lift to positive $\Lambda\otimes{\rm id}_A$, 
the foremost example being the transposition.
\bigskip

\noindent
{\bf Examples 2.1}
\begin{enumerate}
\item
In $M_2(\mathbb{C})$, let $\mathbb{T}_2:\pmatrix{a&b\cr c&d}\mapsto\pmatrix{a&c\cr b&d}$; 
the transposition map $\mathbb{T}_2$ does not alter the spectrum and therefore
is a positive linear map. Let 
$\vert 0\rangle=\pmatrix{1\cr0}$, $\vert 1\rangle=\pmatrix{0\cr1}$ be an
orthonormal basis in $\mathbb{C}^2$ and consider the vector
$$
\vert \Psi^{(2)}_+\rangle=\frac{1}{\sqrt{2}}\Bigl(\vert 0\rangle \otimes
\vert 0\rangle\ +\ \vert 1\rangle\otimes\vert1\rangle\Bigr)
\in \mathbb{C}^2\otimes\mathbb{C}^2\ .
$$
The corresponding projection 
$P^{(2)}_+=\vert\Psi^{(2)}_+\rangle\langle\Psi^{(2)}_+\vert\in M_2(\mathbb{C})
\otimes M_2(\mathbb{C})$,
\begin{eqnarray}
\nonumber
P^{(2)}_+
&=&\frac{1}{2}
\Bigl[\vert 0\rangle\langle 0\vert\otimes\vert0\rangle\langle0\vert+
\vert 1\rangle\langle1\vert\otimes\vert1\rangle\langle1\vert
+\vert0\rangle\langle1\vert\otimes\vert0\rangle\langle1\vert+
\vert1\rangle\langle0\vert\otimes\vert1\rangle\langle0\vert\Bigr]\\
\nonumber
&&\\
\label{sings}
&=&
\frac{1}{2}\pmatrix{1&0&0&1\cr0&0&0&0\cr0&0&0&0\cr1&0&0&1}\ ,
\end{eqnarray}
has eigenvalues $0,1$ and gets transformed by 
$\mathbb{T}_2\otimes{\rm id}_2$ into
\begin{eqnarray*}
\mathbb{T}_2\otimes{\rm id}_2[P^{(2)}_+]&=&
\frac{1}{2}
\Bigl(\vert 0\rangle\langle 0\vert\otimes\vert0\rangle\langle0\vert+
\vert 1\rangle\langle1\vert\otimes\vert1\rangle\langle1\vert\\
&&\hskip 4cm
+\vert1\rangle\langle0\vert\otimes\vert0\rangle\langle1\vert+
\vert0\rangle\langle1\vert\otimes\vert1\rangle\langle0\vert\Bigr)\\
&&\\
&=&\frac{1}{2}\pmatrix{1&0&0&0\cr0&0&1&0\cr0&1&0&0\cr0&0&0&1}\ ,
\end{eqnarray*}
which has eigenvalues $\pm1/2$ and is no longer a positive matrix.
Therefore, $\mathbb{T}_2$ is positive, but not completely positive.
\hfill\break\hfill
\item
The vector state $\vert\Psi^{(2)}_+\rangle$ generalizes 
to the totally symmetric state
\begin{equation}
\label{symst}
\vert\Psi^{(n)}_+\rangle\equiv\frac{1}{\sqrt{n}}\sum_{j=1}^n\,\vert
j\rangle\otimes\vert j\rangle\in\mathbb{C}^n\otimes\mathbb{C}^n\ ,
\end{equation}
where $\vert j\rangle$, $j=1,2,\ldots,n$, is a reference orthonormal
basis in $\mathbb{C}^n$.
Then, with
$P^{(n)}_+=\vert\Psi^{(n)}_+\rangle\langle\Psi^{(n)}_+\vert$, 
and $\mathbb{T}_n$ the transposition on $M_n(\mathbb{C})$,
$$
\mathbb{T}_n\otimes{\rm id}_n[P^{(n)}_+]=\frac{1}{n}\sum_{j,k=1}^n\,
\vert k\rangle\langle j\vert\otimes\vert j\rangle\langle k\vert\ ,
$$
and one checks that the sum in the right hand side of the second
equality is the \textit{flip operator} $V$ on
$\mathbb{C}^n\otimes\mathbb{C}^n$, such that
$V\bigl(\vert\psi\rangle\otimes\vert\phi\rangle\bigr)=
\vert\phi\rangle\otimes\vert\psi\rangle$,
$\forall\, |\psi\rangle, |\phi\rangle\in\mathbb{C}^n$.
It follows that $V$ has eigenvalue $-1$ on any anti-symmetric state
of $\mathbb{C}^n\otimes\mathbb{C}^n$ and thus the transposition is never
completely positive in any dimension. $\Box$
\end{enumerate}
\bigskip

\noindent
{\bf Remark 2.6}\quad
{
Transposition is associated with time-reversal;\cite{thirring} 
as its lifting $\mathbb{T}_n\otimes{\rm id}_n$ sends the state 
$P^{(n)}_+$ out of the space of states, it can not correspond to a
physical operation.
This means that time-reversal 
has to be applied globally and not locally, that is not on 
subsystems alone. $\Box$
}
\bigskip

\noindent
{\bf Examples 2.2}\quad
Let $S$ be a quantum system described by a $2$-dimensional Hilbert
space $\mathbb{C}^2$; the algebra of observables $M_2(\mathbb{C})$ is
linearly spanned by the identity matrix and the three Pauli matrices
$$
\sigma_0=\pmatrix{1&0\cr0&1}\ ,\quad
\sigma_1=\pmatrix{0&1\cr1&0}\ ,\quad
\sigma_2=\pmatrix{0&-i\cr i&0}\ ,\quad
\sigma_3=\pmatrix{1&0\cr0&-1}\ ,
$$
namely, $X=\sum_{\mu=0}^3\,X_\mu\,\sigma_\mu$ with 
$X_\mu={1\over 2}{\rm Tr}(X\sigma_\mu)$.
The linear map 
$$
\mathbb{S}_\alpha:M_2(\mathbb{C})\mapsto M_2(\mathbb{C})\ ,\quad
X\mapsto \mathbb{S}_\alpha[X]=\sigma_\alpha X\,\sigma_\alpha\ ,
$$ 
are written in the Kraus-Stinespring form~(\ref{cp2}), and are thus
completely positive (although not unital).
\begin{enumerate}
\item
Consider the following two maps,
\begin{equation}
\label{idtransp}
X\mapsto\frac{1}{2}\sum_{\alpha=0}^3\mathbb{S}_\alpha[X]\ , \quad
X\mapsto \frac{1}{2}\sum_{\alpha=0}^3\varepsilon_\alpha
\mathbb{S}_\alpha[X]\ ,
\end{equation}
where $\varepsilon_\alpha=1$ when $\alpha\neq 2$, whereas
$\varepsilon_2=-1$.
Using the algebraic relations 
$$
\sigma_\alpha\,\sigma_\beta\,\sigma_\alpha=
\eta_{\alpha\beta}\,\sigma_\beta\ ,\quad\hbox{where}\quad
\eta_{\alpha\beta}=\hbox{\begin{tabular}{c||r|r|r|r}
$\beta\backslash\alpha$&0&1&2&3\\ \hline\hline
0&\phantom{-}1&1&1&1\\ \hline
1&1&1&-1&-1\\ \hline
2&1&-1&1&-1\\ \hline
3&1&-1&-1&1\\ 
\end{tabular}}\ ,
$$
it can be checked that the first map amounts to 
$X\mapsto{\rm Tr}(X)\,\sigma_0$ and is completely positive since it is
in Kraus-Stinespring form,
while the second one is the transposition which indeed is such that
$\mathbb{T}_2[\sigma_\alpha]=\varepsilon_\alpha\sigma_\alpha$, 
therefore positive, but not completely positive.
\item
Given $\Lambda:M_2(\mathbb{C})\mapsto M_2(\mathbb{C})$ in the
form~(\ref{cp4}), by expanding the ${\cal W}_\alpha$ along the Pauli matrices 
$\sigma_\alpha$ it follows that the map can be rewritten as
\begin{equation}
\label{cp5}
X\mapsto\Lambda[X]=\sum_{\alpha,\beta=0}^3 C_{\alpha\beta}\,
\sigma_\alpha\,X\,\sigma_\beta\ ,
\end{equation}
and is completely positive if and only if the matrix
$[C_{\alpha\beta}]$ is positive definite. $\Box$
\end{enumerate}
\bigskip

\noindent
The different structure of positive and
completely positive maps on $M_n(\mathbb{C})$ is better exposed by
means of the state $P^{(n)}_+$ of Example 2.1.2; one has:\cite{horodecki}
\bigskip

\vbox{
\noindent
{\bf Theorem 2.2}\quad
{\it
\begin{itemize}
\item
$\Lambda:M_n(\mathbb{C})\mapsto M_n(\mathbb{C})$ is positive if and only if
\begin{equation}
\label{pos}
\langle\psi\otimes\phi\vert\,\Lambda\otimes{\rm id}_n[P^{(n)}_+]\,
\vert\psi\otimes\phi\rangle\geq 0\ ,\quad\forall
\psi,\phi\in\mathbb{C}^n\ .
\end{equation}
\item
$\Lambda:M_n(\mathbb{C})\mapsto M_n(\mathbb{C})$ is completely positive if
and only if 
\begin{equation}
\label{cpos}
\langle\Psi\vert\,\Lambda\otimes{\rm id}_n[P^{(n)}_+]\,
\vert\Psi\rangle\geq 0\ ,\quad\forall
\Psi\in\mathbb{C}^n\otimes\mathbb{C}^n\ .
\end{equation}
\end{itemize}
}}
\bigskip

\noindent
The request in~(\ref{cpos}) is that the matrix 
$\Lambda\otimes{\rm id}_n[P^{(n)}_+]\in M_{n^2}(\mathbb{C})$ be
positive (see Definition 2.1), while the request in~(\ref{pos})
involves mean values with respect to product vector states, only.
These latter do not exhaust the Hilbert space, as there are 
non-product states as $\vert\Psi^{(n)}_+\rangle$.

\subsection{Complete Positivity and Quantum Entanglement} 
\label{subs2.3}

The projector $P^{(n)}_+$ is the typical instance of an \textit{entangled
pure state}, the general notion of entanglement as opposed to that of
\textit{separability} being as follows.
\bigskip

\noindent
{\bf Definition 2.4}\quad
{\it
Let ${\cal S}(S_1)$ and ${\cal S}(S_2)$ be the state-spaces of two
$n$-level quantum systems $S_{1,2}$; within the state-space 
${\cal S}(S_1+S_2)$
one distinguishes the (convex) subset of \textit{separable states} of the
form
\begin{equation}
\label{seps}
\rho_{S_1+S_2}=\sum_{ij}\lambda_{ij}\ \rho^1_i\otimes \rho^2_j\ ,\
\lambda_{ij}\geq0\ ,\ \sum_{ij}\lambda_{ij}=1\ ,\ \rho^{1,2}_{i,j}
\in\mathcal{S}(S_{1,2})\ .
\end{equation}
All those states which can not be
written as linear convex combinations of product states as in~(\ref{seps})
are called entangled.}
\bigskip

\vbox{
\noindent
{\bf Remarks 2.7}
{
\begin{enumerate}
\item
The first two terms in the first line 
of~(\ref{sings}) represent separable states: it is the presence of the
remaining two interference terms that makes $P^{(2)}_+$ entangled.
\item
It is often stated that while separable states only carry the classical 
correlations embodied in the weights $\lambda_{ij}$ in~(\ref{seps}),
entangled states carry instead non-classical
correlations, that is they are characterized by statistical properties
which are not describable by classical probability theory.
This is best seen by considering the projector 
$P^{(n)}_+=\vert\Psi^{(n)}_+\rangle\langle\Psi^{(n)}_+\vert$, its von Neumann
entropy $S(P^{(n)}_+)=0$.
The partial trace of $P^{(n)}_+$ over either the first or the second
system in $S+S$ gets the totally mixed state 
$\tau_n={\bf 1}_n/n$.
This state has maximal entropy $S(\tau_n)=\log n$,
despite $S$ being a subsytem of $S+S$, which has zero entropy;
instead, in classical probability theory, knowledge of the whole
means having full information on all its parts. $\Box$
\end{enumerate}
}
}
\bigskip

At first glance one may think that
the positivity of linear maps $\Lambda$ on $\mathcal{S}(S)$ may 
be sufficient to guarantee a consistent description of
all possible physical transformations of the states
of $S$: after all, if a map $\Lambda$ preserves the trace and the positivity of
the spectrum of all density matrices, then it maps the state-space
${\cal S}(S)$ into itself as physical transformations surely do.
In fact, a minimal consistency request is 
\medskip

\centerline{
\begin{minipage}[t]{12cm}
\textsf{\noindent
Physically consistent transformations on ${\cal S}(S)$ 
must preserve the
interpretation of the eigenvalues of all density matrices as
probabilities and thus respect the positivity of their spectra.
}
\end{minipage}
}
\vspace{.3cm}

One can now appreciate the important physical
consequence of Theorem 2.2; indeed, from~(\ref{pos}) and~(\ref{cpos})
it turns out that if 
$\Lambda:M_n(\mathbb{C})\mapsto M_n(\mathbb{C})$ is positive, but not
completely positive, then $\Lambda\otimes{\rm id}_n[P^{(n)}_+]$ 
{\it is not} a positive matrix.

Given an $n$-dimensional ancilla $A$, statistically coupled to $S$,
operating the transformation 
$\Lambda$ on ${\cal S}(S)$ amounts to act with 
$\Lambda\otimes{\rm id}_A$ on the state-space
${\cal S}(S+A)$ of the compound system $S+A$.
Now, among the states of $S+A$, the totally symmetric projector
$P^{(n)}_+$ is a physically plausible initial state, which is not
transformed into a state by $\Lambda\otimes{\rm id}_A$.
For this to be true, $\Lambda$ has to be completely positive.
Positivity alone suffices for physical consistency on $S$, but not on $S+A$;
in other words, positive maps can not properly describe physical 
transformations, only completely positive maps may do.  Hence,
\medskip

\centerline{
\begin{minipage}[t]{12cm}
\textsf{\noindent
Complete positivity of maps is necessary to ensure
their physical consistency against possible entanglement with ancillas}
\end{minipage}
}
\bigskip

\noindent
{\bf Remark 2.8}\quad
{
Admittedly, though logically stringent, the explanation of why  
consistent state-transformations must be completely positive is scarcely
appealing from a physical point of view.
Indeed, the ancilla system $A$ is remote from $S$ and inert; a more
concrete scenario will be offered in Section~\ref{subs3.1} when we deal
with dissipative time-evolutions.
On the contrary, couplings to ancillas are natural tools
in quantum information, although, in this case, one deals
with quantum channels and not with continuous time-evolutions. $\Box$
}
\bigskip

If positive maps do not consistently describe physical
transformations, they play however a fundamental role in the detection
of entangled states.  Due to the recent fast advances in quantum
information theory, entanglement has turned from a
quantum riddle with epistemological overtones into a practical
physical resource that must be detected, quantified and
manipulated.  For instance, bipartite entangled pure states as $P^{(2)}_+$
are essential ingredients in quantum cryptography, quantum
teleportation and quantum computation.\cite{nielsen}

\subsubsection{Entanglement Detection}
\label{2.2.1}

From Definition 2.4, entangled states can generically be
mixed.
While, it is easy to spot entangled pure states, it is much harder
in the case of density matrices.
\bigskip

\noindent
{\bf Example 2.3}
As a first simple indication of these difficulties, consider the
totally depolarized state of two $2$-dimensional systems $\tau_2={\bf 1}_4/4$.
The identity matrix can be written in terms of the projectors onto the
orthonormal basis of the so-called \textit{Bell states},
$$
\tau_2=\frac{1}{4}\Bigl(
P^{(2)}_+\,+\,P^{(2)}_-\,+\,Q^{(2)}_+\,+\,Q^{(2)}_-\Bigr)\ ,
$$
where $P^{(2)}_+$ is as in Example 2.1.1 while 
$P^{(2)}_-=\vert\Psi^{(2)}_-\rangle\langle\Psi^{(2)}_-\vert$,
$Q^{(2)}_\pm=\vert\Phi^{(2)}_\pm\rangle\langle\Phi^{(2)}_\pm\vert$ and
$$
\vert\Psi^{(2)}_-\rangle=\frac{1}{\sqrt{2}}\Bigl(
\vert 00\rangle-\vert11\rangle\Bigr)\ ,\
\vert\Phi^{(2)}_\pm\rangle=\frac{1}{\sqrt{2}}\Bigl(
\vert 01\rangle\pm\vert10\rangle\Bigr)\ .
$$
As a combination of entangled projectors, $\tau_2$ may be deemed
entangled. However, mixtures of entangled states have less
entanglement than their constituents; indeed,
$$
\tau_2=\frac{1}{4}\Bigl(
\vert00\rangle\langle00\vert\,+\,\vert01\rangle\langle01\vert\,+\,
\vert10\rangle\langle10\vert\,+\,\vert11\rangle\langle11\vert\Bigr)\ ,
$$
and thus, as a combination of separable states, $\tau_2$ is indeed separable. 
$\Box$
\bigskip

A substantial help in detecting entanglement comes from the
transposition as defined in Examples 2.1, where one looked at the effects
of the action of the transposition on only one of the factors of a
bipartite system $S+S$, the so-called \textit{partial transposition}
$\mathbb{T}_n\otimes{\rm id}_n$.
Also, based on the fact that separable states make a closed convex subset of 
the state-space which can be geometrically separated from any
given entangled state by a suitable (hyper) plane, 
one deduces the so-called \textit{Peres-Horodecki criterion}
for separability.\cite{horodecki}
\bigskip

\noindent
{\bf Theorem 2.3}\quad
{\it
Let $S$ be an $n$-dimensional system. Then,
\begin{itemize}
\item
a state $\rho\in{\cal S}(S+S)$ is entangled if it does not remain
positive under partial transposition
$\mathbb{T}_n\otimes{\rm id}_n$.\cite{peres1}
\end{itemize}
\noindent
More in general: 
\begin{itemize}
\item
a state $\rho\in{\cal S}(S+S)$ is entangled only if 
$\exists$ $\Lambda:M_n(\mathbb{C})\mapsto M_n(\mathbb{C})$ positive such that
$\Lambda\otimes{\rm id}_n[\rho]$ is not positive.\cite{horodecki1}
\end{itemize}
}
\bigskip

\noindent
The sufficient condition easily follows from the fact that the action
of any positive map $\Lambda$, lifted on separable states $\rho_{sep}$
as in (\ref{seps}),
\begin{equation} 
\label{seppos}
\Lambda\otimes{\rm id}_n[\rho_{sep}]=
\sum_{i,j}\,\lambda_{ij}\,\Lambda[\rho^1_i]\otimes\rho^2_j\ ,
\end{equation}
keeps their positivity, for $\Lambda$ keeps the positivity of all
$\rho^1\in{\cal S}(S_1)$ by assumption.
Therefore, if a positive map like the transposition, is such that once lifted 
it does not preserve the positivity of $\rho\in{\cal S}(S+S)$ then the 
state must be entangled: such a positive map is called an
\textit{entanglement detector} for $\rho$.

Of course, entanglement detectors can not be completely positive maps
since, according to Definition 2.3, they are exactly those positive
maps on $M_n(\mathbb{C})$ such that, once lifted to act on 
$M_n(\mathbb{C})\otimes M_n(\mathbb{C})$, they remain positive.

In line of principle, the transposition is a good entanglement
detector for some states and not for others; one may thus suspect that, in
general, all positive maps have to be checked in order to detect the
entanglement of a generic $\rho$.
Fortunately, in the case of a bipartite system consisting of two
$2$-level systems, or more in general of one 2-level and one 3-level system,
the property of not remaining positive under partial transposition
is also a necessary condition for their states to be entangled. 
In other words,\cite{woronovicz}
\bigskip

\noindent
{\bf Theorem 2.4}\quad
{\it
Consider a bipartite system $S_1+S_2$, with $S_1$ a 2-level system
and $S_2$ either a 2-level or a 3-level system; 
a state $\rho\in{\cal S}(S_1+S_2)$ is
entangled if and only if $\mathbb{T}_2\otimes{\rm id}_2[\rho]$,
respectively $\mathbb{T}_2\otimes{\rm id}_3[\rho]$, 
is not positive. 
}
\bigskip

\noindent
{\bf Example 2.4}\quad
Consider the $n^2\times n^2$ matrix:
$$
\rho_F=\alpha\,{\bf 1}_n\otimes{\bf 1}_n\,+\,\beta\, V\in
M_n(\mathbb{C})\otimes M_n(\mathbb{C})\ ,\quad
\alpha,\beta\in\mathbb{R}\ ,
$$
where $V$ is the flip operator introduced in Example 2.1.2,
\begin{equation}
\label{Ws0}
V=n\,\mathbb{T}_n\otimes{\rm id}_n[P^{(n)}_+]\ ,
\end{equation}
and the parameter $F$ is defined by 
\begin{equation}
\label{Ws1}
F\equiv{\rm Tr}\bigl[\rho_F\,V\bigr]=\alpha\, n\,+\,\beta\, n^2\ .
\end{equation}
The second equality follows from the fact that 
$V^2={\bf 1}_n\otimes{\bf 1}_n$ and that, by choosing a basis 
$\vert i,j\rangle=\vert i\rangle\otimes\vert
j\rangle\in\mathbb{C}^n\otimes\mathbb{C}^n$, then
$V\vert i,j\rangle=\vert j,i\rangle$.
We want $\rho_F$ to be a density matrix.
Since $V$ has eigenvalues $\pm1$, $\rho_F$ has eigenvalues 
$\alpha\pm\beta$, positivity and normalization are guaranteed by  
\begin{equation}
\label{Ws2}
-\alpha\leq\beta\leq \alpha\ ,\quad
{\rm Tr}(\rho_F)=\alpha\,n^2\,+\,\beta\, n\,=\,1\ .
\end{equation}
Conditions (\ref{Ws1}) and (\ref{Ws2}) give
$$
\alpha=\frac{n-F}{n(n^2-1)}\ ,\quad\beta=\frac{n\,F\,-1}{n(n^2-1)}\
,\quad -1\leq F\leq 1\ .
$$
The states of the one-parameter family
\begin{equation}
\label{Ws}
\rho_F\,=\, \frac{n-F}{n(n^2-1)}\,{\bf 1}_n\otimes{\bf 1}_n\,+\,
\frac{n\,F\,-1}{n(n^2-1)}\, V\ ,
\end{equation}
are called \textit{Werner states}.\cite{werner} They are
all and the only states that commute with all unitary transformation on
$\mathbb{C}^n\otimes\mathbb{C}^n$ of
the form $U\otimes U$ with $U$ any unitary operator on $\mathbb{C}^n$.
By performing the partial transposition on $\rho_F$, using~(\ref{Ws0})
one gets
$$
\mathbb{T}_n\otimes{\rm id}_n[\rho_F]=\, 
\frac{n-F}{n(n^2-1)}\,{\bf 1}_n\otimes{\bf 1}_n\,+\,
\frac{n\,F\,-1}{(n^2-1)}\, P^{(n)}_+\ ,
$$
with an $(n^2-1)$-degenerate eigenvalue $\displaystyle\frac{n-F}{n(n^2-1)}$ 
which is never negative,
and a nondegenerate eigenvalue
$$
\frac{n-F}{n(n^2-1)}\,+\,\frac{n\,F\,-1}{(n^2-1)}\,=\,
\frac{F}{n}\ ,
$$
which is negative when $F<0$.
Therefore, in agreement with the preceding discussion, when $F<0$, the
Werner states are necessarily entangled, otherwise the partial
transposition could not spoil the positivity of their spectrum.

It turns out that $F<0$ is not only sufficient, but also necessary for
$\rho_F$ to be entangled:\cite{werner} in other words, $\rho_F$ is separable if and
only if $F\geq0$.
Since $F\geq 0$ means remaining positive under partial transposition,
the latter is actually an exhaustive entanglement-detector relative to 
the family of Werner states also in dimension larger than two. $\Box$
\bigskip

In the case of bipartite $2$-dimensional systems, there exists a 
quantitative method to determine whether a state
$\rho\in\mathcal{S}(S+S)$ is entangled or not, based on the notion of
\textit{concurrence}.\cite{wootters1,wootters2,wootters3}
This is defined as follows: with the tensor products of Pauli matrices 
$\sigma_2\otimes\sigma_2$, one constructs 
\begin{equation}
\label{conc0}
R\equiv
\rho\,\sigma_2\otimes\sigma_2\,\rho^*\,\sigma_2\otimes\sigma_2\ ,
\end{equation}
where $\rho^*$ denotes $\rho$ with complex-conjugated entries.
The matrix $R$ turns out to have positive eigenvalues; 
let $R_i$, $i=1,2,3,4$, be the positive
square roots of its eigenvalues in decreasing order.
\bigskip

\noindent
{\bf Theorem 2.5}\quad
{\it
A state $\rho\in\mathcal{S}(S+S)$, with $S$ two-dimensional, is entangled
if and only if its concurrence
\begin{equation}
\label{conc1}
\mathcal{C}(\rho)\equiv\max\Bigl\{R_1\,-\,R_2\,-\,R_3\,-\,R_4,\, 0\Bigr\}\ ,
\end{equation}
is strictly positive.}
\bigskip

\noindent
{\bf Example 2.5}\quad
We consider the family of Werner states studied in the previous
Example and set $n=2$.
Using the Bell states introduced in Example 2.3, the flip operator
explicitly reads
$$
V=P^{(2)}_+\,+\,P^{(2)}_-\,+Q^{(2)}_+\,-\,Q^{(2)}_-\,=\,
\pmatrix{1&0&0&0\cr0&0&1&0\cr0&1&0&0\cr0&0&0&1}\ ,
$$
whence
$$
\rho_F=\frac{1}{6}\pmatrix{1+F&0&0&0\cr0&2-F&2F-1&0\cr
0&2F-1&2-F&0\cr0&0&0&1+F}\ .
\nonumber
$$
Further, one cheks that $\sigma_2\otimes\sigma_2\, \rho_F\, \sigma_2\otimes\sigma_2=\rho_F$;
then (\ref{conc0}) gives $R=\rho_F^2$; therefore, the eigenvalues $R_i$ are
$(1+F)/6$ 
(three times degenerate)
and $(1-F)/2$, which are non-negative, since $-1\leq F\leq 1$; thus, the difference
$R_1\,-\,R_2\,-\,R_3\,-\,R_4$ in~(\ref{conc1}) can assume only two
expressions: $-F$ and $(F-2)/3$.
As a consequence, the concurrence is
$\mathcal{C}(\rho_F)=-F>0$, if and only if $F<0$, which,
as already seen in the previous Example, is a
necessary and sufficient condition for $\rho_F$ to be entangled.~$\Box$
\bigskip

\section{Reduced Dynamics and Markov Approximations}
\label{sec3}

According to the previous considerations, the completely positive maps 
$\mathbb{G}_t$ in~(\ref{cp1})
constitute a physically consistent description of the
irreversible dynamics of a system $S$ in interaction with its environment $E$.
The only necessary assumption is
that the initial state of $S+E$ do not carry either classical or
quantum correlations and be of the form $\rho_S\otimes\rho_E$.

However, the dissipative and noisy effects due to $E$ are 
hidden within the operators ${\cal V}_\alpha(t)$ and quite
difficult to be read off;
in particular, the one-parameter family of maps
$\mathbb{G}_t$ contains memory effects.
On the other hand, if the interaction between $S$
and $E$ is sufficiently weak, one expects that, on a typical
time-scale, the dynamics of $S$ might be disentangled from that of the
total system and efficiently described by a one-parameter semigroup of maps 
$\gamma_t$, $t\geq 0$, satisfying the  forward in time
composition law $\gamma_t\circ\gamma_s=\gamma_{t+s}$, $t,s\geq 0$.

The memory effects present in $\mathbb{G}_t$ are best revealed by
writing the formal integro-differential evolution equation of which
the maps $\mathbb{G}_t$ are solutions.
Its standard derivation is via the so-called projection
technique%
\footnote{Originally introduced in Refs.[\refcite{nakajima},
\refcite{zwanzig}], it is widely used in non-equilibrium
statistical mechanics.\cite{kubo}}
which we shall review in some detail by 
presenting and comparing several Markov approximations, leading to 
\textit{master equations} of the form
\begin{equation}
\label{masteq1}
\partial_t\rho_S(t)=\,(\mathbb{L}_H\,+\,\mathbb{D})[\rho_S(t)]\ ,
\end{equation}
where $\mathbb{L}_H$ acts on the state-space $\mathcal{S}(S)$ as ($-i$ times) 
the commutator with an effective Hamiltonian $H=H^\dagger\in M_n(\mathbb{C})$
as in~(\ref{Liouv11}), whereas 
$\mathbb{D}$ is a linear operator on ${\cal S}(S)$, not in
the form of a commutator, that 
effectively accounts for the dissipative and noisy effects due to $E$.

The solutions $\gamma_t$, $t\geq 0$, to~(\ref{masteq1}) will describe
the \textit{reduced dynamics} of $S$ as a semigroup of linear maps on 
${\cal S}(S)$, obtained by exponentiating the generator:
\begin{equation}
\label{masteq2}
\gamma_t=\exp\Bigl[t\,\bigl(\mathbb{L}_H+\mathbb{D}\bigr)\Bigr]\ .
\end{equation}
\medskip

The arguments developed in the previous sections about positivity and complete
positivity lead to the conclusion that the latter property is necessary to
ensure physical consistency against couplings to ancillas.
Therefore, one can
state in full generality that
\medskip

\centerline{
\begin{minipage}[t]{12cm}
\textsf{\noindent
any physically consistent Markovian approximation must yield
semigroups $\gamma_t$ consisting of completely positive maps.
}
\end{minipage}
}
\bigskip 

In the following, we shall not abide by this request and expose the
inconsistencies thereby arising; before this, however, we discuss  
the general form that time-continuous semigroups of completely positive
maps must have.

\subsection{Abstract Form of Generators of Dynamical Semigroups}
\label{sub3.2.1}

Semigroups of the form~(\ref{masteq2}) fulfill
 $\lim_{t\to0}\gamma_t={\rm id}$; more precisely
$$
\lim_{t\to 0}\|\gamma_t[\rho]-\rho\|_1=0\ , \forall 
\rho\in\mathcal{S}(S)\ ,
$$ 
where $\|X\|_1\equiv{\rm Tr}\sqrt{X^\dagger X}$, $X\in M_n(\mathbb{C})$.

Time-continuity of a semigroup of maps $\gamma_t$ guarantees the
existence of a generator and an exponential structure as
in~(\ref{masteq2}).\cite{alicki-fannes}
We now proceed by imposing one by one those requests that are
deemed necessary to physical 
consistency.\cite{alicki-lendi,gorini1,spohn}
 
The first constraint is that the hermiticity of density matrices be
preserved; we shall also ask for
trace preservation.
Physically speaking, this means that the
overall probability is constant; in other words, we shall not be
concerned with phenomena like particle decays that are characterized
by loss of probability.%
\footnote{These can nevertheless be included in the formalism:
for instance, see Refs.[\refcite{alicki-lendi}, \refcite{bf5}] 
and \hbox{Remark 3.3.3} below.}

Interestingly, hermiticity and probability preservation suffice to partially
fix the form of the generator.\cite{gorini2}
\bigskip

\noindent
{\bf Remark 3.1}\quad
{
Instead of directly referring to transformations on the state-space, 
the following theorem, as present in the literature, deals with
semigroups $\gamma^*_t$ 
acting on the algebra of observables $M_n(\mathbb{C})$
($*$ denotes dual transformations on observables,
namely in the Heisenberg picture).
Hermiticity and trace preservation correspond to unitality, 
$\gamma^*_t[1]=1$, and 
$\bigl(\gamma^*_t[X]\bigr)^\dagger=\gamma^*_t[X^\dagger]$ for all
$X\in M_n(\mathbb{C})$.
Time-continuity is given with respect to
the norm-topology:
$\lim_{t\to0}\|\gamma^*_t[X]-X\|=0$, for all \hbox{$X\in M_n(\mathbb{C})$,}
where 
$$
\|X\|=\sup\Bigl\{\sqrt{\langle\psi\vert\,X^\dagger\,X\,\vert\psi\rangle}\
,\ \|\psi\|=1\,\Bigr\}\ .
$$
Upon passing from the Heisenberg to the Schr\"odinger picture, one gets
semigroups acting on states by means of the duality relation~(\ref{Heis}). $\Box$
}
\bigskip

\noindent
{\bf Theorem 3.1}\quad{\it
\label{qdsth}
Let $\gamma^*_t:M_n(\mathbb{C})\mapsto M_n(\mathbb{C})$, $t\geq0$, 
form a time-continuous semigroup of unital, hermiticity-preserving
linear maps.
Then, the semigroup has the form
$\gamma^*_t=\exp(t(\mathbb{L}^*_H+\mathbb{D}^*))$ with
generator consisting of
\begin{eqnarray}
\label{GKF1}
\mathbb{L}^*_H[X]&=& i\bigl[H\,,\,X\bigr]\ ,\\
\label{GKF2}
\mathbb{D}^*[X]&=&\sum_{i,j=1}^{n^2-1}\,
C_{ij}\biggl(F_i\,X\,F^\dagger_j\,-\,
\frac{1}{2}\Bigl\{F_i F^\dagger_j\,,\,X\Bigr\}\biggr)\ ,
\end{eqnarray}
where the matrix of coefficient $C_{ij}$ (\textit{Kossakowski matrix}) 
is hermitian, $H=H^\dagger$, and 
the $F_j$ are such that $F_{n^2}={\bf 1}_n/\sqrt{n}$ and 
${\rm Tr}(F^\dagger_j\,F_k)=\delta_{jk}$, $0\leq j,k\leq n^2$,
while $\{\ ,\, \}$ represents anticommutation. 
}
\bigskip

The third constraint on $\gamma_t^*$ is that they be positive maps,
namely that they transform positive matrices into positive matrices.
Indeed, as a consequence of the duality relation~(\ref{Heis}),  
the $\gamma^*_t$ are positive maps if and only if their dual maps
preserve the positivity of states $\rho\in\mathcal{S}(S)$; more precisely
$$
0\leq\,{\rm Tr}\Bigl(\,\gamma_t[\rho]\,X\,\Bigr)\Leftrightarrow
{\rm Tr}\Bigl(\,\rho\,\gamma_t^*[X]\,\Bigr)\,\geq 0\ .
$$

It has been stressed that positivity preservation is necessary for 
physical consistency, but it is not sufficient against 
couplings with generic ancillas: for that the stronger request of
complete positivity is compulsory.

The Hamiltonian contribution to the generator
guarantees complete positivity since it gives rise to the standard
unitary evolution $\mathbb{U}_t$ that turns out to be automatically 
of the simplest Kraus-Stinespring form.
Evidently, positivity and complete positivity both depend on the
properties of the Kossakowski matrix $C_{ij}$.   
\bigskip

\noindent
{\bf Remark 3.2}\quad
{
Asking for positivity preservation results in quite an intricate
algebraic problem; in fact, {\it no general necessary
conditions}, but only sufficient ones are so far available  
on the coefficient $C_{ij}$ such that they give rise to
positivity preserving semigroups $\gamma_t$ (for more details,
see Refs.[\refcite{kossakowski1}-\refcite{bfp2}]).
On the contrary, the condition that the $\gamma_t$ be completely
positive is far more stringent. $\Box$
}
\bigskip

\noindent
{\bf Theorem 3.2}\quad
{\it
The semigroup $\{\gamma_t\}_{t\geq 0}$ consist of completely positive
maps if and only if the Kossakowski matrix is positive definite.
}
\bigskip

\vbox{
\noindent
{\bf Remarks 3.3}
\vspace{-.2cm}
{
\label{qdsrem}
\begin{enumerate}
\item
The proof of Theorem 3.2 was given in Ref.[\refcite{gorini2}] 
for finite-dimensional
systems and in Ref.[\refcite{lindblad}] in any dimension, 
under the assumption of boundedness of the generator.
Generators with positive Kossakowski matrix are known
as Kossakowski-Lindblad generators and the
resulting semigroups are known as quantum dynamical semigroups.
\item
Using the duality relation~(\ref{Heis}), one gets the following
master equation on the state-space:
\begin{equation}
\label{lindblad}
\mathbb{L}[\rho]=
-i\bigl[H\,,\,\rho\bigr]\,+\,\sum_{i,j=1}^{d^2-1}\,
C_{ij}\biggl(F^\dagger_j\,\rho\,
F_i\,-\,\frac{1}{2}\Bigl\{F_i F_j^\dagger\,,\,\rho\Bigr\}\biggr)\ .
\end{equation}
The generated dual maps $\gamma_t$ preserve the hermiticity and the
trace of $\rho$; if the Kossakowski matrix $[C_{ij}]$ is positive definite
then $\gamma_t\otimes{\rm id}_A$ preserves the positivity of all
states of $S$ and of the compound systems $S+A$ for any choice of 
finite-dimensional ancillas $A$.
\item
By setting $K\equiv\sum_{i,j=1}^{d^2-1}C_{ij}F_i F_j^\dagger$, together 
with the Hamiltonian the anticommutator in (\ref{lindblad})
can be incorporated in a pseudo-commutator:
$$
-i\bigl[H\,,\,\rho\bigr]\,-\,\frac{1}{2}\,\sum_{i,j=1}^{d^2-1}\,
C_{ij}\Bigl\{F_i F_j^\dagger\,,\,\rho\Bigr\}=-i\bigg(H-\frac{i}{2}K\bigg)\,\rho\,
+\,i\,\rho\,\bigg(H+\frac{i}{2}K\bigg)\ .
$$
The latter is the typical phenomenological expression for the
generator of the time-evolution of a decaying system, where $K$
describes loss of probability that is irreversibly transferred from
the system $S$ to the decay products.
\item
Beside the pseudo commutator, the remaining contribution to dissipation comes
from a term that, in the case of completely positive $\gamma_t$, 
can be put in Kraus-Stinespring form as done in~(\ref{cp4}).
Such a term corresponds to what in classical Brownian motion is the 
diffusive effect of white-noise.\cite{gardiner2}
It is of the form of the wave-packet reduction mechanism~(\ref{red1}) and 
it is interpreted as quantum noise.~$\Box$
\end{enumerate}}
}
\bigskip

\noindent
The consequences of the previous two theorems are best exposed
in the two-dimensional setting as showed in the next example.
\bigskip

\noindent
{\bf Example 3.1}\quad
\label{qdsex}
Let $n=2$ as in Examples 2.2; in such a case, by choosing the
orthonormal basis of Pauli matrices $F_j=\sigma_j/\sqrt{2}$,
$j=1,2,3$, the dissipative contribution to the semigroup generator
reads
\begin{equation}
\label{ex3.1.1}
\mathbb{D}[\rho]=
\sum_{i,j=0}^{3}C_{ij}\biggl(\sigma_j\rho\,\sigma_i-
{1\over 2}\Bigl\{\sigma_i\sigma_j,\, \rho\Bigr\}\biggr)\ ,
\end{equation}
where a factor $1/2$ has been absorbed into the coefficients $C_{ij}$.
In the two-dimensional case, it proves convenient to adopt a
vector-like representation; we write density matrices in the form
\begin{equation}
\label{rho}
\rho=\frac{1}{2}\Big({\bf 1}_2+\vec{\rho}\cdot\vec{\sigma}\Big)=
\frac{1}{2}\pmatrix{
1+\rho_3&\rho_1-i\rho_2\cr
\rho_1+i\rho_2&1-\rho_3}\ ,\
0\leq{\rm Det}[\rho]=\frac{1}{4}\biggl(1-\sum_{j=1}^3\rho_j^2\biggr)\ ,\
\end{equation}
where $\vec{\sigma}=(\sigma_1,\sigma_2,\sigma_3)$ and
$\vec{\rho}$ is a vector in $\mathbb{R}^3$, of unit length if and
only if $\rho$ is a pure state; it is usually referred to
as the coherence\cite{alicki-lendi} or Bloch\cite{louisell} vector.
By representing $\rho$ as a $4$-vector 
$\vert\rho\rangle
\equiv(1,\rho_1,\rho_2,\rho_3)$, any linear operation
$\rho\mapsto \Lambda[\rho]$ corresponds to a $4\times 4$ matrix 
$\mathcal{L}=[\mathcal{L}_{\mu\nu}]$ acting on $\vert\rho\rangle$.
Thus, the evolution equation~(\ref{masteq1}) can be recast in a
Schr\"odinger-like form (the $-2$ in front is for sake of convenience)
\begin{equation}
\label{ex3.1.3}
\partial_t\vert\rho_t\rangle=\,-2\, (\mathcal{H}+\mathcal{D})\, 
\vert\rho_t\rangle\ ,
\end{equation}
where the $4\times 4$ matrices $\mathcal{H}$ and $\mathcal{D}$ correspond 
to the commutator $\mathbb{L}_H$ and to the dissipative contribution 
$\mathbb{D}$.  

It is no restriction to take the Hamiltonian of the form
$H=\vec{\omega}\cdot\vec{\sigma}$, with $\omega_0=0$ and
$\vec{\omega}=(\omega_1,\omega_2,\omega_3)\in\mathbb{R}^3$; using
the algebraic relations
$\bigl[\sigma_i\,,\,\sigma_j\bigr]=\,2i\sum_{k=1}^3\epsilon_{ijk}\sigma_k$,
$i,j=1,2,3$,
it follows that
\begin{equation}
\mathcal{H}=\pmatrix{0&0&0&0\cr
0&0&\omega_3&-\omega_2\cr
0&-\omega_3&0&\omega_1\cr
0&\omega_2&-\omega_1&0}\ .
\label{ham1}
\end{equation}
This is the typical anti-symmetric form of the action of commutators
$-i[H\,,\,\cdot]$ when represented as a matrix acting on $\vert\rho\rangle$.

Concerning the dissipative matrix $\mathcal{D}$, the requests of trace
and hermiticity preservation impose $\mathcal{D}_{0j}=0$,
$j=1,2,3$, and $\mathcal{D}_{\mu\nu}\in\mathbb{R}$.
By splitting $\mathcal{D}$ into the sum of a symmetric and antisymmetric
matrix, after incorporating the latter into $\mathcal{H}$, one
remains with
\begin{equation}
\label{pauli6}
\mathcal{D}=\pmatrix{0&0&0&0\cr u&a&b&c\cr v&b&\alpha&\beta\cr
w&c&\beta&\gamma}\ ,
\end{equation}
where the nine real parameters depend on the phenomenology of the
system-environment interaction.
By exponentiation, one gets a
semigroup of $4\times 4$ matrices
$\mathcal{G}_t={\rm e}^{-2t(\mathcal{H}+\mathcal{D})}$; its correspondence
with the semigroup
$\{\gamma_t\}_{t\geq 0}$ on the state-space $\mathcal{S}(S)$ is given by
$$
\rho\mapsto\rho(t)=\gamma_t[\rho]=\sum_{\mu=0}^3\,\rho_\mu(t)\,\sigma_\mu
\ ,
$$
where $\rho_\mu(t)$ are the components of the 4-vector
$|\rho_t\rangle$.

Since the trace is preserved at all times, 
checking positivity preservation amounts to checking whether
${\rm Det}[\rho(t)]\geq0$ for all $t\geq0$ and for all initial $\rho$.
The contributions of the anti-symmetric $\mathcal{H}$ cancel out, thus
only the dissipative term remains and the 
time-derivative of the determinant reads
\begin{equation}
\label{det}
\dot{D}[\rho]\equiv\frac{{\rm d}{{\rm Det}[\rho(t)]}}{{\rm d} t}\Bigl|_{t=0}=
2\,\Biggl[\sum_{i,j=1}^3\,\mathcal{D}_{ij}\rho_i\rho_j\,
+\, \sum_{j=1}^3\mathcal{D}_{j0}\rho_j
\Biggr]\ .
\end{equation}
Let $\rho$ be a pure state: $ P({\vec n})\equiv
\big({\bf 1}_2+\vec{n}\cdot\vec{\sigma}\big)/2$, with $\vec n$ a unit
vector, $\|\vec{n}\|=1$;
then ${\rm Det}[P({\vec n})]=0$. 
For positivity to be preserved, it is
necessary that 
$$
\dot{D}[P({\vec n})]=
2\,\Bigl(\sum_{i,j=1}^3\,\mathcal{D}_{ij}n_in_j\,
+\, \sum_{j=1}^3\mathcal{D}_{j0}n_j
\Bigr)\geq 0\ .
$$
Sending ${\vec n}\mapsto-{\vec n}$, the same argument for the pure state
$P(-{\vec n})$ yields 
$$
\dot{D}[P(-{\vec n})]=
2\,\Bigl(\sum_{i,j=1}^3\,\mathcal{D}_{ij}n_in_j\,-\, 
\sum_{j=1}^3\mathcal{D}_{j0}n_j
\Bigr)\geq0\ .
$$ 
Summing the previous two inequalities and
varying ${\vec n}$ in the unit sphere, it turns out that positivity is
preserved only if
\begin{equation}
\label{pauli11}
\mathcal{D}^{(3)}=\pmatrix{a&b&c\cr b&\alpha&\beta\cr c&\beta&\gamma}\,\geq\,
0\ .  
\end{equation}
The above condition on $\mathcal{D}^{(3)}$ is only necessary to
positivity preservation as $\dot{D}[P({\vec n})]<0$ can follow because
of the presence of the additional contribution
$\sum_{j=1}^3\mathcal{D}_{j0}\rho_j$.
However, it becomes also sufficient when we ask that ${\cal G}_t$
does not decrease the von Neumann entropy 
of any initial state, as this is equivalent to
$u=v=w=0$ in $\mathcal{D}$.
Indeed, the totally depolarized state 
${\bf 1}_2/2$, {\it i.e.} $(1,0,0,0)$ in vectorial notation, 
has maximal von Neumann entropy $\log 2$; if we impose that ${\cal G}_t$
does not decrease its entropy, it can only stay constant, which means
that ${\bf 1}_2/2$ is a stationary state.
In general, the vectorial expression of stationarity reads 
$(\mathcal{H}+\mathcal{D})\vert\rho\rangle=0$, which in the case at
hands implies $u=v=w=0$. 

In the following, we shall restrict to entropy-increasing semigroups; 
then, in terms of the entries $C_{ij}$, the matrix $\mathcal{D}$ 
in~(\ref{pauli6}) reads
\begin{equation}
\label{diss0}
\mathcal{D}=\pmatrix{
0&0&0&0\cr
0&C_{22}+C_{33}&-C_{12}&-C_{13}\cr
0&-C_{12}&C_{11}+C_{33}&-C_{23}\cr
0&-C_{13}&-C_{23}&C_{11}+C_{22}}\ .
\end{equation}
Thus, the positivity of $[C_{ij}]$, which, according to the previous 
discussion, 
is necessary and sufficient for the complete positivity of $\gamma_t$,
results in the inequalities
\begin{eqnarray}
\nonumber
&&2R\equiv\alpha+\gamma-a\geq0\ ,\qquad RS\geq b^2\ ,\\
\nonumber
&&2S\equiv a+\gamma-\alpha\geq0\ ,\qquad RT\geq c^2\ ,\\
\nonumber
&&2T\equiv a+\alpha-\gamma\geq0\ , \qquad ST\geq\beta^2\ ,\\
\label{diss1}
&&RST\geq 2\, bc\beta+R\beta^2+S c^2+T b^2\ .\\
\nonumber
\end{eqnarray}
\vskip -.5cm
\noindent             
These constraints are much stronger than those coming from positivity alone,
that is from  
$\mathcal{D}^{(3)}\geq 0$, which yields
\begin{equation}
\nonumber
\left\{\matrix{a\geq0\cr \alpha\geq 0\cr\gamma\geq0}\right.\ ,\qquad 
\left\{\matrix{a\alpha\geq b^{2}\cr
\label{diss2}
a\gamma\geq c^2\cr\alpha\gamma\geq\beta^2}\right.\ ,\qquad
{\rm Det}\mathcal{D}^{(3)}\geq0\ .
\end{equation}
\hfill $\Box$
\bigskip

The fact that the conditions for complete positivity are stronger than
those for positivity have an important physical consequence in that the
decay-times related to dissipative completely positive semigroups must
obey a definite hierarchy as showed in the following example.
\bigskip

\noindent
{\bf Example 3.2}\quad
A typical relaxation behaviour induced by a heat bath on a two level
system is determined by the following evolution equations for the entries
$\rho_{ij}$ of its density matrix:\cite{slichter}
\begin{eqnarray}
\label{ex32.1a}
&&\frac{{\rm d}\rho_{11}}{{\rm d}t}=-p\ \rho_{11}\,+\,q\ \rho_{22}\ ,\\
\label{ex32.1b}
&&\frac{{\rm d}\rho_{22}}{{\rm d}t}=-q\ \rho_{22}\,+\,p\ \rho_{11}\ ,\\
\label{ex32.1c}
&&\frac{{\rm d}\rho_{12}}{{\rm d}t}=
-\bigl(i\omega\,+\,r\bigr)
\rho_{12}\ ,\\
\label{ex32.1d}
&&\frac{{\rm d}\rho_{21}}{{\rm d}t}=
\bigl(i\omega\,-\,r\bigr)
\rho_{21}\ ,
\end{eqnarray}
where $p$, $q$, $r$ and $\omega$ are positive constants, the latter
representing the Hamiltonian contribution.
The above time-evolution equations have been written for the standard
matrix representation 
$\displaystyle
\rho=\pmatrix{\rho_{11}&\rho_{12}\cr\rho_{21}&\rho_{22}}$; going to
the vectorial representation introduced in the previous example
one gets 
\begin{equation}
\label{ex32.2}
\partial_t\vert{\rho_t}\rangle=-\left[
\pmatrix{0&0&0&0\cr
0&0&\omega&0\cr
0&-\omega&0&0\cr
0&0&0&0}
+
\pmatrix{0&0&0&0\cr
0&r&0&0\cr
0&0&r&0\cr
p-q&0&0&p+q}
\right]\vert\rho_t\rangle\ .
\end{equation}
The choice $p=q$ induces entropy-increase, so that
positivity-preservation is
equivalent to the positivity of
$\mathcal{D}^{(3)}$ which only requires $r,p\geq 0$.
On the contrary, from~(\ref{diss1}), complete positivity asks for 
$r\geq p$, {\it i.e.} the
so-called \textit{phase-relaxation} $1/T_2\equiv r$ must be larger or equal
than $1/(2T_1)$, where $1/T_1$ is the so-called
\textit{population-relaxation}.~$\Box$
\bigskip

\noindent
{\bf Remark 3.4}\quad
{
The hierarchy between characteristic decay times induced by the
request of complete positivity is often refused or challenged in the
physical literature ({\it e.g.} see Refs.[\refcite{pechukas},
\refcite{budimir1}-\refcite{wielkie}]).
Indeed, as already stressed, it arises because of the possibility of an uncontrollable
statistical coupling of the experimentally accessible system with an
inert entity, the ancilla.
Such abstract eventuality seems to be too weak to constrain the 
physical properties of dissipative quantum systems.
We shall later show how the true role of complete positivity is far
from being an abstract mathematical non-sense, but has to do with the
possibility of entangled bi-partite states evolving dissipatively in
a same environment. $\Box$
}
\bigskip

\subsection{Master Equation}
\label{subs3.1}

In this Section, we describe how from the global time-evolution equation
\begin{equation}
\label{reddyn3}
\partial_t\rho_{S+E}(t)=\mathbb{L}_{S+E}\big[\rho_{S+E}(t)\big]\ ,
\end{equation}
by performing certain Markov approximations, one can obtain master 
equations as in~(\ref{masteq1}).\cite{alicki-lendi,gorini1,spohn}
We shall be particularly interested in comparing them with the general
form in~(\ref{lindblad}).

The first step consists in defining a projection 
$P\,:\,{\cal S}(S+E)\mapsto{\cal S}(S+E)$ on the state-space of
$S+E$ which, by partial tracing, extracts the state of $S$:
\begin{equation}
\label{reddyn1}
P[\rho_{S+E}]={\rm Tr}_E(\rho_{S+E})\otimes\rho_E\ ,
\end{equation}
where $\rho_E$ is a chosen environment reference state.
It follows that $P^2\equiv P\circ P=P$ and, 
with $Q$ such that $P+Q={\bf 1}_{S+E}$,
\begin{equation}
\label{reddyn2}
P[\rho_S\otimes\rho_E]=\rho_S\otimes\rho_E\ ,\quad
Q[\rho_S\otimes\rho_E]=0\ .
\end{equation}

The second step uses $P$ and $Q$ to split the evolution equation 
into the two coupled equations
\bigskip
\begin{eqnarray}
\label{reddyn4a}
\partial_t P[\rho_{S+E}(t)]&=& \mathbb{L}^{PP}_{S+E}\big[\,P[\rho_{S+E}(t)]\,\big]\,+\,
\mathbb{L}^{PQ}_{S+E}\big[\,Q[\rho_{S+E}(t)]\,\big]\ ,\\
\nonumber
\hfill\\
\label{reddyn4b}
\partial_t Q[\rho_{S+E}(t)]&=& \mathbb{L}^{QP}_{S+E}\big[\,P[\rho_{S+E}(t)]\,\big]\,+\,
\mathbb{L}^{QQ}_{S+E}\big[\,Q[\rho_{S+E}(t)]\,\big]\ ,
\end{eqnarray}
\medskip

\noindent
where $\mathbb{L}^{PP}_{S+E}\equiv P\circ \mathbb{L}_{S+E}\circ P$, 
$\mathbb{L}^{PQ}_{S+E}\equiv P\circ \mathbb{L}_{S+E}\circ Q$,
$\mathbb{L}^{QQ}_{S+E}\equiv Q\circ \mathbb{L}_{S+E}\circ Q$ and 
$\mathbb{L}^{QP}_{S+E}\equiv Q\circ \mathbb{L}_{S+E}\circ P$.

Formal integration of the second equation yields: 
\begin{equation}
\label{reddyn5}
Q[\rho_{S+E}(t)]={\rm e}^{t\,\mathbb{L}^{QQ}_{S+E}}\big[\,Q[\rho_{S+E}(0)]\,\big]\,+\,
\int_0^t{\rm d}s\, {\rm e}^{(t-s)\mathbb{L}^{QQ}_{S+E}}\circ
\mathbb{L}^{QP}_{S+E}\big[\,P[\rho_{S+E}(s)]\,\big]\ .
\end{equation}
We shall now assume an uncorrelated initial state of the tensor
product form $\rho_S\otimes\rho_E$.
Thus, the first term on the right hand side vanishes because 
of~(\ref{reddyn2}); once 
inserted into~(\ref{reddyn4a}), the solution~(\ref{reddyn5}) leads to
the following evolution equation
\begin{equation}
\label{reddyn6}
\partial_t\rho_{S}(t)\otimes\rho_E=\mathbb{L}^{PP}_{S+E}[\rho_{S}(t)
\otimes\rho_E]\,+\,
\int_0^t{\rm d}s\, \mathbb{L}^{PQ}_{S+E}\circ{\rm e}^{(t-s)
\mathbb{L}^{QQ}_{S+E}}\circ\mathbb{L}^{QP}_{S+E}[\rho_{S}(s)\otimes\rho_E]\ .
\end{equation}
The tensorized $\rho_E$ on the left hand side also appears on the
right hand side because of~(\ref{reddyn2})--(\ref{reddyn4b}); it can
thus be traced away yielding a master equation for 
$\rho_S(t)\equiv{\rm Tr}_E[\rho_{S+E}(t)]$.

At this point, one assumes the reference state $\rho_E$ to be an
equilibrium state of the dynamics of the environment alone,
\begin{equation}
\label{reddyn7}
\mathbb{L}_E[\rho_E]=-i\bigl[\,H_E\,,\,\rho_E\,\bigr]=0\ .
\end{equation}
Such a condition is physically plausible because typical environments are
modelled as very large heat baths or thermostats whose equilibrium states 
are not perturbed by the interaction with $S$.
As a consequence, $\big({\bf 1}_S\otimes\mathbb{L}_E\big)P=0$; 
similarly, $P\big({\bf 1}_S\otimes\mathbb{L}_E\big)=0$
follows from the cyclicity of the trace operation and expresses
the conservation of probability in the environment.  

Further, in~(\ref{global1}), one assumes an interaction Hamiltonian 
of the form
\begin{equation}
\label{reddyn8}
\widetilde{H}'=\sum_\alpha\, V_\alpha\otimes \widetilde{B}_\alpha\ ,
\end{equation} 
where $V_\alpha$ and $\widetilde{B}_\alpha$ are hermitian operators
acting on the Hilbert spaces
$\mathbb{C}^n$ of $S$, respectively $\mathfrak{H}$ of $E$.
It proves convenient to define \textit{centered environment operators}  
$B_\alpha\equiv \widetilde{B}_\alpha-{\rm Tr}_E(\widetilde{B}_\alpha)$;
in this way one gets a new Hamiltonian for $S$ and a new interaction term 
\begin{equation}
\label{reddyn9}
H^\lambda_S=H_S\,+\,\lambda\underbrace{
\sum_\alpha\,V_\alpha\,{\rm
Tr}_E(\widetilde{B}_\alpha)}_{H_S^{(1)}}\ ,
\quad
H'=\sum_\alpha\,V_\alpha\otimes B_\alpha\ ,\quad
{\rm Tr}\big[\rho_E\,B_\alpha\big]=0\ .
\end{equation}
\bigskip

\noindent
{\bf Remark 3.5}\quad
{
The redefinition of $H_S$ amounts to a Lamb shift of the energy levels
due to a mean-field, first order (in $\lambda$) approximation of the 
interaction with $E$. $\Box$
}
\bigskip

\noindent
Denoting by $\mathbb{L}^\lambda_S$ ($-i$ times) the commutator with respect to 
$H^\lambda_S$
and by $\mathbb{L}'$ ($-i$ times) the commutator with respect to
the new interaction term $H'$, the total Liouvillian
in~(\ref{reddyn3}) 
becomes
$$
\mathbb{L}\equiv\mathbb{L}_{S+E}=\mathbb{L}^\lambda_S\otimes{\bf 1}_E\,
+\,{\bf 1}_S\otimes\mathbb{L}_E\,+\,
\lambda\,\mathbb{L}'\ .
$$ 
Moreover, by means of~(\ref{reddyn7}) and of the last relation 
in~(\ref{reddyn9}) one gets
\begin{equation}
\label{reddyn10a}
P\circ (\mathbb{L}^\lambda_S\otimes{\bf 1}_E)=(\mathbb{L}^\lambda_S
\otimes{\bf 1}_E)\circ P
\ ,\quad P\circ \mathbb{L}'\circ P=0 \ .
\end{equation}
Using the latter relations and computing the partial trace ${\rm
Tr}_E$ with respect to the
orthonormal basis of eigenvectors of $\rho_E$, further yield
\begin{eqnarray}
\label{reddyn10b}
&&\mathbb{L}^{PP}_{S+E}[\rho_S(t)\otimes\rho_E]=
\bigl(\mathbb{L}^\lambda_S\otimes{\bf 1}_E\bigr)^{PP}[\rho_S(t)\otimes\rho_E]
=\mathbb{L}^\lambda_S[\rho_S(t)]\otimes\rho_E\ ,\\
\label{reddyn10d}
&&\mathbb{L}^{PQ}_{S+E}[\rho_{S+E}(t)]
=\lambda\, \big(\mathbb{L}'\big)^{PQ}[\rho_{S+E}(t)]=
\lambda\,{\rm Tr}_E\Bigl(
\mathbb{L}'\circ Q[\rho_{S+E}(t)]\Bigr)\otimes\rho_E\ ,\qquad\\ 
\label{reddyn10c}
&&\mathbb{L}^{QP}_{S+E}[\rho_S(t)\otimes\rho_E]=
\lambda\,\big(\mathbb{L}'\big)^{QP}[\rho_S(t)\otimes\rho_E]=
\lambda\, \mathbb{L}'[\rho_S(t)\otimes\rho_E]\ , 
\end{eqnarray}
where the last equality follows from
\begin{equation}
\label{reddyn10e}
Q\,\mathbb{L}'[\rho_S\otimes\rho_E]=\mathbb{L}'[\rho_S\otimes\rho_E]\ .
\end{equation}
Inserting these expressions into~(\ref{reddyn6}) and getting rid of the 
appended factor $\rho_E$, one finally gets
the \textit{master equation}
\begin{equation}
\label{reddyn11}
\partial_t\rho_S(t)=\mathbb{L}^\lambda_S[\rho_S(t)]\,+\,
\lambda^2\,\int_0^t{\rm d}s\, {\rm Tr}_E
\Bigl(\mathbb{L}'\circ{\rm e}^{(t-s)\,\mathbb{L}^{QQ}}\circ
\mathbb{L}'\bigl[\rho_S(s)\otimes\rho_E\bigr]\Bigr)\ .
\end{equation}

The second term in the right hand side 
of~(\ref{reddyn11}) compactly comprises the effects due to
the environment; these are described by two nested commutators
with respect to the interaction Hamiltonian $H'$, with the full
time-evolution  present in \hbox{$\exp((t-s)\mathbb{L}^{QQ})$,} 
and through a final overall trace over the environment degrees of freedom.
Finding the concrete form of these effects would require the
solution of the time-evolution of the total system $S+E$.
 
In order to proceed to a more manageable equation and to the
elimination of all memory effects, one first notices 
that ${\rm Tr}_E$ makes the dissipative term ultimately depend on two-point
correlation functions of the environment operators $B_\alpha$.
Memory effects are thus expected to become negligible on a
time-scale much longer than the typical decay time $\tau_E$ 
of correlations in the environment. 

That indeed this is the case can be illustrated with the help 
of the following simple, exactly solvable model of 
system-environment coupling.\cite{palma}
\bigskip

{\bf Example 3.3} \quad
In the notation of Example 3.1, consider a two-level system
immersed in a bosonic thermal bath at inverse temperature
$\beta=T^{-1}$. The Hamiltonian for the total system
contains three contributions as in (\ref{global1}):
\begin{equation}
H=\underbrace{ \frac{\Omega}{2}\sigma_3}_{H_S}\otimes\, {\bf 1}_E
+{\bf 1}_S\otimes\underbrace{ \sum_k\omega_k\, a^\dagger_k a_k}_{H_E}
+\underbrace{\sigma_3\otimes\sum_k\Big(\lambda_k\, a_k^\dagger
+\bar\lambda_k\, a_k\Big)}_{H'}\ ,
\label{3.78-1}
\end{equation}
where, for simplicity, the bath modes are assumed to be one-dimensional.
The first two pieces, giving together the free Hamiltonian $H_0$, drive the
uncoupled motion: $\Omega$ represents the splitting between the two system 
levels, while $\omega_k$ the energy of the bath modes, labelled by the
discrete index $k$ (the bath is for the moment assumed to be confined
in a finite ``box''); the corresponding creation, $a^\dagger_k$, and
annihilation, $a_k$, operators are taken to satisfy the standard
oscillator algebra: $[a^\dagger_k, a_{k'}]=\delta_{k,k'}$,
$[a^\dagger_k, a^\dagger_{k'}]=[a_k, a_{k'}]=\,0$.

The coupling between the subsystem and the bath degrees of freedom is described
by the third contribution in (\ref{3.78-1}), with $\lambda_k$, $\bar\lambda_k$
playing the role of mode-dependent coupling constants. Note that the interaction
Hamiltonian $H'$ is diagonal in the subsystem Hilbert space: it is this
property that allows deriving an exact analytic expression for the
total evolution operator $U_t={\rm e}^{-it H}$.

In order to get rid of the free motion, it is convenient to work 
in the interaction picture. Then, after
conjugation with the free evolution operator $U_t^{(0)}={\rm e}^{-it H_0}$,
$U_t$ can be obtained through the time-ordered exponentiation of
the following time-dependent interaction Hamiltonian:
\begin{equation}
H'(t)=\sum_k\Big[\lambda_k\, {\rm e}^{i\omega_k t}\, A_k^\dagger
+\bar\lambda_k\, {\rm e}^{-i\omega_k t}\, A_k\Big]
\equiv \sum_k H'_k(t)\ .
\label{3.78-2}
\end{equation}
Since the operators $A^\dagger_k\equiv \sigma_3\otimes a^\dagger_k$,
$A_k\equiv \sigma_3\otimes a_k$ obey the same algebra as
$a^\dagger_k$, $a_k$, this exponentiation can be computed in closed
form using standard Lie algebraic manipulations.\cite{louisell,puri}
As a result, the time-evolution
operator in the interaction picture reads ($\cal T$ represents
time ordering):
\begin{eqnarray}
\nonumber
U(t_0,t)&\equiv& {\cal T} {\rm e}^{-i\int_{t_0}^t d\tau H'(\tau)}=
\Pi_k {\cal T} {\rm e}^{-i\int_{t_0}^t d\tau H'_k(\tau)}\\
&=&{\rm e}^{i\varphi(t-t_0)}\, \Pi_k {\rm e}^{-i\int_{t_0}^t d\tau H'_k(\tau)}=
{\rm e}^{i\varphi(t-t_0)}\, {\rm e}^{\sigma_3\otimes K(t_0,t)}\ ,
\label{3.78-3}
\end{eqnarray}
where
\begin{eqnarray}
\nonumber
&&\varphi(t-t_0)=\sum_k\frac{|\lambda_k|^2}{\omega_k^2}\Big[
\omega_k(t-t_0)-\sin\omega_k(t-t_0)\Big]\ ,\\
&&K(t_0,t)=\sum_k\Big[f_k(t_0,t)\, a^\dagger_k-\bar f_k(t_0,t)\, a_k\Big]\ ,
\label{3.78-4}
\end{eqnarray}
and
\begin{equation}
f_k(t_0,t)=-i\int_{t_0}^t {\rm d}\tau\lambda_k\, {\rm e}^{i\omega_k\tau}=
\frac{\lambda_k}{\omega_k}\Big({\rm e}^{i\omega_k t_0} - {\rm e}^{i\omega_k t}\Big)\ .
\label{3.78-5}
\end{equation}
At the initial time $t_0$, subsystem and environment are assumed to be uncorrelated,
with the bath in a thermal equilibrium state described by the density matrix
$\rho_E={\rm e}^{-\beta H_E}/{\rm Tr}[{\rm e}^{-\beta H_E}]$. The evolution in time 
of the reduced density matrix $\rho(t)$ pertaining to the subsystem
is then given by:
\begin{equation}
\rho(t_0)\to\rho(t)={\rm Tr}_E\Big[U(t_0,t)\, \rho(t_0)\otimes\rho_E\, 
U^\dagger(t_0,t)\Big]\ .
\label{3.78-6}
\end{equation}
A convenient basis in the two-dimensional subsystem Hilbert space is provided
by the eigenvectors of $H_S$, $\sigma_3 |i\rangle=(-1)^{i+1} |i\rangle$,
$i=0,1$. Using (\ref{3.78-3}), from (\ref{3.78-6}) one then deduces the evolution
of the corresponding matrix elements $\rho_{ij}(t)=\langle i|\rho(t)\, |j\rangle$:
\begin{eqnarray}
\nonumber
&&\rho_{00}(t)=\rho_{00}(t_0)\ ,\qquad \rho_{11}(t)=\rho_{11}(t_0)\ ,\\
&&\rho_{10}(t)={\rm Tr}_E\Big[ {\rm e}^{2K(t_0,t)}\,\rho_E\Big]\ \rho_{10}(t_0)=
{\rm e}^{-\Gamma(t_0,t)}\, \rho_{10}(t_0)\ ,
\label{3.78-7}
\end{eqnarray}
with
\begin{equation}
\Gamma(t_0,t)=4\sum_k\frac{|\lambda_k|^2}{\omega^2_k}\big[1-\cos\omega_k(t-t_0)\big]\,
\coth\bigg(\frac{\beta\omega_k}{2}\bigg)\ .
\label{3.78-8}
\end{equation}
Only the off-diagonal matrix elements are affected by the interaction with the bath:
in this case, the environment is responsible for the loss of quantum coherence,
without affecting the population of the two levels; this is a consequence of the
form of interaction Hamiltonian $H'$, which is diagonal in the chosen system basis.

In order to analyze in detail this decoherence phenomenon, it is useful to pass
to the thermodynamical limit of the bath system: in this way, the mode label $k$
becomes a continuous variable. The damping function (\ref{3.78-8}) can then
be expressed as an integral over the frequency variable 
($\sum_k\to(1/2\pi)\int_0^\infty d\omega$):
\begin{equation}
\Gamma(t_0,t)\equiv\Gamma(t-t_0)=\frac{2\lambda^2}{\pi}\int_0^\infty
{\rm d}\omega \bigg[\frac{1-\cos\omega(t-t_0)}{\omega}\bigg]\,
\coth\bigg(\frac{\beta\omega}{2}\bigg)\, {\rm e}^{-\varepsilon\omega}\ ,
\label{3.78-9}
\end{equation}
where a typical $\sqrt{\omega}$ dependence of the coupling constants on frequency
has been assumed, together with a massless dispersion relation.
The introduction of the damping exponential term is necessary to make
the integral converge at infinity: physically, it is justified by assuming
the presence of a characteristic scale $1/\varepsilon$ below which 
the system-bath coupling decreases rapidly, as found in many realistic
models.

The function in (\ref{3.78-9}) can be conveniently split as
$\Gamma(t)=\Gamma_0(t) + \Gamma_\beta(t)$ into a vacuum, $\Gamma_0(t)$,
and thermal, $\Gamma_\beta(t)$, contributions, that can be separately
evaluated. The vacuum piece,
\begin{equation}
\Gamma_0(t)\equiv\frac{2\lambda^2}{\pi}\int_0^\infty
{\rm d}\omega \bigg[\frac{1-\cos\omega t}{\omega}\bigg]\,
\bigg[\coth\bigg(\frac{\beta\omega}{2}\bigg)-1\bigg]\, {\rm e}^{-\varepsilon\omega}=
\frac{\lambda^2}{\pi}\ln\Big[1+\big(t/\varepsilon\big)\Big]\ ,
\label{3.78-10}
\end{equation}
is temperature independent and describes how the bath vacuum fluctuations
affect the quantum coherence of the two-level system. On the other hand,
the action of the bath thermal fluctuations on the subsystem dynamics
is determined by\cite{prudnikov}
\begin{equation}
\Gamma_\beta(t)\equiv\frac{4\lambda^2}{\pi}\int_0^t {\rm d}s\int_0^\infty
{\rm d}\omega \sin\omega t\ \frac{{\rm e}^{-\varepsilon\omega}}{{\rm e}^{\beta\omega}-1}
=-\frac{4\lambda^2}{\pi}\ln\Bigg[
\frac{\big|\Gamma\big(1+\varepsilon/\beta+it/\beta\big)\big|}
{\Gamma\big(1+\varepsilon/\beta\big)}\Bigg]\ .
\label{3.78-11}
\end{equation}

The existence in the problem of different time scales allows to
identify, through the explicit results (\ref{3.78-10}) and
(\ref{3.78-11}), three time regimes in the two-level
system dynamics:
\begin{enumerate}
\rm
\item
short times, $t\ll\varepsilon$, for which
$\Gamma(t)\sim (\lambda^2/\pi) (t/\varepsilon)^2$; in this case the bath
fluctuations hardly affect the subsystem free evolution;
\item
intermediate times, $\varepsilon<t<\beta$, where
$\Gamma(t)\sim(2\lambda^2/\pi)\ln(t/\varepsilon)$;
here the main cause of decoherence is due to the vacuum
fluctuations;
\item
long time regime $t\gg\beta$, where
$\Gamma(t)\sim (2\lambda^2/\beta) t$.%
\footnote{This result follows from the following asymptotic behaviour
of the modulus of $\Gamma(x+iy)$ as $|y|\to\infty$, for $x$ and $y$ real:
$|\Gamma(x+iy)|\sim\sqrt{2\pi}\, 
|y|^{x-1/2}\, {\rm e}^{-\pi|y|/2}$.\cite{erdelyi}
}
This is the Markovian
regime, which holds for times much longer than the
characteristic decay time of the bath correlations,
which is proportional to the inverse temperature, $\tau_E\simeq\beta$.
The loss of coherence follows an exponential
decreasing law, with a lifetime
\hbox{$T_2=\beta/2\lambda^2$;} this corresponds exactly to the phase relaxation
parameter introduced in Example 3.2. As already mentioned,
the population relaxation parameter $T_1$ is here vanishing
due to the particularly simple choice of subsystem-bath coupling.
\end{enumerate}
\hfill$\Box$
\bigskip

\subsection{Markovian Approximations}

Obtaining a physically consistent Markovian approximation for the master
equation in (\ref{reddyn11}) is notoriously 
tricky.\cite{alicki-lendi,gorini1,spohn,lendi1} 
Indeed, with respect to the previous Example, the situation
is complicated by the dynamics of the system $S$ alone,
generated by the Hamiltonian $H_S$, which in general
does not commute
with the interaction with the bath. This
introduces in the problem another timescale $\tau_S$, 
and the need of a hierarchy condition $\tau_E\ll\tau_S$,
that allow a clear separation between subsystem
and environment. Following Ref.[\refcite{dumcke-spohn}], 
we shall now present the most commonly used Markovian
limits of (\ref{reddyn11}), focusing in particular on their physical
consistency, especially in relation to the property of complete
positivity.

\subsubsection{Second Order Approximation}

One of the most simple assumptions allowing an immediate identification
of subsystem and environment is the hypothesis of {\it weak coupling}
between the two. In this case, the ratio $\tau_E/\tau_S$ is small
because $\tau_S\to\infty$, while $\tau_E$ remains finite, and further
$\lambda\ll 1$. The master equation (\ref{reddyn11}) can be more explicitly
rewritten as
\begin{eqnarray}
\nonumber
\partial_t\rho_S(t)&=&-i\Bigl[H_S+\lambda H^{(1)}_S\,,\,
\rho_S(t)\Bigr]\\
\label{reddyn12a}
&-&\lambda^2\,\int_0^t{\rm d}s\, {\rm Tr}_E\Bigl(
\Bigl[H'\,,\,{\rm
e}^{(t-s)\,\mathbb{L}^{QQ}_{S+E}}
\Bigl[H'\,,\,\rho_S(s)\otimes\rho_E\Bigr]\Bigr]\Bigr)\ ,
\end{eqnarray}
making clear that dissipation becomes relevant on a
\textit{slow time-scale} of order $\lambda^{-2}$.

A first Markovian approximation to~(\ref{reddyn12a}) can be euristically
obtained as follows; integration of (\ref{reddyn12a}) yields:
\begin{eqnarray}
\nonumber
\rho_S(t)&=&{\rm e}^{t\mathbb{L}^\lambda_S}[\rho_S(0)]\,-\,
\lambda^2\,\int_0^t{\rm d}v\,\,\int_0^v{\rm d}u\ {\rm
e}^{(t-v)\mathbb{L}^\lambda_S} 
\times\\
\label{wcl00}
&&\hskip 2cm
\times {\rm Tr}_E\Bigl(
\Bigl[H'\,,\,{\rm
e}^{(v-u)\,\mathbb{L}^{QQ}_{S+E}}\Bigl[H'\,,\,\rho_S(u)
\otimes\rho_E\Bigr]\Bigr]\Bigr)\ .
\end{eqnarray}
Then, by changing the integration order and by introducing the new
integration variable $w\equiv v-u$, (\ref{wcl00}) becomes:
\begin{eqnarray}
\nonumber
\rho_S(t)&=&{\rm e}^{t\mathbb{L}^\lambda_S}[\rho_S(0)]\,-\,
\lambda^2\,\int_0^t{\rm d}u\,{\rm e}^{(t-u)\mathbb{L^\lambda_S}}
\,\times\\
\label{wcl01}
&\times&
\bigg\{\int_0^{t-u}{\rm d}w\, {\rm e}^{-w\mathbb{L}^\lambda_S} 
{\rm Tr}_E\Bigl(
\Bigl[H'\,,\,{\rm e}^{w\,
\mathbb{L}^{QQ}}\Bigl[H'\,,\,\rho_S(u)\otimes\rho_E\Bigr]
\Bigr]\Bigr)\bigg\}\ .
\end{eqnarray}

In view of the assumed
smallness of the coupling constant $\lambda$, the dissipative character
of the dynamics emerges on the slow time scale $\tau=\lambda^2t$.
A first instance of Markovian approximation consists in 1) substituting
$\tau/\lambda^2$ for $t$ in the curly bracket and 2) in letting $\lambda\to0$
in the upper limit of the integration in $w$ and in the 
generators $\mathbb{L}^\lambda_S$ and $\mathbb{L}_{S+E}$.
As $\mathbb{L}_{S,E}$ commute with $Q$ and because
of~(\ref{reddyn10e}), this finally gets
\begin{eqnarray}
\nonumber
\rho_S(t)&=&{\rm e}^{t\mathbb{L}^\lambda_S}[\rho_S(0)]\,-\,
\lambda^2\,\int_0^t{\rm d}u\,\,{\rm e}^{(t-u)\mathbb{L}^\lambda_S}
\times\\ 
&\times&
\int_0^{\infty}{\rm d}w\,{\rm Tr}_E\Big({\rm e}^{-w\mathbb{L}_S}
\Bigl[H'\,,\,{\rm e}^{w(\mathbb{L}_{H_S}+\mathbb{L}_E)}
\Bigl[H'\,,\,\rho_S(u)\otimes\rho_E\Bigr]
\Bigr]\Big)\ ,
\end{eqnarray}
which is the formal solution to
\begin{equation}
\label{wcl0}
\partial_t\rho_S(t)=-i\Bigl[H_S+\lambda H^{(1)}_S\,,\,
\rho_S(t)\Bigr]+\lambda^2\,\mathbb{D}_1[\rho_S(t)]\ ,
\end{equation}
where
\begin{equation}
\label{wcl1}
\mathbb{D}_1[\rho_S]=\,-\,\int_0^{\infty}{\rm d}s\, 
{\rm Tr}_E\Bigl({\rm e}^{-s\mathbb{L}_{H_S}}
\Bigl[H'\,,\,{\rm e}^{s\,(\mathbb{L}_S+\mathbb{L}_E)}
\Bigl[H'\,,\,\rho_S\otimes\rho_E\Bigr]\Bigr]\Bigr)\ .
\end{equation}
Master equations of the form~(\ref{wcl0}) with the dissipative
contribution as in~(\ref{wcl1}) are known as 
\textit{Redfield equations}.\cite{slichter}
Given the form~(\ref{reddyn9}) of the $S-E$
interaction, we set
\begin{equation}
\label{wcl1.1}
V_\alpha(t)\equiv{\rm e}^{-t\mathbb{L}_S}[V_\alpha]=
{\rm e}^{itH_S}\,V_\alpha\,{\rm e}^{-itH_S}\ ,\quad
B_\alpha(t)\equiv{\rm e}^{-t\mathbb{L}_E}[B_\alpha]=
{\rm e}^{itH_E}\,B_\alpha\,{\rm e}^{-itH_E} \ .
\end{equation}
Further, we introduce the environment \textit{two-point correlation
functions}
\begin{equation}
\label{corrfunc}
G_{\alpha\beta}(s)\equiv{\rm Tr}\Bigl[\rho_E\,
B_\alpha(s)B_\beta\Bigr]={\rm Tr}\Bigl[\rho_E\,
B_\alpha B_\beta(-s)\Bigr]\ .
\end{equation}
The dissipative term $\mathbb{D}_1[\rho_S(t)]$ exhibits
the following structure of the environment induced second 
order effects:
\begin{eqnarray}
\nonumber
\mathbb{D}_1[\rho_S(t)]&=&\,-\,
\sum_{\alpha,\beta}\int_0^{\infty}{\rm d}s\Biggl\{
G_{\alpha\beta}(s)\,
\Bigl[V_\alpha(s)\,,\,V_\beta\,\rho_S(t)\Bigr]\\
\label{wcl2}
&&\hskip 4cm
+G_{\beta\alpha}(-s)\,
\Bigl[\rho_S(t)\,V_\beta\ ,\,V_\alpha(s)\Bigr]\Biggr\}\ .
\end{eqnarray}

In order to compare them with the Kossakowski-Lindblad 
generator~(\ref{lindblad}), it is convenient to introduce the
operators
\begin{equation}
\label{Redf0}
C^S_\beta\equiv\sum_\alpha\int_0^{\infty}{\rm d}s\,
G_{\alpha\beta}(s)\,V_\alpha(s)\ ,
\end{equation} 
and rewrite
\begin{eqnarray}
\nonumber
\mathbb{D}_1[\rho_S(t)]&=&
\sum_\beta\Bigl\{B^S_\beta\,\rho_S(t)\,C^S_\beta\,+\,
\left(C^S_\beta\right)^\dagger\,\rho_S(t)\,B^S_\beta\\
\label{Redf1}
&&\hskip 2cm 
-C^S_\beta\,B^S_\beta\,\rho_S(t)\,-\,
\rho_S(t)\,B^S_\beta\,\left(C_\beta^S\right)^\dagger\Bigr\}\ .
\end{eqnarray}
It can be verified that trace and hermiticity are preserved, namely
that ${\rm Tr}\Bigl(\mathbb{D}_1[\rho_S(t)]\Bigr)=0$ and that
$\Bigl(\mathbb{D}_1[\rho_S(t)]\Bigr)^\dagger=\mathbb{D}_1[\rho_S(t)]$.
Despite a certain similarity to~(\ref{lindblad}), in order to check
whether the semigroup generated by the Redfield
equations consists of completely positive maps one has to extract 
from~(\ref{Redf1}) the corresponding Kossakowski matrix and ensure its 
positivity.
Instead of doing this, in the following example we shall show that, in
general, the semigroups generated by the Redfield equations do not even
preserve positivity.

Up to this point we have considered any equilibrium environment state
$\rho_E$; however, typical environments are heat baths and $\rho_E$
thermal states at inverse temperature $\beta\equiv T^{-1}$.
They satisfy the so-called 
\textit{Kubo-Martin-Schwinger (KMS) conditions}\cite{thirring}
\begin{equation}
\label{KMS}
G_{\alpha\beta}(t)=G_{\beta\alpha}(-t-i\beta)\ .
\end{equation}
These equalities express the analiticity properties of thermal
correlation functions with respect to time; they are easily derived
when $H_E$ has discrete spectrum, but survive the
thermodynamical limit and then hold for truly infinite environments.
\medskip
 
\noindent
{\bf Example 3.4}\quad
As in the previous Example, consider a two-level system immersed into a
free Bose-gas in equilibrium at temperature $T=\beta^{-1}$.
This time, however, we choose an interaction Hamiltonian which does not
commute with the system one:
\begin{equation}
\label{ex3.2.1}
H_{S+E}=\underbrace{\frac{\Omega}{2}\sigma_3}_{H_S}
\otimes{\bf 1}_E\,+\,{\bf 1}_S\otimes
\underbrace{\sum_k\omega_k
a^\dagger_k a_k}_{H_E}\,+\,\lambda\,\underbrace{\sigma_1\otimes B}_{H'}\ ;
\end{equation}
here, $B=B^\dagger$ is any observable of $E$ with ${\rm Tr}[\rho_E\,B]=0$,
where, as in Example 3.3, the environment reference state is given by
\begin{equation}
\label{ex3.2.2}
\rho_E=\frac{{\rm e}^{-\beta H_E}}{{\rm Tr}[{\rm e}^{-\beta H_E}]}\ .
\end{equation}
Using the fact that
$$
\sigma_1(s)={\rm e}^{i s\Omega\sigma_3/2}\,\sigma_1\,
{\rm e}^{-i s\Omega\sigma_3/2}=\cos(s\Omega)\,\sigma_1\,-\,
\sin(s\Omega)\,\sigma_2\ ,
$$
the resulting master equation derived by means of the same Markov 
approximation used for~(\ref{wcl2}) reads, with 
$G(s)\equiv\hbox{Tr}[\,\rho_E\,B(s)\,B\,]$,
\begin{equation}
\label{ex3.2.3}
\partial_t\rho_S(t)=
i\frac{\Omega}{2}\Bigl[\sigma_3\,,\,\rho_S(t)\Bigr]+
\lambda^2\,\mathbb{D}_1[\rho_S(t)]\ ,
\end{equation}
where
\begin{eqnarray}
\nonumber
\mathbb{D}_1[\rho_S(t)]&=&\int_0^{\infty}{\rm d}s\biggl\{
G(s)\Bigl(\cos(s\Omega)\,\Bigl[\sigma_1\,,\,\sigma_1\,\rho_S(t)\Bigr]
\,-\,\sin(s\Omega)\,\Bigl[\sigma_2\,,\,\sigma_1\,\rho_S(t)\Bigr]\Bigr)\\
&+&G(-s)\Bigl(\cos(s\Omega)\,\Bigl[\rho_S(t)\,\sigma_1\,,\,\sigma_1
\Bigr]
\,-\,\sin(s\Omega)\,\Bigl[\rho_S(t)\,\sigma_1\,,\,\sigma_2\Bigr]\Bigr)
\biggr\}\ .
\label{ex3.2.4}
\end{eqnarray} 
In order to discuss the implication of this Markov approximation, it
proves convenient to adopt the vector representation~(\ref{ex3.1.3}).
Therefore, we insert the expansion of $\rho_S(t)$ along the Pauli
matrices as in~(\ref{rho}); this yields 
the following expression
for the matrix $\mathcal{H}$ and $\mathcal{D}$:
\begin{equation}
\mathcal{H}=\pmatrix{
0&0&0&0\cr
0&0&\Omega/2&0\cr
0&-\Omega/2&0&0\cr
0&0&0&0}
\ ,\qquad
{\mathcal{D}}_1=\pmatrix{
0&0&0&0\cr
0&0&b&0\cr
0&0&\alpha&0\cr
d&0&0&\alpha}\ ,
\label{3.106}
\end{equation}
where
\begin{eqnarray}
\label{ex3.2.5}
\nonumber
\alpha&=&\lambda^2\int_0^{\infty}{\rm d}s\,\cos(s\Omega)\Bigl(
G(s)+G(-s)\Bigr)\\
\label{ex3.2.6}
&=&\frac{\lambda^2}{2}\int_{-\infty}^{+\infty}{\rm d}s\,{\rm e}^{i\Omega s}\,
G(s)\,+\,\frac{\lambda^2}{2}\int_{-\infty}^{+\infty}{\rm d}s\,
{\rm e}^{-i\Omega s}\,G(s)\ ,\\
\label{ex3.2.7}
b&=&\lambda^2\int_0^{\infty}{\rm d}s\,\sin(s\Omega)\Bigl(G(s)+G(-s)\Bigr)\ ,\\
\nonumber
d&=&i\,\lambda^2\,\int_0^{\infty}{\rm d}s\,\sin(s\Omega)\Bigl(
G(s)-G(-s)\Bigr)\\
\label{ex3.2.8}
&=&\frac{\lambda^2}{2}\int_{-\infty}^{+\infty}{\rm d}s\,{\rm
e}^{i\Omega s}\,
G(s)\,-\,\frac{\lambda^2}{2}\int_{-\infty}^{+\infty}{\rm d}s\,
{\rm e}^{-i\Omega s}\,G(s)\ .
\end{eqnarray}
Because of $b$, the $3\times 3$ matrix ${\mathcal{D}}_1^{(3)}=
\pmatrix{
0&b&0\cr
0&\alpha&0\cr
0&0&\alpha}$
is not hermitian. 
However, it splits into a symmetric and anti-symmetric part:
\begin{equation}
\label{split}
{\mathcal{D}}_1=\underbrace{\pmatrix{
0&0&0&0\cr
0&0&b/2&0\cr
0&-b/2&0&0\cr
0&0&0&0}}_{\mathcal{H}^{(2)}}+\underbrace{\pmatrix{
0&0&0&0\cr
0&0&b/2&0\cr
0&b/2&\alpha&0\cr
d&0&0&\alpha}}_{\mathcal{D}}\ ;
\end{equation}
the contribution $\mathcal{H}^{(2)}$ can be absorbed into $\mathcal{H}$ 
as a second order correction to the free Hamiltonian.
Now the corresponding matrix $\mathcal{D}^{(3)}$ is hermitian, but, again because of
$b$, apparently non-positive.
Suppose $b=0$, in order to check that the generated 
semigroup is positivity preserving, according to~(\ref{diss2}), it is 
necessary that $\alpha\geq0$. 
This is indeed so: in fact, from the KMS conditions~(\ref{KMS}) applied
to the thermal state $\rho_E$, it follows that
\begin{equation}
\label{ex3.2.9}
\alpha\,+\,d=\lambda^2\int_{-\infty}^{+\infty}{\rm d}s\,{\rm e}^{is\Omega}\,
G(s)\ ,\quad
\alpha\,-\,d={\rm e}^{-\beta\Omega}(\alpha+d)\ .
\end{equation}
If we thus show that $\alpha+d\geq0$, then $\alpha\geq 0$; this is
achieved by
explicitly computing the two-point correlation function within the
simplifying framework where $\rho_E$ is effectively a density matrix,
with orthonormal basis 
$H_E\vert\varepsilon_j\rangle=\varepsilon_j\vert\varepsilon_j\rangle$.
With $\mathcal{Z}_\beta$ the normalization of $\rho_E$, 
\begin{eqnarray}
\nonumber
\alpha+d&=&\mathcal{Z}_\beta
\int_{-\infty}^{+\infty}{\rm d}s\,{\rm e}^{is\Omega}\,
{\rm Tr}\biggl({\rm e}^{-\beta H_E}{\rm e}^{isH_E}\, B\,
{\rm e}^{-isH_E}B\biggr)\\
\nonumber
&=&\mathcal{Z}_\beta\sum_{j,k}{\rm e}^{-\beta \varepsilon_j}\,
\vert\langle\varepsilon_j\vert
B\vert\varepsilon_k\rangle\vert^2\int_{-\infty}^{+\infty}
{\rm d}s\,{\rm e}^{is(\Omega+\varepsilon_j-\varepsilon_k)}\\
\label{positivity}
&=&2\pi\,\mathcal{Z}_\beta\sum_{j,k}{\rm e}^{-\beta \varepsilon_j}\,
\vert\langle\varepsilon_j\vert
B\vert\varepsilon_k\rangle\vert^2\,
\delta(\Omega+\varepsilon_j-\varepsilon_k)\geq 0\ .
\end{eqnarray}

We are now in position to study the consequences of the presence of
$b$ and $d$. We shall use the same argument developed in Example 3.1;
at time $t=0$, (\ref{det}) gives
$$
\frac{{\rm d}}{{\rm
d}t}\hbox{Det}(\rho)=2b\rho_1\rho_2+\alpha\rho_2^2+d\rho_3+\alpha\rho_3^2\ .
$$
One checks that 
$$
\rho_1=\frac{1}{2}\sqrt{\frac{4\alpha^2-d^2}{\alpha^2+b^2}}\ ,\quad
\rho_2=-\frac{b}{2\alpha}\sqrt{\frac{4\alpha^2-d^2}{\alpha^2+b^2}}\ ,\quad 
\rho_3=-\frac{d}{2\alpha}\ ,
$$
fulfil $\rho_1^2+\rho_2^2+\rho_3^2=1$ and thus give a pure state
$\rho=P$ ($\hbox{Det}(P)=0$). Moreover, unless $b=d=0$,
$$
\frac{{\rm d}}{{\rm d}t}\hbox{Det}(P)=
-\frac{\alpha(4b^2+d^2)}{4(\alpha^2+b^2)}<0\ .
$$
Because of~(\ref{KMS}), even if $b=0$, $d=0$ is possible only at 
infinite temperature, that is when $\beta=0$.
It thus follows, that at finite temperature, the Markov approximation
discussed in this Section does not preserve positivity, since a state
like $P$ is immediately turned into an operator with negative determinant,
thus signalling the emergence of negative eigenvalues in its spectrum.

Despite this physical inconsistency, the semigroup $\gamma_t$
generated by $\mathcal{H}+\mathcal{D}$ in~(\ref{3.106}) is
nevertheless often used in phenomenological applications 
on the basis of its good asymptotic behaviour.
In fact, it has a unique equilibrium state $\rho_{eq}$ which is easily
obtained from the condition 
$(\mathcal{H}+\mathcal{D})\vert\rho_{eq}\rangle=0$, yielding
$\rho_1=\rho_2=0$ and $\rho_3=-d/\alpha$.
Using~(\ref{ex3.2.9}), this gives:
\begin{equation}
\label{rhoeq} 
\rho_{eq}=\frac{1}{{\rm e}^{\beta\Omega/2}+{\rm e}^{-\beta\Omega/2}}
\pmatrix{{\rm e}^{-\beta\Omega/2}&0\cr0&{\rm e}^{\beta\Omega/2}}=
\frac{{\rm e}^{-\beta H_S}}{{\rm Tr}[{\rm e}^{-\beta H_S}]}\ .
\end{equation}
Therefore, as expected, 
the bath drives the embedded open system at equilibrium at
its own temperature. Nevertheless, as we shall see next,
there is no need to use an unphysical treatment in order to get 
a Markovian evolution leading to
asymptotic relaxation to equilibrium. $\Box$
\bigskip

\noindent
{\bf Remark 3.6}\quad
{
The origin of the lack of positivity-preservation lies in the way the free
time-evolution generated by $H_S$ interferes with the various
truncations. Indeed, if $\Omega=0$, so that there is no free evolution
of $S$, then $d=b=0$.
Observe, however, that in this case, according to~(\ref{diss1}), 
the dynamical maps $\gamma_t$ turn out to be completely
positive and not only positive.
In the next Section, we shall see a particular way to deal with the
free evolution that automatically leads to a semigroup of
completely positive maps. $\Box$}

\subsubsection{Weak Coupling Limit}
\label{subsub3.1.2}

In order to obtain a physically consistent Markovian approximation
of (\ref{reddyn12a}) in the limit of weak couplings, a more careful derivation
is needed.\cite{dumcke-spohn}
The formal integration of equation~(\ref{wcl0}), 
\begin{equation}
\label{cpwcl0}
\rho_S(t)={\rm e}^{t\mathbb{L}^\lambda_S}[\rho_S(0)]\,+\,
\lambda^2\int_0^t{\rm d}s\,{\rm e}^{(t-s)\mathbb{L}^\lambda_S}\circ 
\mathbb{D}_1[\rho_S(s)]\ ,
\end{equation}
can be rewritten as
\begin{equation}
\label{cpwcl1}
{\rm e}^{-t \mathbb{L}_S^\lambda}\rho_S(t)=
\rho_S(0)+\lambda^2\int_0^t{\rm d}s\,
\biggl({\rm e}^{-s\mathbb{L}^\lambda_S}\circ 
\mathbb{D}_1\circ{\rm e}^{sL^\lambda_S}\biggr)
\circ {\rm e}^{-s\mathbb{L}^\lambda_S} [\rho_S(s)]\ .
\end{equation}
Upon passing to the slow time $\tau=\lambda^2 t$, it reads
\begin{eqnarray}
\nonumber
&&
{\rm e}^{-(\tau/\lambda^2) \mathbb{L}_S^\lambda}\rho_S(\tau/\lambda^2)=
\rho_S(0)\\
\label{cpwcl1.0}
&&\hskip 1cm
+\int_0^\tau{\rm d}s\,
\Biggl({\rm e}^{-(s/\lambda^2)\mathbb{L}^\lambda_S}\circ \mathbb{D}_1\circ
{\rm e}^{(s/\lambda^2)L^\lambda_S}\Biggr)\circ
{\rm e}^{-(s/\lambda^2)L^\lambda_S}  [\rho_S(s/\lambda^2)]\ .
\end{eqnarray}
Using~(\ref{wcl1.1}) and~(\ref{wcl2}) one has
\begin{eqnarray}
\nonumber
&&
{\rm e}^{-(s/\lambda^2)\mathbb{L}^\lambda_S}\circ \mathbb{D}_1\circ
{\rm e}^{(s/\lambda^2)\mathbb{L}^\lambda_S}[\rho_S]=
\,-\,\sum_{\alpha,\beta}\int_0^{\infty}{\rm d}u\\
\nonumber
&&
\times\,\Biggl\{
G_{\alpha\beta}(u)\Biggl[{\rm e}^{(is/\lambda^2)H_S^\lambda}\,
V_\alpha(u)\,{\rm e}^{-(is/\lambda^2)H_S^\lambda}\,,\,
{\rm e}^{(is/\lambda^2)H^\lambda_S}\,V_\beta\,
{\rm e}^{-(is/\lambda^2)H^\lambda_S}\,\rho_S\Biggr]\\
&&
\label{wcl2.1}
+G_{\beta\alpha}(-u)
\Biggl[\rho_S\,{\rm e}^{(is/\lambda^2)H^\lambda_S}\,V_\beta\,
{\rm e}^{-(is/\lambda^2)H^\lambda_S}
\ ,\,{\rm e}^{(is/\lambda^2)H_S^\lambda}\,V_\alpha(u)\,
{\rm e}^{-(is/\lambda^2)H_S^\lambda}\Biggr]\Biggr\}\ .\qquad
\end{eqnarray}
For sake of simplicity, we assume now the effective Hamiltonian
$H_S^\lambda$ to have discrete spectrum
$H_S^\lambda=\sum_{a=1}^n\varepsilon^\lambda_a\,P_a^\lambda$.
By inserting  
$\sum_{a=1}^nP^\lambda_a={\bf 1}_n$ into~(\ref{wcl2.1}), we get
\begin{eqnarray}
\nonumber
&&
{\rm e}^{-(s/\lambda^2)\mathbb{L}^\lambda_S}\circ \mathbb{D}_1\circ
{\rm e}^{(s/\lambda^2)\mathbb{L}^\lambda_S}[\rho_S]=\,-\,
\sum_{\alpha,\beta}\sum_{a,b,c,d=1}^n\int_0^{\infty}{\rm d}u\,
{\rm e}^{is\big[\varepsilon^\lambda_a-
\varepsilon^\lambda_b-(\varepsilon^\lambda_d-
\varepsilon^\lambda_c)\big]\big/\lambda^2}\times\\
&&\hskip 2cm 
\nonumber
\times
\Biggl\{
G_{\alpha\beta}(u)\,
\Biggl[P^\lambda_aV_\alpha(u)P_b^\lambda\,,\,
P^\lambda_c\,V_\beta\,P^\lambda_d\,\rho_S\Biggr]\\
&&\hskip 4cm
+G_{\beta\alpha}(-u)\,
\Biggl[\rho_S\,P^\lambda_a\,V_\beta\,P^\lambda_b
\ ,\,P^\lambda_c\,V_\alpha(u)\,P_d^\lambda\Biggr]\Biggr\}\ .
\label{wcl2.3}
\end{eqnarray}

When $\lambda\to0$, only the exponentials 
with $\varepsilon^\lambda_a-\varepsilon^\lambda_b=
\varepsilon^\lambda_d-\varepsilon^\lambda_c$ 
are expected to effectively contribute, since otherwise the
the term within parenthesis in~(\ref{cpwcl1.0}) rapidly oscillates,
giving vanishingly small contributions;
this approximation is known as \textit{rotating-wave
approximation}.
Also, under suitable assumptions, when $\lambda\to 0$, the eigenvalues
$\varepsilon_a^\lambda$ and the eigenprojections
$P^\lambda_a$ are expected to go into eigenvalues $\varepsilon_a$
and eigenprojectors $P_a$ of $H_S$.

In more mathematically precise terms, one can show
that the oscillating term
can be substituted by the 
following ergodic average:\cite{davies1,davies2}
\begin{equation}
\label{cpwcl3}
\widetilde{\mathbb{D}}
\equiv\lim_{T\to+\infty}\frac{1}{2T}\int_{-T}^{+T}{\rm d}s\,
{\rm e}^{-s\mathbb{L}_{H_S}}\circ \mathbb{D}_1\circ
{\rm e}^{s\mathbb{L}_S}\ .
\end{equation}
Further, when $\lambda\to 0$, $\widetilde{\mathbb{D}}$ commutes with 
${\rm e}^{-(s/\lambda^2)\mathbb{L}^\lambda_S}$, so that,
one finally gets the Markov approximation
\begin{equation}
\label{cpwcl4}
\rho_S(t)={\rm e}^{t\mathbb{L}^\lambda_S}[\rho_S(0)]\,+\,
\lambda^2\int_0^t{\rm d}s\,
{\rm e}^{(t-s)\mathbb{L}^\lambda_S}\circ\widetilde{\mathbb{D}}[\rho_S(s)]\ ,
\end{equation}  
which solves the master equation
\begin{equation}
\label{cpwcl5}
\partial_t\rho_S(t)=\mathbb{L}^\lambda_S[\rho_S(t)]\,+\,
\lambda^2\widetilde{\mathbb{D}}[\rho_S(t)]\ .
\end{equation} 
Before discussing in details the form of the dissipative term
$\widetilde{\mathbb{D}}$, we reconsider Example 3.4.
\bigskip

\noindent
{\bf Example 3.5}\quad
In the vectorial representation of Example 3.1, the free
time-evolution amounts to the $4\times 4$ matrix
$$
\mathcal{U}^S_s=\pmatrix{
1&0&0&0\cr
0&\cos(s\Omega)&-\sin(s\Omega)&0\cr
0&\sin(s\Omega)&\cos(s\Omega)&0\cr
0&0&0&1}\ ,
$$
which yields the ergodic average
\begin{equation}
\label{ex3.3.1}
\lim_{T\to+\infty}\frac{1}{2T}\int_{-T}^{+T}{\rm d}s\,
\left(
\mathcal{U}_s^S\pmatrix{0&0&0&0\cr
0&0&b&0\cr
0&0&\alpha&0\cr
d&0&0&\alpha}\mathcal{U}_{-s}^S\right)
=\pmatrix{0&0&0&0\cr
0&\alpha/2&b/2&0\cr
0&-b/2&\alpha/2&0\cr
d&0&0&\alpha}\ .\
\end{equation}
From the averaged matrix we extract the antisymmetric part 
\begin{equation}
\label{ex3.3.2}
\mathcal{H}^{(2)}\equiv\pmatrix{0&0&0&0\cr
0&0&b/2&0\cr
0&-b/2&0&0\cr
0&0&0&0}\ ,
\end{equation}
which gives a second order correction to $\mathcal{H}$
in~(\ref{ex3.2.5}), and the dissipative term
\begin{equation}
\label{ex3.3.3}
\mathcal{D}_2=\pmatrix{0&0&0&0\cr
0&\alpha/2&0&0\cr
0&0&\alpha/2&0\cr
d&0&0&\alpha}\ .
\end{equation}
A part from the entry $d$, we see that the $3\times 3$ matrix 
$\mathcal{D}_2^{(3)}$ satisfies the conditions~(\ref{diss1}); 
also, the presence of $d$ does not spoil
the complete positivity of the generated map $\gamma_t$, for, as we
shall show, the ergodic average automatically yields completely
positive semigroups.
We conclude the example by noticing that the equilibrium state,
determined by
$
(\mathcal{H}+\mathcal{H}^{(2)}+{\mathcal{D}}_2)\vert\rho_{eq}\rangle=0
$
is as in~(\ref{rhoeq}). $\Box$
\bigskip

In order to prove the claim at the end of the preceding Example, we
now study in detail the structure of~(\ref{cpwcl3}) and show that
it amounts to a generator as in~(\ref{GKF1})--(\ref{GKF2}) with
positive Kossakowski matrix.

We set $\omega=\varepsilon_a-\varepsilon_b$ with $\varepsilon_a$ running
over the spectrum of $H_S$; by means of the decompositions
\begin{equation}
\label{Kraus1}
V_\alpha(\omega)\equiv
\sum_{\varepsilon_a-\varepsilon_b=\omega}P_a\,V_\alpha\,P_b\  ,\qquad
V_\beta(-\omega)\equiv\sum_{\varepsilon_d-\varepsilon_c=\omega}
P_c\,V_\beta\,P_d=V_\beta^\dagger(\omega)\ ,
\end{equation}
from (\ref{wcl2.3}) and (\ref{cpwcl3}), we explicitly get
\begin{eqnarray}
\nonumber
&&
\widetilde{\mathbb{D}}[\rho_S]=\,-\,
\sum_{\alpha,\beta}
\sum_\omega\,
\int_0^{\infty}{\rm d}u\,
\Biggl\{
{\rm e}^{iu\omega}\,G_{\alpha\beta}(u)\,
\Biggl[V_\alpha(\omega)\,,\,V^\dagger_\beta(\omega)\,\rho_S\Biggr]\\
\label{wcl2.4}
&&\hskip 4.5cm
+\ {\rm e}^{-iu\omega}\,G_{\beta\alpha}(-u)\,
\Biggl[\rho_S\,V_\beta(\omega)\ ,\,V^\dagger_\alpha(\omega)\Biggr]\Biggr\}\ .
\end{eqnarray}
\noindent
Further, it follows from standard Fourier analysis that\cite{alicki-lendi}
\begin{equation}
\label{Kraus3}
\int_0^{\infty}{\rm d}t\,{\rm e}^{it\omega}\,
G_{\alpha\beta}(t)=\frac{h_{\alpha\beta}(\omega)}{2}
\,+i\,s_{\alpha\beta}(\omega)\ ,
\end{equation}
where
\begin{equation}
\label{Kraus4}
h_{\alpha\beta}(\omega)\equiv\int_{-\infty}^{+\infty}{\rm d}t\,
{\rm e}^{it\omega}\,G_{\alpha\beta}(t)=
h^*_{\beta\alpha}(\omega)
\end{equation}
is a positive matrix,%
\footnote{This can be proved by applying to $h_{\alpha\beta}(\omega)$
considerations similar to the ones employed 
in~(\ref{positivity}) of Example 3.4 that use the spectral
decomposition of the environment time-evolution.}
while
\begin{equation}
\label{Kraus5}
s_{\alpha\beta}(\omega)\equiv\frac{1}{2\pi}\mathcal{P}\int_{-\infty}^{+\infty}
{\rm d}w\,
\frac{h_{\alpha\beta}(\omega)}{w-\omega}=s^*_{\beta\alpha}(\omega)\ ,
\end{equation}
and $\cal P$ indicates principal value.
In terms of the Fourier transforms of the two-point correlation
functions of the environment, one then rewrites
\begin{eqnarray}
\label{Kraus6}
\widetilde{\mathbb{D}}[\rho_S]&=&\, -i\Biggl[
\underbrace{\sum_{\alpha,\beta}\sum_\omega\,s_{\alpha\beta}(\omega)
V_\alpha(\omega)V_\beta^\dagger(\omega)}_{H_E^{(2)}}\,,\,
\rho_S\Biggr]\\
&+&\underbrace{\sum_{\alpha,\beta}\sum_\omega\,h_{\alpha\beta}(\omega)\Biggl(
V^\dagger_\beta(\omega)\,\rho_S\,V_\alpha(\omega)\,-\,
\frac{1}{2}\Biggl\{V_\alpha(\omega)V_\beta^\dagger(\omega)\,,\,\rho_S
\Biggr\}\Biggr)}_{\mathbb{D}_2}\ .
\label{Kraus7}
\end{eqnarray}
We thus arrive at the following explicit form of the generator~(\ref{cpwcl5})
\begin{equation}
\label{cpwclf}
\partial_t\rho_S(t)=-i\biggl[H_S+\lambda H_E^{(1)}+\lambda^2
H_E^{(2)}\,,\,\rho_S(t)\biggr]\,+\,\lambda^2\, \mathbb{D}_2[\rho_S(t)]\ .
\end{equation}
We thus see that, beside the second order dissipative term of the 
form~(\ref{lindblad}),  the environment contributes with a first-order
mean-field term and also a second order term to the redefinition of
an effective Hamiltonian of $S$.
Further, the Kossakowski matrices $[h_{\alpha\beta}(\omega)]$ 
are positive-definite for all $\omega$.
\bigskip

\vbox{
\noindent
{\bf Remarks 3.7}\quad
{\vspace{-.2cm}
\begin{enumerate}
\item
Formally, the last statement above
follows from Bochner's theorem 
(see for instance Appendix A.3 in Ref.[\refcite{alicki-lendi}]) 
which states that the
Fourier transform $h_{\alpha\beta}(\omega)$ of a function of positive
type, as the two-point correlation functions $G_{\alpha\beta}(t)$,
gives rise to a
positive-definite matrix.
\item
The ergodic average elimination of the rapid
oscillations due to the
free system dynamics results in a
semigroup consisting of completely positive maps.
\item
An equivalent derivation of the physically consistent Markovian
master equation (\ref{cpwclf}) can be obtained through the
so-called {\it stochastic limit}:\cite{accardi1,accardi2}
it provides a different, more abstract justification for the ergodic
average (\ref{cpwcl3}). 
\item
A so-called quantum FokkerÐPlanck master equation has been derived 
in [\refcite{vacchini}]; it provides a description of
quantum Brownian motion with additional corrections 
due to quantum statistics. $\Box$
\end{enumerate}
}
}
\bigskip

The mathematical consistency of the previous manipulations rests on
some assumptions on the behaviour of two-point correlation functions.
In particular, as proven in Refs.[\refcite{davies1}, \refcite{davies2}],
the following conditions turn out to be sufficient.
The first one is a condition on the norms of the system
observables $V_\alpha$ which are coupled to the environment, that is
$\sum_\alpha\|V_\alpha\|<\infty$.
The second is an assumption on the environment which is assumed to be
\textit{quasi-free}, namely all its higher
time-correlation functions can be expressed as sums of two-point
correlation functions.
The third regards the decay of two-point correlation functions which
is assumed to be such that
there exists $\eta>0$ for which
\begin{equation}
\int_0^{\infty}{\rm d}t\, 
\left|G_{\alpha\beta}(t)\,\right|(1+t)^\eta< a\ ,
\label{davies-condition}
\end{equation}
with $a$ independent of $\alpha,\beta$.
\bigskip

\noindent
{\bf Remark 3.8}\quad The derivation of the master equation 
(\ref{cpwclf}) is based on the assumption
of a weak coupling between subsystem and external bath;
a master equation of the same kind can also describe a different situation,
in which the interactions between subsystem and environment are rare,
but not necessarily weak. The typical instance is that of an atom, effectively
modeled as an $n$-level system, immersed in a rarefied gas, that represents
the environment. The collisions of the atom with the gas particles are
described by the scattering operator, which involves the full interaction
Hamiltonian; however, these are rare events, since the gas is by hypothesis
very dilute. Using mathematically rigorous techniques, one can then show
that in the so-called {\it low density limit} the atom subdynamics is
generated by a master equation in Kossakowski-Lindblad form.\cite{dumcke,alicki-lendi,alicki}
$\Box$
\bigskip

\subsubsection{Singular Coupling Limit}
\label{subsub3.1.4}

Unlike in the weak coupling limit, in the singular coupling regime,
it is the decay time of correlations in the environment that becomes small,
$\tau_E\to 0$, while the typical variation time
of $\rho_S$, $\tau_S$, stays constant.
Concretely, the two-point correlation functions appearing in the
dissipative contributions to the open dynamics are made tend to Dirac
deltas:\cite{gorini2,gorini3,gorini4}
\begin{equation}
\label{scl00}
G_{\alpha\beta}(t)\to C_{\alpha\beta}\, \delta(t)\ ,
\end{equation}
so that the dissipative effects manifest
themselves on time-scales of order $t$.

In order to implement such a behaviour and derive the corresponding
master equation, we consider~(\ref{global1}) 
and rescale it as follows:\cite{spohn}
\begin{equation}
\label{scl0}
H_{S+E}=H_S\otimes{\bf 1}_E\,+\,\varepsilon^{-2}{\bf 1}_S\otimes
H_E\,+\,\varepsilon^{-1}\,H'\ .
\end{equation}
Here we assume $H'=\sum_\alpha V_\alpha\otimes B_\alpha$, with
centered environment operators $B_\alpha$, {\it i.e.} ${\rm Tr}[B_\alpha \rho_E]=\,0$.
Then, the same steps leading to~(\ref{wcl00}) yield
\begin{eqnarray}
\nonumber
\rho_S(t)&=&{\rm e}^{t\mathbb{L}_{H_S}}[\rho_S(0)]\,-\,
\varepsilon^{-2}\,\int_0^t{\rm d}v\,\,\int_0^v{\rm d}u\ 
{\rm e}^{(t-v)\mathbb{L}_S}\,\times\\
\label{scl101}
&&\hskip 2cm
\times\ 
{\rm Tr}_E\Bigl(\Bigl[H'\,,\,
{\rm e}^{u\,\mathbb{L}^{QQ}}\Bigl[H'\,,\,\rho_S(v-u)\otimes\rho_E\Bigr]
\Bigr]\Bigr)\ .
\end{eqnarray}
By letting $u=\varepsilon^2 w$ and $\varepsilon\to 0$, one formally
gets
\begin{eqnarray}
\nonumber
\rho_S(t)&=&{\rm e}^{t\mathbb{L}_{H_S}}[\rho_S(0)]\,
-\,\int_0^t{\rm d}v\,\,{\rm e}^{(t-v)\mathbb{L}_{H_S}}\,\times\\
\label{scl111}
&&\hskip 2cm
\times\
\int_0^{\infty}{\rm d}u\, 
{\rm Tr}_E\Bigl(\Bigl[H'\,,\,{\rm e}^{u\,\mathbb{L}_{H_E}}
\Bigl[H'\,,\,\rho_S(v)\otimes\rho_E\Bigr]\Bigr]\Bigr)\ .
\end{eqnarray}
Explicitly, the above expression solves the evolution equation
\begin{equation}
\partial_t\rho_S(t)=-i\Bigl[H_S\,,\,\rho_S(t)\Bigr]\, +\, 
\widehat{\mathbb{D}}[\rho_S(t)]\ ,
\end{equation}
where
\begin{eqnarray}
\nonumber
\widehat{\mathbb{D}}[\rho_S(t)]&=&\,-\,
\sum_{\alpha,\beta}\int_0^{\infty}{\rm d}s\Biggl\{
G_{\alpha\beta}(s)\Bigl[V_\alpha,\, V_\beta\, \rho_S(t)\Bigr]\\ 
\label{scl2}
&&\hskip 4cm
+G_{\beta\alpha}(-s)\Bigl[\rho_S(t)\,V_\beta\ ,\
V_\alpha\Bigr]
\Biggr\}\ .
\label{sing3}
\end{eqnarray}
As was done in the weak coupling regime, one writes 
$$
\int_0^{\infty}G_{\alpha\beta}(s)=
\frac{h_{\alpha\beta}}{2}+is_{\alpha\beta}\ ,
$$
so that (\ref{sing3}) splits into a second-order Hamiltonian contribution,
with Hamiltonian
\begin{equation}
H^{(2)}=\sum_{\alpha,\beta} s_{\alpha\beta} V_\alpha V_\beta\ ,
\end{equation}
and a true dissipative term of the form:
\begin{equation}
\mathbb{D}_3[\rho_S]=
\sum_{\alpha,\beta}h_{\alpha\beta}\biggl(
V_\beta\,\rho_S\,V_\alpha\,-\,
\frac{1}{2}\Bigl\{V_\alpha V_\beta\,,\,\rho_S
\Bigr\}\biggr)\ .
\end{equation}
Altogether, this yields the following semigroup generator:
\begin{equation}
\nonumber
\partial_t\rho_S(t)=-i\Bigl[H_S+H_E^{(2)},\,\rho_S(t)\Bigr]\,+\,
\mathbb{D}_3[\rho_S(t)]\ .
\label{cpscl1.1}
\end{equation}
\bigskip

\noindent
{\bf Remarks 3.9}\quad

{\begin{enumerate}
\item
The singular coupling limit, as well as the weak coupling limit, leads
to a generator of the Lindblad form~(\ref{lindblad}) with a positive
Kossakowski matrix and thus to a dissipative semigroup consisting of
completely positive maps.
Notice that, unlike in the weak coupling limit case, the operators
contributing to the dissipative term $\mathbb{D}_3$ are
hermitian.
\item
By going to the slow time $\tau=\varepsilon^2t$, the global 
Hamiltonian~(\ref{scl0}) becomes
\begin{equation}
\label{scwl}
H_{S+E}=\varepsilon^2\,H_S\otimes{\bf 1}_E\,+\,{\bf 1}_S\otimes
H_E\,+\,\varepsilon\,H'\ .
\end{equation}
This shows that the singular coupling limit amounts to a weak coupling
limit where the free motion generated by the
Hamiltonian $H_S$ is of the same order of the
dissipative effects.\cite{palmer,spohn}
\item
Unlike the weak coupling limit, thermal Bose or Fermi heat baths 
provide frameworks suited to the singular coupling limit only if their 
temperature is infinite.\cite{gorini1,gorini3,gorini4}
Indeed, the two-point correlation functions may tend to Dirac deltas
in time only if their Fourier transforms tend to a constant,
but this is in contradiction with the KMS-conditions
(\ref{KMS}), unless $\beta=0$
and thus $T=\infty$.
\item
The infinite temperature limit is equivalent to add to the system
Hamiltonian $H_S$ an external Gaussian stochastic
potential $V(t)$.%
\footnote{This approximation is very useful in many phenomenological
applications, in particular for treating decoherence effects in the
propagation of elementary particles in matter;
for details, see \hbox{Refs.[\refcite{bf13}, \refcite{bf14}].}}
In such a case, the partial trace ${\rm Tr}_E$ with respect to
$\rho_E$, is substituted by the average with respect to the Gaussian
noise driving $V(t)$.\cite{gorini3}
The resulting generator is as in~(\ref{cpscl1.1}) with a real-symmetric
Kossakowski matrix $[h_{\alpha\beta}]$ (for a specific instance,
see Example 3.8 below); consequently, the totally
depolarized state $\tau_n={\bf 1}_n/n$ is an invariant equilibrium state
and corresponds to the state (\ref{rhoeq}) 
with $\beta=0$, that is at infinite temperature. $\Box$
\end{enumerate}
}
\medskip

\noindent
{\bf Example 3.6}\quad
We now reexamine Example 3.4 in the light of the singular coupling
limit; thus, we consider the rescaled total Hamiltonian
\begin{equation}
\label{ex3.5.1}
H_{S+E}=\frac{\Omega}{2}\sigma_3\otimes{\bf 1}_E\,+\,{\bf 1}_S\otimes
\varepsilon^{-2}\sum_k\omega_k
a^\dagger_k a_k\,+\,\varepsilon^{-1}\,\sigma_1\otimes B\ ,
\end{equation}
With respect to~(\ref{3.106}) and~(\ref{ex3.3.3}), in the vectorial
representation, the singular coupling limit gives the following
dissipative matrix
\begin{equation}
\label{ex3.5.2}
\mathcal{D}_3=\pmatrix{0&0&0&0\cr
0&0&0&0\cr
0&0&\alpha&0\cr
0&0&0&\alpha}\ ,\quad\hbox{where\ now}\quad
\alpha=2\,\mathcal{R}e\Biggl(
\int_0^{\infty}{\rm d}t\,G(t)\Biggr)\ .
\end{equation}
This result directly follows from~(\ref{ex3.2.6})--(\ref{ex3.2.8}) by setting
$\Omega=0$ and thus neglecting the free system time-evolution whence
$b=d=0$.
By the same argument developed in Example 3.4, it turns out
that $\alpha\geq0$ and the generated semigroup consists of
completely positive maps. $\Box$

\subsubsection{A Convolution-Free Master Equation}
\label{subsub3.1.3}

The Markov approximations presented in the previous three sections all
start from the convolution master equation~(\ref{reddyn11}).
There is however a derivation which is convolution-free and also leads to the
Redfield-type equations as in~(\ref{wcl0}).\cite{budimir1}

Within the approximation of weak coupling between system and 
environment ($\lambda\ll 1$), one starts from the complete master
equation for the compound system
$T=S+E$,
$$
\partial_t\rho_T(t)=(\mathbb{L}_0+\lambda\, \mathbb{L}')[\rho_T(t)]\ ,
$$
where $\mathbb{L}_0$ comprises commutators with respect to system and
environment Hamiltonians and $\mathbb{L}'$ the commutator with an
interaction Hamiltonian as in~(\ref{reddyn9}), with centered environment
observables.
One then passes to the interaction representation 
$\widetilde{\rho}_T(t)\equiv\exp(-t\mathbb{L}_0)[\rho_T(t)]$ and solves 
iteratively the resulting time-dependent evolution equation
$$
\partial_t\widetilde{\rho}_T(t)=\lambda \mathbb{L}'(t)
[\widetilde{\rho}_T(t)]\
,\quad
\mathbb{L}'(t)\equiv{\rm e}^{-t\mathbb{L}_0}\mathbb{L}'{\rm e}^{t\mathbb{L}_0}\ .
$$
With initial condition $\widetilde{\rho}_T(0)=\rho_T(0)$, 
the solution can be expressed as a series expansion in the
small parameter $\lambda$:
$$
\widetilde{\rho}_T(t)=\sum_{k=0}^\infty\lambda^k\, \mathbb{N}_k(t)
[\widetilde{\rho}_T(0)]\ ,
$$
where
$$
\mathbb{N}_k(t)[\widetilde{\rho}_T(0)]\equiv\int_0^t{\rm
d}s_1\int_0^{s_1}{\rm d}s_2\cdots
\int_0^{s_{k-1}}{\rm d}s_k\,\mathbb{L}'(s_1)\circ\mathbb{L}'(s_2)\cdots
\circ\mathbb{L}'(s_k)[\rho_T(0)]\ ,
$$
and $\mathbb{N}_0(t)[\rho_T(0)]\equiv\rho_T(0)$.

Under the hypothesis of a factorized initial state,
$\rho_T(0)=\rho_S\otimes\rho_E$, with $\rho_E$ an environment equilibrium
state, $\mathbb{L}_{H_E}[\rho_E]=0$,  using the
projection introduced in (\ref{reddyn1}), one gets
$$
P[\widetilde{\rho}_T(t)]=\underbrace{
{\rm Tr}_E\big[\widetilde{\rho}_T(t)\big]
}_{\widetilde{\rho}_S(t)}\otimes\rho_E
=\sum_{k=0}^\infty\lambda^k\, {\rm Tr}_E\Bigl(\mathbb{N}_k(t)
[\rho_S\otimes\rho_E]\Bigr)\otimes\rho_E\ .
$$
Further, setting 
\begin{equation}
\label{stochexp}
\mathcal{M}_k(t)[\rho_S]\equiv{\rm Tr}_E\Bigl(\mathbb{N}_k(t)
[\rho_S\otimes\rho_E]\Bigr)\ ,\qquad
\Phi_t\equiv\sum_{k=0}^\infty\lambda^k\mathcal{M}_k(t)\ ,
\end{equation}
one writes $\widetilde{\rho}_S(t)=\Phi_t[\rho_S(0)]$ and,
by a formal inversion of $\Phi_t$, one gets 
$$
\partial_t\widetilde{\rho}_S(t)
=\dot{\Phi}_t[\rho_S(0)]=\dot{\Phi}_t\circ\Phi_t^{-1}[\widetilde{\rho}_S(t)]\ .
$$
Due to the ${\rm Tr}_E$ operation, the operators $\mathcal{M}_k(t)$
contain $k$-time correlation functions of the environment. 
At this point one makes the assumption of a thermal environment for which only
even correlation functions matter and, 
according to the weak coupling assumption, expand to lowest order in 
$\lambda^2$,
$\dot{\Phi}_t\circ\Phi_t^{-1}\simeq\lambda^2\dot{\mathcal{M}}_2(t)$.
Returning to the Schr\"odinger picture and explicitly inserting the
environment two-point correlation functions one gets the
Redfield-type equations
\begin{equation}
\partial_t\rho_S(t)=-i\Bigl[H_S\,,\,\rho_S(t)\Bigr]\
\,+\,\mathbb{D}_4[\rho_S(t)]\ ,
\end{equation}
where
\begin{eqnarray}
\nonumber
\mathbb{D}_4[\rho_S(t)]&=&
\sum_{\alpha\beta}\int_0^{\infty}{\rm d}s\biggl\{
G_{\alpha\beta}(s)\Bigl[V_\beta(-s)\,\rho_S(t)\,,\,
V_\alpha\Bigr]\\
&&\hskip 3.5cm
+G_{\beta\alpha}(-s)\Bigl[V_\alpha\,,\,\rho_S(t)
\,V_\beta(-s)\Bigr]\biggr\}\ .
\label{cf3}
\end{eqnarray}
The dissipative term $\mathbb{D}_4$ in the above 
time-evolution
equation is rather similar to~(\ref{wcl2}) and thus suffers from the
same problems in relations to positivity-preservation.
This is explicitly shown by the following example.
\bigskip

\noindent
{\bf Example 3.7}\quad
Let us consider Example 3.4 in the light of the convolutionless
approach; in the vectorial
representation, it gives the following
dissipative matrix
\begin{equation}
\label{ex3.7.1}
\widetilde{\mathcal{D}}_4=\pmatrix{0&0&0&0\cr
0&0&0&0\cr
0&-b&\alpha&0\cr
d&0&0&\alpha}\ ,
\end{equation}
where $\alpha$, $b$ and $d$ are as in Example 3.4.
As done there, we split the $3\times 3$ matrix 
$\widetilde{\mathcal{D}}^{(3)}_4=\pmatrix{0&0&0\cr
-b&\alpha&0\cr
0&0&\alpha}$ into symmetric and anti-symmetric part.
The first term provides a second order correction $\mathcal{H}^{(2)}$
to the free-Hamiltonian term, while the second one gives rise to a 
dissipative term
$$
\mathcal{D}_4=\pmatrix{0&0&0&0\cr
0&0&-b/2&0\cr
0&-b/2&\alpha&0\cr
d&0&0&\alpha}\ ,
$$
of the same form as $\mathcal{D}_1$ in Example 3.4, see the
second piece in (\ref{split}), with $b\to-b$.
Therefore, unless $b=d=0$, there exist initial states that, as
four-dimensional vectors, are mapped
by the semigroup 
$$
\mathcal{G}_t=
\exp\Bigl[t(\mathcal{H}+\mathcal{H}^{(2)}+\mathcal{D}_4)\Bigr]
$$
into vectors, corresponding to matrices with negative eigenvalues. $\Box$
\bigskip

\noindent
Further elaborations 
are then needed to make physically acceptable the dynamics 
obtained by the convolutionless method.
\bigskip

\noindent
{\bf Remark 3.10}\quad
A cure against lack of complete positivity and even of positivity has
been put forward which is based on the so-called \textit{slippage of
initial conditions} (see, \hbox{Refs.[\refcite{gnutzmann}-\refcite{wielkie}]}).
The argument is that any Markov approximation neglects a certain
initial span of time, namely a transient, during which no semigroup 
time-evolution is possible due to unavoidable memory effects.
The transient dynamics can be effectively depicted as a slippage
operator that, out of all possible initial states, selects those which
can not conflict with the Markov time-evolution when it sets in.
As we shall see in the next Section, the slippage-argument may cure the
positivity-preserving problem, but misses the point about the complete
positivity issue which emerges in full only when dealing with bi- and
multi-partite open quantum systems and with the existence of entangled
states.\cite{bf15} $\Box$
\bigskip

The convolutionless approach is often advocated in situations where,
instead of a weak coupling to an environment, the open system is
affected by a weak stochastic potential.\cite{budimir1,budimir2,budimir3}
As already observed in Remark 3.9.4, one substitutes the trace with respect
to the environment degrees of freedom by an average over the noise.
Experimental contexts where one can simulate external noisy
potentials seem to be feasible; by modulating the noise one can
enhance the non-positivity preserving character of the Redfield-type 
equations and thus expose the physical inconsistency of the Markovian
approximation on which they are based.\cite{bf3,bf6,bf12}

In particular, neutron interferometry, which has proved to be 
an extremely powerful tool to investigate gravitational, inertial and 
phase-shifting effects occurring inside the interferometer,%
\footnote{For an account of relevant results, see {\it e.g.}
Refs.[\refcite{sears,rauch}].}
may be used to investigate the notion of completely positive open system dynamics.
In order to do that, we consider below the case in which neutrons while propagating
inside the interferometric apparatus are 
subjected to weak time-dependent, stochastic magnetic fields coupled to their
spin degree of freedom. In this case, the noisy effects result 
from the classical stochastic
character of the fields and not from an interaction with
a quantum environment.
\bigskip

\eject

\noindent
{\bf Example 3.8}\quad
The states of the neutrons inside the interferometer
can then be described by a
$2\times 2$ density matrix $R(t)$.
Its dynamics is generated by a time-dependent
Hamiltonian that takes the form
\begin{equation}
H(t)=\frac{\Omega}{2}\sigma_3 + \vec\sigma\cdot\vec{\bf B}(t)\ ,
\label{st1}
\end{equation}
where the system Hamiltonian $H_S$ is as in Example 3.3, while
${\bf B}(t)=(B_1(t),B_2(t),B_3(t))$ 
represents a Gaussian stochastic magnetic field.
We assume ${\bf B}(t)$ to have  zero mean, 
$\langle{\bf B}(t)\rangle=0$, and a
stationary, real, positive-definite covariance matrix 
$G_{ij}(t)$
with entries 
\begin{equation}
G_{ij}(t-s)=\langle B_i(t)B_j(s)\rangle=G^*_{ij}(t-s)=G_{ji}(s-t)\ .
\label{st2}
\end{equation}

The time-dependent evolution equation for the 
density matrix $R$ describing the 
spin degree of freedom splits into
\begin{eqnarray}
\label{st3}
&&\partial_t R(t)=(\mathbb{L}_0+\mathbb{L}_t)[R(t)]\ ,\\
&&\mathbb{L}_0[R(t)]=-i\biggl[{\Omega\over 2}\sigma_3\,,\,R(t)\biggr]\ ,\quad
\mathbb{L}_t[R(t)]=-i\Bigl[{\bf B}(t)\cdot\vec{\sigma}\,,\,R(t)\Bigr]\ .
\label{st4}
\end{eqnarray}

Because of the stochastic field $\vec{\bf B}(t)$, the solution
$R(t)$ of (\ref{st3}) is also stochastic;
an effective spin density matrix $\rho(t)\equiv\langle R(t)\rangle$
is obtained by averaging over the noise.
At time $t=0$ we may suppose spin and noise to
decouple so that  the initial state is
$\rho\equiv\langle R(0)\rangle=R(0)$.

The effective time-evolution of $\rho(t)$ can naturally be obtained
following the techniques of the previous Section, where the
trace over the environment degrees of freedom is substituted
by the average over the classical noise, ${\rm Tr}_E\to \langle\ \rangle$.
The resulting master equation has the form (\ref{cf3}):
\begin{equation}
\partial_t\rho_S(t)=-i\biggl[\frac{\Omega}{2}\sigma_3\,,\,\rho(t)\biggr]\
\,+\,\widetilde{\mathbb{D}}_4[\rho(t)]\ ,
\label{st5}
\end{equation}
where
\begin{equation}
\widetilde{\mathbb{D}}_4[\rho(t)]=
-\sum_{i,j}\int_0^{\infty}{\rm d}s\
G_{ij}(s)\Bigl[\sigma_i\, ,\Bigl[ \sigma_j(-s), \,\rho(t)
\Bigr]\Bigr]\ .
\label{st6}
\end{equation}
The time dependent $\sigma_j(-s)$ explicitly reads:
$$
\nonumber
\sigma_j(s)={\rm e}^{is H_S}\, \sigma_j\, {\rm e}^{-i s H_S}\equiv
\sum_{k=1}^3U_{jk}(s)\sigma_k\ ,
$$
where
$$
\nonumber
U_{jk}(s)=\pmatrix{\cos\Omega s&-\sin\Omega s&0\cr
\sin\Omega s&\cos\Omega s&0\cr
0&0&1}\ .
$$
Inserting these expressions in (\ref{st6}), and extracting
the symmetric and antisymmetric contributions, the master equation
results of the form:

\begin{equation}
\partial_t\rho(t)=-i\Bigl[H_S+H^{(2)}\ ,\ \rho(t)\Bigr]\ 
+\ \mathbb{D}_4[\rho(t)]\ ,
\label{st7}
\end{equation}
where $H^{(2)}=\sum_{i,j,k=1}^3\epsilon_{ijk}\, \kappa_{ij}\,\sigma_k$,
and $\mathbb{D}_4$ is as in (\ref{ex3.1.1}),
with
\begin{eqnarray}
&&\kappa_{ij}=\frac{1}{2}\sum_{k=1}^3\int_0^\infty {\rm d}s\, \Big[G_{ik}(s)\, U_{kj}(-s)
-U_{ik}(s)\, G_{kj}(-s)\Big]\ ,\\
&&C_{ij}=\sum_{k=1}^3\int_0^\infty {\rm d}s\, \Big[G_{ik}(s)\, U_{kj}(-s)
+U_{ik}(s)\, G_{kj}(-s)\Big]\ .
\end{eqnarray}

We now study the positivity and complete positivity of the 
generated time-evolution in relation to the decay properties
of the stochastic magnetic field correlation matrix $G_{ij}(t)$.
We will consider two representative cases.
\bigskip

\noindent
$\bullet$\quad {\it White Noise}
\medskip

\noindent
Let the stochastic magnetic field have
white-noise correlations
$G_{{ij}}(t-s)=G_{ij}\,\delta(t-s)$,
where the matrix $G_{ij}$ is time-independent, symmetric and
by assumption positive-definite. Hence, $\kappa_{ij}=\,0$, and
$C_{ij}=G_{ij}$, so that the Kossakowski matrix results
positive definite and the generated semigroup is completely
positive. 
Further, note that, due to the delta-correlations, all terms
${\cal M}_k(t)$, $k\geq 3$, in the expansion (\ref{stochexp}) 
identically vanish, so that the evolution equation (\ref{st7}) is in this case
exact.\cite{gorini3}
\bigskip

\noindent
$\bullet$\quad {\it Single Component Covariance Matrix}
\medskip

\noindent
Consider a stochastic magnetic field along the $x$-direction,
$\vec{\bf B}(t)=(B(t),0,0)$, with
$$
\langle B(t)B(s)\rangle=B^2\,{\rm e}^{-\lambda|t-s|}\ .
$$
Then, 
$$
h_{ij}={\Omega B^{2}\over 2(\lambda^2+\Omega^{2})}\pmatrix{
0&1&0\cr
-1&0&0\cr0&0&0}\ ,\quad
C_{ij}={B^{2}\over\lambda^{2}+\Omega^{2}}
\pmatrix{
2\lambda&\Omega&0\cr
\Omega&0&0\cr
0&0&0}\ .
$$
Unless $\Omega=\,0$ the matrix $C_{ij}$ is not positive definite,
and therefore the generated semigroup is surely not completely positive.
Actually, adopting the vectorial formulation of Example 3.1 and setting:
$$
\alpha={2 B^2\lambda\over\lambda^2+\Omega^2}\ ,\qquad
b={ B^{2}\Omega\over
\lambda^{2}+\Omega^{2}}\  ,
$$
the relevant $3\times 3$ dissipative matrix (\ref{pauli11})
reads
$$
{\cal D}^{(3)}=\pmatrix{0&-b&0\cr
-b&\alpha&0\cr
0&0&\alpha}\ ,
$$
which is of the form of the previous Example 3.7; hence,
even positivity is not preserved.

Although apparently formal, these results are not purely academic:
indeed, the entries of $\rho(t)$ are directly accessible 
to experiments, where, by modulating a background magnetic field,
one may get close to the stochastic properties 
of the two situations just analyzed. $\Box$

\subsection{Why Completely Positive Semigroups?}
\label{subs3.2}

In the previous sections we have analyzed three Markov approximations that
lead to reduced dynamics consisting of semigroups of trace and
hermiticity preserving linear maps $\gamma_t$, $t\geq 0$, 
acting on the state-space of an open $n$-level system $S$.
As emphasized before, any mathematical effective description
of actual dissipative dynamics must in the first place preserve the 
positivity of any initial density matrix.
Only in this case, the time-evolution turns out to be consistent with
the interpretation of the state-eigenvalues as probabilities.

In the second place, one has also to care of possible couplings with
inert and remote ancillas $A$ and therefore to guarantee the positivity-preserving 
character of the semigroup of maps $\gamma_t\otimes{\rm id}_A$, too; 
this is only assured if $\gamma_t$ is completely positive.

The weak coupling limit with the ergodic average
prescription and the singular coupling limit give semigroups of
completely positive maps and therefore fully physically consistent.
On the other hand, without ergodic average or through the
convolutionless procedure one is led to Redfield-type equations which generate
dynamical maps that are, in general, not even positivity preserving.
\bigskip

\noindent
{\bf Remark 3.11}
{
One can always construct semigroups of linear
maps that are only positive and not completely positive, as seen in
Example 3.1 ({\it cf.} Eq.~(\ref{diss2})).
The real problem is the derivation of such kind of dissipative
dynamics from a microscopic interaction with the environment.
One knows that these semigroups would not be acceptable in the case of
couplings with ancillas: it is this inconsistency that essentially
forbids their construction from 
microscopic interactions.\cite{davies1,dumcke-spohn}
Indeed, all known rigorous
microscopic derivations of an open quantum reduced dynamics 
lead to completely positive semigroups. $\Box$
}
\bigskip

It has also been remarked that complete positivity guarantees full
physical consistency at the price of an order relation among typical 
life-times, which, in the two dimensional case is expressed by the
necessary inequalities~(\ref{diss1}) (see also \hbox{Example 3.2}).
It has been observed that from a strictly physical point of view, it
appears doubtful that possible couplings to remote and inert ancillas
might have concrete consequences on actual experimental contexts.
In view of this, complete positivity is often refused as a
mathematical artifact, with nice structural consequences as the
Kossakowski-Lindblad form of the generators in~(\ref{lindblad}), but
without solid physical justifications.

Leaving aside the difficulty of deriving a
positivity-preserving semigroup from fundamental 
system-environment interactions,
one may argue that positivity preservation might suffice to guarantee
physical consistency.
Indeed, it is enough to ensure the probabilistic
interpretation of the spectra of evolving states, while the danger of
possible couplings to uncontrollable ancillary systems is a far too abstract
speculation to be of any physical importance in actual experimental contexts.

However, we have also seen that the true meaning of complete
positivity can mainly be appreciated in relation to quantum entanglement of
which it is just another facet.
Quantum entanglement requires at least two systems, whereas, until
recently, the literature on open quantum systems primarily has been dealing with
single systems immersed in heat baths.

It appears then clear why the fundamental role of complete positivity
does not emerge in full in usual open quantum system contexts.
On the contrary,
the whole perspective radically changes when one considers two systems in a
same environment, as, for instance, two atoms in
a heat bath at some finite temperature.
The main point is that two such systems can be prepared in an
entangled initial state; the usual unitary time-evolution is
completely positive by its very construction and thus gentle to
entangled states, while some dissipative time-evolution may not be
completely positive and thus potentially conflicting with initial
entangled states, making negative eigenvalues emerge in the course of
the time-evolution.

More concretely, suppose the reduced dynamics of just one system $S$ 
in the environment $E$
to be given in terms of a semigroup of positive maps $\gamma_t$.
Consider then two such systems (both denoted by $S$) embedded into $E$,
without direct interaction between themselves. 
In first approximation, the states of the compound system $S+S$ 
will evolve according to the dissipative semigroups of maps
$\Gamma_t\equiv\gamma_t\otimes\gamma_t$.

We stress the difference between the two contexts: the one just
depicted, where the maps $\gamma_t\otimes\gamma_t$ 
describe the dynamics of two
systems $S$ in the same environment $E$, and the other, where the maps 
$\gamma_t\otimes{\rm id}_A$ drive a single open quantum system
$S$ coupled to a remote and inert ancilla $A$
that has nothing to do with it apart for
possible statistical correlations (entanglement) in the initial state
of $S+A$.

Suppose now that complete positivity is
a mathematical artifact and may thus be dispensed with in actual physical
contexts, where no ancilla effectively intervenes.
Since we want that at least the positivity of the states
be preserved, the crucial question is then the following:
\medskip

\centerline{
\begin{minipage}[t]{12cm}
\textsf{\noindent
Which are the conditions on $\gamma_t$ 
such that $\Gamma_t=\gamma_t\otimes\gamma_t$ be 
positivity-preserving on the state-space of the open
system $S+S$?} 
\end{minipage}
}
\medskip

\noindent
For $n$-level systems the answer is given by the following Theorem.\cite{bf16}
\medskip

\noindent
{\bf Theorem 3.3}\quad
{\it 
If a semigroup of positive maps 
$\gamma_t$ acing on $\mathcal{S}(S)$ describes the
dynamics of an $n$-level system $S$, then the semigroup 
$\Gamma_t=\gamma_t\otimes\gamma_t$ acting on $\mathcal{S}(S+S)$ 
is positivity-preserving if and only if
$\gamma_t$ is completely positive.}
\medskip

The conclusion which is to be drawn from the previous theorem is 
the following.
If $\gamma_t$ is only positivity-preserving but not completely
positive, then $\Gamma_t=\gamma_t\otimes\gamma_t$ can not be 
positivity-preserving.
Therefore, there are states of $S+S$ whose spectrum does not remain
positive under $\Gamma_t$ and which thus loose their probabilistic 
interpretation in the course of time.
Such states are necessarily entangled as $\Gamma_t$ preserves in any
case the positivity of separable states as in~(\ref{seps}). In fact,
$$
\Gamma_t[\rho_{sep}]=\sum_{ij}\lambda_{ij}\ \gamma_t[\rho^1_i]\otimes 
\gamma_t[\rho^2_j]
$$
and $\gamma_t$ is positivity-preserving by assumption.
It is thus the existence of entangled states that makes the request of
complete positivity necessary.
As already observed, unitary time-evolutions are automatically
completely positive, while dissipative semigroups are not
in general and their physical consistency depends on the way Markov
approximations are implemented.
\bigskip

\noindent
{\bf Remarks 3.12}\quad
{
\begin{enumerate}
\item
The argument of Theorem 3.3 gives a concrete explanation of why
complete positivity is necessary by hinging on the necessity that at
least the positivity of the time-evolution of bipartite systems be guaranteed. 
If $\Gamma_t=\gamma_t\otimes\gamma_t$ is positive, then $\gamma_t$
must be completely positive and $\Gamma_t$ is completely positive, too. 
Indeed, by using the Kraus-Stinespring representation (\ref{cp2}),
one checks that the composition of two completely positive maps
is again in Kraus-Stinespring form, and thus completely positive.
In the case above,
$\Gamma_t=(\gamma_t\otimes{\rm id})\circ({\rm id}\otimes\gamma_t)$.
\item
If one considers semigroups of maps of the form 
$\Gamma_t=\gamma_t^{(1)}\otimes\gamma_t^{(2)}$ with two different 
single-system semigroups, then $\Gamma_t$ can be positivity preserving
without $\gamma_t^{(1,2)}$ being both completely positive.\cite{bf17}
In this case, $\Gamma_t$  is not completely positive; in fact,
its generator has a Kossakowski matrix which consists of the orthogonal sum
of those of the generators of $\gamma^{(1)}_t$ and $\gamma^{(2)}_t$, and one of
them is not positive. Then, $\Gamma_t$ is not robust
against the coupling of $S+S$ to ancillas.
\item
An important ingredient in the proof of Theorem 3.3 is time-continuity
and the resulting existence of a generator; without continuity, there are 
counterexamples to its conclusions.
For instance, take the transposition map $\mathbb{T}_n$,
which can not be continuously connected to the identity.
It is not completely
positive, but it is easy to check that 
$\mathbb{T}_n\otimes\mathbb{T}_n$ is nevertheless a positive map.

\item
With reference to the slippage of initial conditions argument briefly
mentioned in Remark 3.10, its usage in the bipartite context would lead
to argue that the transient effects prior to the Markovian regime
are such that all entangled states dangerous to a semigroup of
non-positive $\Gamma_t=\gamma_t\otimes\gamma_t$ should somehow be
eliminated by the slippage operator.
Such an operator might plausibly be constructed, but it would
be adapted to the bipartite contexts; by going to multipartite systems
it should then be re-adjusted, thus making the whole procedure
become {\it ad hoc} and 
lose the general applicability it should \hbox{possess.\cite{bf15} $\Box$}
\end{enumerate}
}

\section{Open Quantum Systems and Entanglement Generation}

In this Section we shall apply the results and considerations discussed
so far to the study of a specific instance of open quantum system,
that of atoms interacting with a bath made of quantum fields.
As we shall see, this model, besides having an intrinsic theoretical
interest, underlays many phenomenological applications
in chemistry, quantum optics and condensed matter physics. 

As remarked before, the use of the open system paradigm in modelling
specific physical situations requires an {\it a priori} unambiguous
separation between subsystem and environment.
In the weak coupling regime, this is typically achieved when
the correlations in the environment decay much faster than the 
characteristic evolution time of the subsystem: in this case,
the changes in the
evolution of the subsystem occur on time-scales that are very long, so large
that the details of the internal environment dynamics result irrelevant.
It is precisely this ``coarse grained'' evolution that is captured by the
Markovian master equations studied in Section 3.

Independent atoms immersed in external quantum fields and weakly
coupled to them constitute one of the most common and interesting
examples in which the open system paradigm finds its concrete 
realization. Indeed, in typical situations
the differences between the atom internal energy levels ($\simeq \tau_S^{-1}$)
result much smaller
than the field correlation decay constants ($\simeq \tau_E^{-1}$)
so that a clear distinction between
subsystem, the non-interacting atoms, and environment, the external fields,
is automatically achieved.

The atoms are usually treated in a non-relativistic approximation,
as independent $n$-level systems, with zero size, while the environment
is described by a set of quantum fields, often the electromagnetic field,
in a given quantum state, typically either a temperature state or simply 
the vacuum state. The interaction of the fields with the atoms
is taken to be of dipole type, a well justified approximation
within the weak coupling assumption. Even in this simplified setting,
that ignores all intricacies related to the internal atom
structure and to the full coupling with the electromagnetic field,
the model is of great relevance both theoretically and
phenomenologically: indeed, with suitable adaptations,
it is able to capture the main features of the dynamics 
of very different physical systems, like ions in traps,
atoms in optical cavities and fibers, impurities in phonon 
fields (many concrete examples are presented
in Refs.[\refcite{alicki-lendi}-\refcite{weiss}]).

Despite this ample range of possible applications and the 
attention devoted to it in the recent literature,
no particular care is generally adopted in the various derivations of 
consistent subdynamics. As a result, time-evolutions 
that are not even positive, of the type discussed in Examples 3.3 and
3.8, have often been adopted
in order to describe the subsystem physical properties: we have already
pointed out that physical inconsistencies might be the outcome
of such ill-defined dynamics.

As a relevant application of the theory of open quantum systems
dynamics, in the following we shall
explicitly derive and study in detail the reduced time-evolution
obtained by tracing over the unobserved (infinite)
field degrees of freedom. For simplicity, we shall restrict our
attention to two-level atoms in interaction with a collection
of independent, free, scalar fields in arbitrary $d$ space-time
dimensions, assumed to be in a state at temperature $T=1/\beta$.

The time-evolution of a single atom will be first analyzed. 
Following the steps outlined in Section 3,
the weak coupling limit procedure will be adopted;
it leads to a mathematically sound and physically consistent
expression for the atom subdynamics in terms of completely
positive quantum dynamical semigroups. Not surprisingly,
the atom is found to be subjected to dissipative and decohering
effects that asymptotically drive its state to an equilibrium 
configuration that is exactly thermal, at temperature $T$.

The case of a subsystem composed by more than one atom
can be similarly treated. Again for simplicity we shall limit
the discussion to the case of two, independent two-level atoms
immersed in the same environment of thermal quantum fields.
The corresponding master equation that, in the weak coupling limit,
describes the reduced time-evolution of the atoms can be obtained
by generalizing the techniques used in the single atom case.
It again generates a semigroup of completely positive maps,
that can be studied in detail. In particular, the asymptotic
equilibrium state can be computed explicitly, and in general
it turns out to be an entangled state of the two atoms, 
even in the case of a totally separable initial state.

This remarkable conclusion might appear at first sight rather surprising.
As observed before, the interaction with an environment
usually leads to decoherence and noise, mixing enhancing
phenomena typically going against entanglement.
\bigskip

\noindent
{\bf Example 4.1}\quad
As a drastic instance of mixing-enhancing behaviour, let us consider the
so-called depolarizing channel affecting a two-level system; it is
described by the following Kossakowski-Lindblad master equation 
$$
\partial_t\rho_t=\sigma_1\rho_t\sigma_1+\sigma_2\rho_t\sigma_2+
\sigma_3\rho_t\sigma_3\,-\,3\rho_t\ .
$$
Its solution is 
$\displaystyle\rho_t=(1-{\rm e}^{-4t})\,{\bf 1}_2/{2}+{\rm
e}^{-4t}\rho$, 
so that any initial state $\rho$, even a pure one, with zero von Neumann
entropy, goes asymptotically
into the totally depolarized state ${\bf 1}_2/{2}$
with maximal entropy $\log2$. $\Box$
\bigskip

Since the closer to the totally depolarized state, the less entangled
a state is, one generally expects that when a bipartite
system is immersed in an environment, quantum correlations
that might have been created before by a direct interaction between the
two subsystems actually disappear. And this could occur even in
finite time.\cite{diosi,bfp1} The degrading of entanglement by an 
external environment is clearly a curse in quantum information,
where one tries to maintain quantum correlations long enough to
complete a quantum computation. Various error correcting strategies
have been devised precisely 
to counteract environment decoherence.\cite{nielsen}

However, an external environment can also provide an indirect
interaction between otherwise totally decoupled subsystems
and therefore a means to correlate 
them.\cite{braun,kim,milburn,ficek}
This phenomenon has
been first established in exactly solvable models: there,
correlations between the two subsystems take place during a
short time transient phase, where the reduced dynamics
of the subsystems contains memory effects.

Remarkably, entanglement generation may occur also in the
long time, Markovian regime, through a purely noisy
mechanism:\cite{bf19} it is precisely this situation that is relevant
in the analysis of the dynamics of two independent atoms
interacting with the same vacuum quantum field.

\subsection{Single Atom Systems: Master Equation}

We shall first deal with a single atom in weak interaction
with a collection of free scalar fields at temperature $T$,
in $d$ space-time dimensions.
As mentioned before, we are not interested in the details
of the atom internal dynamics. We shall therefore model it,
in a nonrelativistic way, as a simple two-level system,
which can be fully described in terms of a two-dimensional
Hilbert space.

In absence of any interaction with the external scalar fields,
the atom internal dynamics will be driven by a $2\times 2$
Hamiltonian matrix $H_S$, that in the chosen basis can be taken
to assume the most general form ({\it cf.} Examples 2.2, 3.1)
\begin{equation}
H_S={\omega\over 2}\, \vec n\cdot\vec\sigma\ ,
\label{4.1}
\end{equation}
where $\sigma_i$, $i=1,2,3$ are again the Pauli matrices, $n_i$, $i=1,2,3$
are the components of a unit vector, while $\omega$ represents the
gap between the two energy eigenvalues. 

To conform with the standard open system paradigm
(see~(\ref{global1})), the interaction of the atom 
with the external scalar fields is assumed to be weak and described by
a Hamiltonian $H'$ that is linear in both atom and field
variables:
\begin{equation}
H'=\sum_{i=1}^3\sigma_i\otimes \Phi_i(x)\ .
\label{4.2}
\end{equation}
As the atom is taken to be an idealized point-particle, without size
(see Remark 4.1.2 below), the interaction is
effective only at the atom position $x^\mu$, $\mu=0, 1,\ldots, d$,
that, without loss of generality, can be chosen to be
the origin of the reference frame. As a consequence,
all coordinates vanish, $x^\mu=\,0$, $\mu=1, 2,\ldots, d$,
except for the time variable, $x^0\equiv t$. The operators
$\Phi$ represent the external quantum fields,
and are taken to be spinless and massless for simplicity. 
They evolve in time as free relativistic fields with an Hamiltonian $H_\Phi$
whose explicit expression need not be specified in detail.
Further, we shall assume the atom and the fields to be prepared
at time $t=\,0$ in an uncorrelated state, with the fields in
the temperature state $\rho_\beta$, as in (\ref{ex3.2.2}), and the atom in a
generic initial state $\rho(0)$.

\bigskip
\noindent
{\bf Example 4.2}\quad 
Although apparently very simplified, the model
presented above is of relevance in various phenomenological applications.
As an instance, let us discuss the case of a two-level atom in interaction
with the electromagnetic field in an optical fiber
through the standard dipole coupling.\cite{leclair1} 
In this case, working
in three-dimensional space, the interaction Hamiltonian
takes the form:
\begin{equation}
H'=-{\vec d}\cdot{\vec E}({\vec x})\ ,
\label{4.3}
\end{equation}
where $\vec d$ denotes the operator representing the
atom dipole moment, while $\vec E$
is the electric field at the atom position $\vec x$.

In absence of the field, the atom dynamics can again be taken to be driven
by the generic two-level Hamiltonian (\ref{4.1}), with eigenstates $|-\rangle$
and $|+\rangle$, while the electromagnetic
field is quantized using the Maxwell action:
\begin{equation}
{\cal S}=-\frac{1}{4}\int {\rm d}^3\vec x\, dt\, F_{\mu\nu} F^{\mu\nu}\ ,
\label{4.4}
\end{equation}
where $F_{\mu\nu}=\partial_\mu A_\nu-\partial_\nu A_\mu$ is the 
field strength expressed in terms 
of the electromagnetic $4$-vector potential
$A_\mu$. Without loss of generality, we shall choose
to work in the familiar $A_0=\,0$ gauge, so that only the
vector potential $\vec A$ will be relevant.

Since the dipole moment $\vec d$ is a vector operator with odd
parity, one finds: 
$\langle -|\vec d\,|-\rangle=\langle +|\vec d\,|+\rangle=\,0$. 
Furthermore, one can write:
\begin{equation}
\langle +|\vec d\,|-\rangle= \langle -|\vec d\,|+\rangle^*
=|\vec d|\, {\rm e}^{i\theta}\ \vec u\ ,
\label{4.5}
\end{equation}
where $\vec u$ is the unit vector that defines the
orientation of the atom in space, while $\theta$ is a phase.
As a consequence, the representation of the operator $\vec d$
on the two-dimensional atom Hilbert space
has vanishing diagonal elements, so that one can rewrite
(\ref{4.3}) as:
\begin{equation}
H'=-|\vec d|\, E(\vec x)\, \Big[ {\rm e}^{i\theta}\sigma_+
+{\rm e}^{-i\theta}\sigma_-\Big]\ ,
\label{4.6}
\end{equation}
where $E(\vec x)\equiv\vec u\cdot\vec E(\vec x)$ and 
$\sigma_\pm=\sigma_1\pm i\sigma_2$. Further, note that
the phase $\theta$ can be reabsorbed into a redefinition of the
Pauli matrices, $\sigma_\pm\to {\rm e}^{\mp i \theta}\sigma_\pm$,
without affecting their algebraic properties; in other terms,
there is no loss of generality in setting $\theta$ to zero.

Let us now exploit the fact that the atom is inserted in an optical fiber.
One can always choose the direction along the fiber to coincide
with the $z$-axis, the $x$- and $y$-axis representing the 
corresponding transverse directions. The spatial geometry
becomes essentially unidimensional and the vector potential
$\vec A$ can be taken to be independent from the $x$ and $y$ coordinates.
Moreover, in such a geometry the electromagnetic energy flows along the fiber
and thus only the components of $\vec A$ transverse to the fiber are
non-vanishing; further, of these only $A\equiv\vec u\cdot\vec A $
really couples to the atom. 

Taking into account all these conditions,
the Maxwell action (\ref{4.4}) reduces to
\begin{equation}
{\cal S}= \frac{\Sigma}{2}\int {\rm d}t\, {\rm d}z \Big[\partial_t A\, \partial_t A -
\partial_z A\, \partial_z A\Big]\ ,
\label{4.7}
\end{equation}
where $\Sigma=\int dx\, dy$ is the cross section area of the fiber. 
By further noting that $E=-\partial_t A$ and by introducing the dimensionless
rescaled field $\phi(t,z)=A(t,z)\sqrt\Sigma$, the interaction Hamiltonian
(\ref{4.6}) can be rewritten as:
\begin{equation}
H'=\lambda\, \partial_t\phi(0, z)\,\big[\sigma_+ +\sigma_-\big]\ ,
\label{4.8}
\end{equation}
where $\lambda=|\vec d|/\sqrt\Sigma$ plays the role of coupling constant.
It is precisely of the form (\ref{4.2}), with $\Phi_1=\partial_t\phi$
and $\Phi_2=\Phi_3=\,0$, where $\phi$ is a 1+1-dimensional free,
massless scalar field;
its dynamics is generated by
the Hamiltonian $H_\phi$ that corresponds to the action (\ref{4.7}):
\begin{equation}
H_\phi= \frac{1}{2}\int {\rm d}t\, {\rm d}z \Big[\big(\partial_t \phi\big)^2 +
\big(\partial_z \phi\big)^2\Big]\ .
\label{4.9}
\end{equation}

The scattering theory for atoms in an optical fiber 
described by interactions of the form (\ref{4.8}) has been discussed
in Refs.[\refcite{leclair2}, \refcite{leclair3}]
by means of a formal mapping to the anisotropic Kondo model.
Here instead we are interested in analyzing the full reduced dynamics
of the atom and not only its asymptotic properties. We shall do it
by turning to the more general interacting Hamiltonian in (\ref{4.2}).
$\Box$
\bigskip

The total Hamiltonian $H$ describing the complete system, the two-level atom
together with the external fields $\Phi_i$, can thus be written as
\begin{equation}
H=H_S\otimes {\bf 1}_\Phi+ {\bf 1}_S\otimes H_\Phi+ \lambda H'\ .
\label{4.10}
\end{equation}
It is precisely of the form discussed in the previous Section,
with centered environmental operators, ${\rm Tr}[\rho_\beta \Phi_i]=\,0$.
It generates the time-evolution of the 
total density matrix $\rho_{\rm tot}$, via the Hamiltonian equation
\begin{equation}
{\partial\rho_{\rm tot}(t)\over\partial t}=\, 
\mathbb{L}_H^\lambda[\rho_{\rm tot}(t)]\ ,
\label{4.11}
\end{equation}
starting at $t=\,0$ from the initial
configuration: $\rho_{\rm tot}(0)=\rho(0)\otimes \rho_\beta$.
As explained before, we want to follow the dynamics of the
reduced density matrix $\rho(t)\equiv{\rm Tr}_\Phi[\rho_{\rm tot}(t)]$
on time-scales that are long with respect to the decay time
of correlations in the environment. This is can be obtained by
suitably rescaling the time variable, $t\to t/\lambda^2$
and then taking the limit $\lambda\to 0$, following the 
mathematically precise procedure of the {\it weak coupling limit}
discussed in Section 3.3.2. 

The resulting master equation for $\rho(t)$ takes a
Kossakowski-Lindblad form~(\ref{lindblad})
\begin{equation}
{\partial\rho(t)\over \partial t}= -i \big[H_{\rm eff},\, \rho(t)\big]
 + \mathbb{D}[\rho(t)]\ ,
\label{4.12}
\end{equation}
with $\mathbb{D}$ as in (\ref{ex3.1.1}).
The effective Hamiltonian $H_{\rm eff}$ and the coefficients of the
$3\times 3$ Kossakowski matrix $C_{ij}$ that appear in (\ref{ex3.1.1})
are determined by the Fourier
transform of the field correlations:
\begin{equation}
h_{ij}(\zeta)=\int_{-\infty}^{+\infty} {\rm d}t\, {\rm e}^{i\zeta t} \langle 
\Phi_i(t)\Phi_j(0)\rangle\ .
\label{4.14}
\end{equation}
Since the fields are independent and assumed to obey a free evolution,
one finds:
\begin{equation}
\langle \Phi_i(x)\Phi_j(y)\rangle\equiv {\rm
Tr}\big[\Phi_i(x)\Phi_j(y)
\rho_\beta\big]
=\delta_{ij}\, G_\beta(x-y)\ ,
\label{4.15}
\end{equation}
where $G_\beta(x-y)$ is the standard $d$-dimensional Wightmann function for
a single massless relativistic scalar field 
in a state at inverse temperature $\beta$.\cite{haag}
With the proper $i\varepsilon$ prescription, it can be written as:
\begin{equation}
G_\beta(x)=\int \frac{{\rm d}^d k}{(2\pi)^{d-1}}\, \theta(k^0)\, \delta(k^2)
\Big[\big(1+{\cal N}_\beta(k^0)\big)\, {\rm e}^{-ik\cdot x}+{\cal
N}_\beta(k^0)\, {\rm e}^{ik\cdot x}
\Big]\,{\rm e}^{-\varepsilon k^0}\ ,
\label{4.16}
\end{equation}
where
\begin{equation}
{\cal N}_\beta(k^0)=\frac{1}{{\rm e}^{\beta k^0} -1}\ .
\label{4.17}
\end{equation}
Its Fourier transform can be readily evaluated:
\begin{equation}
{\cal G}_\beta(\zeta)=\int_{-\infty}^{+\infty} {\rm d}t\, {\rm e}^{i\zeta t}\, 
G_\beta(t)=\bigg[\frac{2|\zeta\,|^{d-2}}{(4\pi)^\frac{d-1}{2} 
\Gamma\left(\frac{d-1}{2}\right)}\bigg]\ \frac{\pi}{\zeta}\, 
\frac{1}{1-{\rm e}^{-\beta\zeta}}\ .
\label{4.18}
\end{equation}
%


\bigskip
\noindent
{\bf Remarks 4.1}\hfill\break 
\vspace{-0.4cm}
{
\begin{enumerate}
\rm
\item
As mentioned in Remark 3.7.3, the convergence in the
{\it weak coupling limit} of the evolution equation for the reduced
density matrix $\rho(t)$ to the Markovian limit (\ref{4.12})
stems from the asymptotic behavior of the two-point function
$G_\beta(t)\equiv G_\beta(t, \vec 0\,)$ for large $t$
(see, (\ref{davies-condition})). More precisely,
the combination $|G_\beta(t)|(1+t)^\eta$ should be integrable 
on the positive half real line, for some $\eta>0$.
In the case of a massless field, the integrals in (\ref{4.16}) can be expressed
in terms of the standard Poly-Gamma function, 
$\psi^{(n)}(z)=d^{n+1}\log\Gamma(z)/dz^{n+1}$; explicitly, 
one finds:\cite{prudnikov}
\begin{equation}
G_\beta(t)=\frac{1}{(4\pi)^{\frac{d-1}{2}}\, \Gamma\left(\frac{d-1}{2}\right)}
\bigg\{\frac{2}{(-\beta)^{d-2}}\, {\cal R}e\bigg[\psi^{(d-3)}\bigg(
\frac{\varepsilon-it}{\beta}\bigg)\bigg]-
\frac{(d-3)!}{(\varepsilon-it)^{d-2}}\bigg\}\ .
\label{4.19}
\end{equation}
Since $\psi^{(n)}(z)\sim 1/z^n$ for $|z|\to\infty$, ${\rm Arg}(z)<\pi$,
$G_\beta(t)$ behaves like $1/t^{d-3}$ for large $t$ in $d>4$,
while for $d=4$ it falls off as $1/t^2$ due to a cancellation
arising from taking the real part of $\psi^{(1)}$ in (\ref{4.19}).
Therefore, the conditions for the existence of the {\it weak coupling
limit} are assured, at infinity by the fall off of $G_\beta(t)$,
provided $d\geq4$, and at zero by the $i\varepsilon$ 
prescription.%
\footnote{In four-dimensional spacetime, the asymptotic
behaviour for large space-like separation of the two-point Wightmann 
function has been studied in Ref.[\refcite{bros}] in the case of arbitrary 
interacting fields. In general, $G_\beta(t)$ is found to fall
off at infinity as $1/t^{3/2}$, so that the {\it weak coupling limit}
is assured to exists even in this very general setting.}
Notice that in two dimensions (\ref{4.19}) gives
a logarithmic behaviour for the two point function,
so that in general the {\it weak coupling limit}
is not guaranteed to exist. The limit is however
well-defined for an atom in an optical fiber as
discussed in Example 4.1. There, the field
operators $\Phi$ that couple with the atom variables
involve the time derivative of a two-dimensional, free,
massless scalar field; as a consequence, the relevant
environment correlation $\langle\Phi(t)\Phi(0)\rangle$ actually
behaves as $\partial_t^2 G_\beta(t)$, and integrability
at infinity is still assured.

\item
The $i\varepsilon$ prescription involved in the definition
of the Wightmann function (\ref{4.16}) originates from
causality requirements. In the present setting,
it can be alternatively interpreted as an infinitesimal remnant of the 
size of the atom.\cite{takagi} Indeed, assume that the atom has a spatial
extension, with a profile given by a function $f(\vec x\,)$.
Since the atom-field interaction takes now place on 
the whole region occupied by the atom, the field operator 
entering the Hamiltonian $H'$ in (\ref{4.2}) is now smeared out 
over the atom size,
\begin{equation}
\Phi_i(t)=\int {\rm d}^{d-1} x\, f(\vec x\,)\, \Phi_i(t,\vec x\,)\ ,
\label{4.20}
\end{equation}
instead of being just $\Phi_i(t, \vec 0\,)$, {\it i.e.}
evaluated at the atom position.
As a consequence, the correlation function (\ref{4.15})
involves the Fourier transform 
${\hat f}(\vec k\,)=\int {\rm d}^{d-1}x\, {\rm e}^{i\vec k\cdot\vec x} f(\vec x\,)$
of the shape function. As an illustration, assume
the fields to be in the vacuum, zero-temperature state; then,
(compare with (\ref{4.16})):
\begin{equation}
\langle \Phi_i(t)\Phi_j(0)\rangle=\delta_{ij}\, 
\int \frac{{\rm d}^{d-1} k}{(2\pi)^{d-1}}\ \frac{1}{2 |\vec k\, |}\
\big|{\hat f}(\vec k\,)\big|^2\ {\rm e}^{-i |\vec k\,| t}\ .
\label{4.21}
\end{equation}
Since at the end we want to model a point-like atom, we choose
a shape function approximating a $(d-1)$-dimensional
$\delta$-function:
\begin{equation}
f(\vec x\,)=\prod_{i=1}^{d-1}\frac{1}{2\pi}\frac{\varepsilon}{\big[ (x_i)^2 +
(\varepsilon/2)^2 \big]^2}\ , \qquad \varepsilon>0\ ;
\label{4.22}
\end{equation}
in this way, the atom is viewed as $d-1$-dimensional box with size
$\varepsilon$. The Fourier transform of (\ref{4.22}) can be easily
computed: ${\hat f}(\vec k\,)={\rm e}^{-|\vec k\,| \varepsilon/2}$;
inserting this result in (\ref{4.21}), one readily recovers 
the standard $i\varepsilon$ prescription for the field correlations.
$\Box$ 
\end{enumerate}
}
\bigskip

As discussed in Section 3.3.2, the explicit expression of the 
Kossakowski matrix $C_{ij}$ involves a sum
over the differences of the energy levels
of the system Hamiltonian $H_S$, here just $\pm \omega/2$; 
it corresponds to the
decomposition of the system operators $\sigma_i$ along the
two energy eigenprojectors
\begin{equation}
P_\pm=\frac{1}{2}\big({\bf 1}_2\pm\vec n\cdot\vec\sigma\big)\ ,
\label{4.23}
\end{equation}
as indicated in~(\ref{Kraus1}), namely:
\begin{equation}
\sigma_i(0)=P_+\sigma_i P_+ + P_- \sigma_i P_-\ ,\qquad
\sigma(\pm)=P_\pm\sigma_i P_\mp\ .
\label{4.24}
\end{equation}
The $2\times2$ auxiliary matrices
$\sigma_i(0)$, $\sigma_i(+)$ and $\sigma_i(-)$ can be further
decomposed along the Pauli matrices themselves,
\begin{equation}
\sigma_i(\xi)=\sum_{j=1}^3 \psi^{(\xi)}_{ij}\, \sigma_j\ ,
\qquad {\rm for}\ \xi= +,-,0\ ,
\label{4.25}
\end{equation}
with the help of the three-dimensional tensor coefficients
\begin{equation}
\psi_{ij}^{(0)}=n_i\, n_j\ ,\qquad 
\psi_{ij}^{(\pm)}={1\over 2}\big(\delta_{ij} - n_i\, n_j\pm
i\epsilon_{ijk} 
n_k\big)\ .
\label{4.26}
\end{equation}
Recalling (\ref{4.14}), one can then write (compare with~(\ref{Kraus7})):
\begin{eqnarray}
\label{4.27}
C_{ij}&=&\sum_{\xi=0,\pm}\ \sum_{k,l=1}^3\ h_{kl}(\xi\omega)\,
\psi_{ki}^{(\xi)}\,
\psi_{lj}^{(-\xi)}\\
\label{4.28}
&=&{\cal G}_\beta(0)\, \psi^{(0)}_{ij}+{\cal G}_\beta(\omega)\,\psi^{(-)}_{ij}
+{\cal G}_\beta(-\omega)\,\psi^{(+)}_{ij}\ ,
\end{eqnarray}
where the second line follows from the property: 
$\sum_{k=1}^3\psi^{(\xi)}_{ki}\, \psi^{(-\xi)}_{kj}=\psi^{(-\xi)}_{ij}$.
One can easily check that, being the sum of three positive matrices,
the matrix $C_{ij}$ is also positive, so that the 
dynamical semigroup generated by (\ref{4.12}) consists of
completely positive maps. As discussed before, this is to be expected,
since (\ref{4.28}) is the result of the {\it weak coupling limit},
in the mathematically well-defined formalism developed in 
Section 3.3.2.
On the other hand, let us remark
once more, that direct use of the standard second order perturbative 
approximation
(as adopted for example in Ref.[\refcite{marzlin}]) leads to physically inconsistent
results, giving a finite time-evolution for $\rho(t)$ that in general does 
not preserve the positivity of probabilities.
\bigskip

\noindent
{\bf Example 4.3} \quad
As an explicit illustration of this general framework, it is instructive
to reconsider the simple system discussed in Example 3.3, and analyze
its dynamics in the long-time, Markovian regime using the techniques
just presented. The model describes the behaviour of a two-level
atom immersed in a one-dimensional boson gas at temperature $\beta^{-1}$,
that in the thermodynamical limit can be fully described in terms of
a two-dimensional, free, massless scalar field.

Positioning the atom at the origin, the total Hamiltonian in (\ref{3.78-1}) 
can then be conveniently rewritten as:
\begin{equation}
H=\frac{\Omega}{2}\, \sigma_3\otimes{\bf 1}_\phi +{\bf 1}_S\otimes H_\phi
+\lambda\, \sigma_3\otimes\partial_t\phi(0)\ ,
\label{4.28-1}
\end{equation}
so that we are precisely in the condition of Eq. (\ref{4.1}) and (\ref{4.2})
with $\vec n=(0,0,1)$ and $\Phi_1=\Phi_2=\,0$, $\Phi_3=\partial_t \phi$.
The Fourier transform of the bath correlation functions,
$G_\beta(t)={\rm Tr}\big[\partial_t\phi(t)\, \partial_t\phi(0)\, \rho_\beta\big]$,
can be easily evaluated:
\begin{equation}
{\cal G}_\beta(\zeta)=\frac{\zeta}{1-{\rm e}^{-\beta\zeta}}\ ;
\label{4.28-2}
\end{equation}
through (\ref{4.27}), one then obtains the expression for the Kossakowski
matrix $C_{ij}$. It turns out that only the entry $C_{33}$
is actually nonvanishing. As a consequence, the Markovian master equation
describing the reduced dynamics of the two-level system takes, in
the interaction picture, the particularly simple form:
\begin{equation}
\partial_t\rho(t)=\frac{\lambda^2}{2\beta}\big[\sigma_3\,\rho(t)\,\sigma_3
-\rho(t)\big]\ .
\label{4.28-3}
\end{equation}
In the basis of eigenvectors of $\sigma_3$, $\sigma_3 |i\rangle=(-1)^{i+1}|i\rangle$,
$i=0,1$, one then recovers the exponential damping of the off-diagonal entries
of $\rho(t)$ already discussed in Example~3.3:
\begin{eqnarray}
\nonumber
&&\langle i|\rho(t)|i\rangle=\langle i|\rho(0)|i\rangle\ ,\\
\label{4.28-4}
&&\langle 1|\rho(t)|0\rangle={\rm e}^{-2\lambda^2 t/\beta}\ \langle 1|\rho(0)|0\rangle\ .
\end{eqnarray}
Although in a simplified settings, this result constitutes a direct evidence
that the Markovian master equation (\ref{4.28-3}), obtained using the
mathematically well defined {\it weak coupling limit} procedure,
correctly reproduces the long time behaviour of the exact reduced dynamics.
$\Box$
\bigskip 

The Lamb shift correction $H_E^{(2)}$ to the effective Hamiltonian
$H_{\rm eff}=H_S+H_E^{(2)}$ can be similarly computed. It involves the Hilbert
transform of the correlations (\ref{4.14}), as indicated in~(\ref{Kraus5}) 
and~(\ref{Kraus6}). In practice,
recalling (\ref{4.15}), it can be expressed in terms of 
the following integral transform of the scalar Wightmann function
\begin{equation}
{\cal K}_\beta(\zeta)={{\cal P}\over i\pi}\int_{-\infty}^\infty {\rm d}z\ 
{{\cal G}_\beta(z)\over z-\zeta}\ .
\label{4.29}
\end{equation}
Explicitly, one finds:
\begin{equation}
H_E^{(2)}=\frac{1}{2}\sum_{i,j,k=1}^3\epsilon_{ijk}\, \Big[
{\cal K}_\beta(\omega)\,\psi^{(-)}_{jk}
+{\cal K}_\beta(-\omega)\,\psi^{(+)}_{jk}\Big]\ \sigma_i\ .
\label{4.30}
\end{equation}

\bigskip

\noindent
{\bf Remark 4.2}\quad 
{
This result deserves a closer look. 
Recalling (\ref{4.18}), the definition of ${\cal K}_\beta(\zeta)$
in (\ref{4.29}) can be split as:
\begin{eqnarray}
\label{4.31}
{\cal K}_\beta(\zeta)&=&
\frac{2}{i(4\pi)^\frac{d-1}{2} 
\Gamma\left(\frac{d-1}{2}\right)}\Bigg[ {\cal P}
\int_0^\infty {\rm d}z\
\frac{z^{d-3}}{z-\zeta}\\
\label{4.32}
&&\hskip 2.5cm +{\cal P}\int_0^\infty {\rm d}z\ \frac{z^{d-3}}{1-{\rm e}^{\beta z}}
\Bigg(\frac{1}{z+\zeta}-\frac{1}{z-\zeta}\Bigg)\Bigg]\ ,
\end{eqnarray}
into a vacuum and a temperature-dependent piece.
Although not expressible in terms of elementary functions, the temperature
dependent second term is a finite, odd function of $\zeta$, vanishing
as $\beta$ becomes large, {\it i.e.} in the limit of zero temperature.
The first contribution in (\ref{4.31}) is however divergent for $d\geq3$.
As a consequence, despite some cancellations 
that occur in (\ref{4.30}) (see below), the Lamb contribution $H^{(2)}_E$ turns
out in general to be infinite, and its definition requires the introduction 
of a suitable cutoff and a renormalization procedure.

This is a well known fact, and has nothing to do with the
weak coupling assumptions used in deriving the master equation.
Rather, the appearance of the
divergences is due to the non-relativistic treatment of
the two-level atom, while any sensible calculation
of energy shifts would require 
quantum field theory techniques.\cite{milonni}

In our quantum mechanical setting, the procedure needed
to make the Lamb contribution $H^{(2)}_E$ well defined is therefore
clear: perform a suitable
temperature independent subtraction, so that
the expression in (\ref{4.30}) reproduces the correct
quantum field theory result, obtained by considering
the field in its vacuum state.~$\Box$
}
\bigskip

\subsection{Single Atom Dynamics and Decoherence}

From now on we shall work in ordinary four-dimensional Minkowski spacetime
and therefore specialize $d=4$. In this case, the Kossakowski 
matrix $C_{ij}$ in (\ref{4.28}) takes the general form
\begin{equation}
C_{ij}=A\, \delta_{ij}-iB\, \epsilon_{ijk}\, n_k + C\, n_i\, n_j\ ,
\label{4.33}
\end{equation}
where the quantities $A$, $B$ and $C$ depend on the system frequency $\omega$
and the inverse temperature $\beta$:
\begin{eqnarray}
\label{4.34}
A&=&{1\over2}\Big[{\cal G}_\beta(\omega)+{\cal G}_\beta(-\omega)\Big]
={\omega\over 4\pi}
\bigg[{1+{\rm e}^{-\beta\omega}\over 1-{\rm e}^{-\beta\omega}}\bigg]\ ,\\
\label{4.35}
B&=&{1\over2}\Big[{\cal G}_\beta(\omega)-{\cal G}_\beta(-\omega)\Big]
={\omega\over 4\pi}\ ,\\
\label{4.36}
C&=&{1\over2}\Big[2{\cal G}_\beta(0)-{\cal G}_\beta(\omega)
-{\cal G}_\beta(-\omega)\Big]={\omega\over 4\pi}
\bigg[{2\over\beta\omega}-{1+{\rm e}^{-\beta\omega}\over 1-{\rm e}^{-\beta\omega}}\bigg]\ .
\end{eqnarray}
On the other hand, the effective Hamiltonian $H_{\rm eff}$ can be rewritten as
\begin{equation}
H_{\rm eff}={\tilde\omega\over2}\, \vec n\cdot\vec\sigma\ ,
\label{4.37}
\end{equation}
in terms of a redefined frequency:
\begin{equation}
\tilde\omega=\omega+i\big[ {\cal K}_\beta(-\omega) 
- {\cal K}_\beta(\omega)\big]\ .
\label{4.38}
\end{equation}

\bigskip

\noindent
{\bf Remark 4.3}\quad
{ 
As explained above, a suitable temperature
independent subtraction has been implicitly included in the definition
of the combination ${\cal K}(-\omega) - {\cal K}(\omega)$,
which otherwise would have been logarithmically divergent 
(compare with the expression in (\ref{4.31})). It comes
from a ``mass'' renormalization due to virtual vacuum effects
of the full relativistic theory,\cite{milonni} and can not be accounted for
by a mere frequency shift, as the expression (\ref{4.38})
might erroneously suggest.
$\Box$
}
\bigskip

In order to discuss explicitly the properties of the solutions 
of the master equation (\ref{4.12}),
it is convenient to decompose the density matrix $\rho$ in terms
of the Pauli matrices, as introduced in Example 3.1:
\begin{equation}
\rho={1\over2}\Big({\bf 1}_2 + \vec\rho\cdot\vec\sigma\Big)\ .
\label{4.39}
\end{equation}
Then, as shown there,
the evolution equation (\ref{4.12}) can be rewritten 
as a Schr\"odinger-like equation for the coherence (Bloch)
vector $|\rho(t)\rangle$ of components 
$(1,\rho_1(t),\rho_2(t),\rho_3(t))$:
\begin{equation}
{\partial\over\partial t} |\rho(t)\rangle=-2\, \big({\cal H}+{\cal
D}\big)\, |\rho(t)\rangle
\ .
\label{4.40}
\end{equation}
The $4\times4$ matrices $\cal H$ and $\cal D$ correspond to the Hamiltonian 
and dissipative contributions, respectively;
they can be parametrized as in Example 3.1, Eqs.(\ref{ham1}) and (\ref{pauli6}),
in terms of the quantities 
$\omega_i=(\tilde\omega/2)\, n_i$, $i=1,2,3$, coming from $H_{\rm eff}$,
$u=-4B\, n_1$, $v=-4B\, n_2$, $w=-4B\, n_3$, coming from the imaginary
part of the Kossakowski matrix (\ref{4.33}), and
\begin{eqnarray}
\label{4.41}
a=2A+C\big(n_2^2+n_3^2\big)\ ,\qquad b=-C\, n_1 n_2\ ,\\
\label{4.42}
\alpha=2A+C\big(n_1^2+n_3^2\big)\ ,\qquad c=-C\, n_1 n_3\ ,\\
\label{4.43}
\gamma=2A+C\big(n_1^2+n_2^2\big)\ ,\qquad \beta=-C\, n_2 n_3\ ,
\end{eqnarray}
coming from its real part. The $3\times3$, symmetric submatrix 
${\cal D}^{(3)}$ defined in~(\ref{pauli11}) with these last six
parameters as entries, is manifestly positive; its eigenvalues
are in fact $\lambda_1=2A$ and $\lambda_\pm=2A+C$. As a consequence,
for large times $|\rho(t)\rangle$ reaches a unique equilibrium state
$|\rho_{eq}\rangle$.\cite{lendi2} This asymptotic state is determined
by the condition $({\cal H}+{\cal D})|\rho_{eq}\rangle=\,0$,
and therefore is obtained by inverting ${\cal D}^{(3)}$:
\begin{equation}
|\rho_{eq}\rangle=\big\{1, -R\, n_1, -R\, n_2, -R\, n_3\big\}\ ,\qquad
R\equiv{B\over A}={1-{\rm e}^{-\beta\omega}\over 1+{\rm e}^{-\beta\omega}}\ .
\label{4.44}
\end{equation}
Inserting these components in the expansion (\ref{4.39}), one finds that the
asymptotic density matrix $\rho_{eq}$ is a thermal state in
equilibrium with the field environment at the latter temperature
$T=1/\beta$,
\begin{equation}
\rho_{eq}={{\rm e}^{-\beta H_S}\over {\rm Tr}\big[{\rm e}^{-\beta H_S}\big]}\ .
\label{4.45}
\end{equation}

This asymptotic thermalization phenomenon is just one 
of the many effects that the dissipative dynamics generated by
(\ref{4.40}) produces on the two-level atom. 
More details on the thermalization process can be retrieved by 
studying the behaviour of the atom state $\rho(t)$ for finite times;
it is formally given by the exponentiation of (\ref{4.40}): 
\begin{equation}
|\rho(t)\rangle={\rm e}^{-2({\cal H}+{\cal D})t}\ |\rho(0)\rangle\ .
\label{4.46}
\end{equation}
Recalling the definition (\ref{meanv}), the evolution in time of any relevant atom
observable, a Hermitian $2\times2$ matrix $X$, can thus be
explicitly computed:
\begin{equation}
\langle X(t)\rangle={\rm Tr}\left[X\, \rho(t)\right]\ .
\label{4.47}
\end{equation}
When the observable $X$ represents itself an admissible 
atom state $\rho_{{\rm f}}$, the mean value (\ref{4.47}) coincides
with the probability ${\cal P}_{{\rm i}\to {\rm f}}(t)$
that the evolved atom density matrix $\rho(t)$,
initially in $\rho(0)\equiv\rho_{{\rm i}}$,
be found in such a state at time $t$. Using (\ref{4.46}), this probability
can be computed explicitly:
\begin{eqnarray*}
{\cal P}_{{\rm i}\to {\rm f}}(t)&=&{1\over2}\bigg\{1-
\big(\vec\rho_{\rm f}\cdot\vec n\big)\, \Big(1-{\rm e}^{-4At}\Big)\
\bigg[{1-{\rm e}^{-\beta \omega}\over 1+{\rm e}^{-\beta \omega}}\bigg]
+{\rm e}^{-4At}\ \big(\vec\rho_{\rm i}\cdot\vec n\big)\, 
\big(\vec\rho_{\rm f}\cdot\vec n\big)\\
&+&{\rm e}^{-2(2A+C)t}\ \Big(\Big[\big(\vec\rho_{\rm i}\cdot\vec\rho_{\rm f}\big)
-\big(\vec\rho_{\rm i}\cdot\vec n\big)\, 
\big(\vec\rho_{\rm f}\cdot\vec n\big)\Big]\ \cos\tilde\omega t
-\vec n\cdot\big(\vec\rho_{\rm i}\times\vec\rho_{\rm f}\big)\
\sin\tilde\omega t\Big)\bigg\}\ .
\end{eqnarray*}
\begin{equation}
\label{4.48}
\end{equation}
In the above expression, we have expanded the density matrices
$\rho_{\rm i}$, $\rho_{\rm f}$ as in (\ref{4.39}) and used the notations
\hbox{$\big(\vec\rho_{\rm i}\cdot\vec\rho_{\rm f}\big)$}
and $\big(\vec\rho_{\rm i}\times\vec\rho_{\rm f}\big)$
(and similarly with $\vec n$) to
represent scalar and vector products of their corresponding
three-dimensional coherence vectors.
As expected, this expression contains exponentially decaying factors
involving the eigenvalues $\lambda_1$ and $\lambda_\pm$ defined
earlier, modulated
by oscillating terms in the effective frequency $\tilde\omega$.
In other terms, the two-level atom immersed in scalar fields at
temperature $1/\beta$
is subjected to decoherence and dissipation,
both effects being governed by the Planckian factors appearing in the
Kossakowski matrix (\ref{4.33})-(\ref{4.36}).

When $\vec\rho_{\rm i}=-\vec n$ and $\vec\rho_{\rm f}=\vec n$,
the density matrices $\rho_{\rm i}$,
$\rho_{\rm f}$ represent the ground and excited states
of the system Hamiltonian $H_S$ in (\ref{4.1}).
In this case, the expression in (\ref{4.48}) simplifies,
\begin{equation}
{\cal P}_{{\rm i}\to {\rm f}}(t)={1\over 1+{\rm e}^{\beta \omega}}\ 
\Big(1- {\rm e}^{-4At}\Big)\ ,
\label{4.49}
\end{equation}
giving the probability for a spontaneous transition of the atom from
the ground state to its excited state. Note that this phenomenon 
of spontaneous excitation disappears
as $\beta\to\infty$, {\it i.e.} when the environment 
is at zero temperature.

\bigskip

\noindent
{\bf Remark 4.4}\quad
{
 Although the behaviour of ${\cal P}_{{\rm i}\to {\rm f}}(t)$
in (\ref{4.48}) and (\ref{4.49}) is in principle experimentally observable
through the use of suitable interferometric devices,
in the study of the behaviour of atoms in optical cavities
one often limits the
discussion to the spontaneous excitation rate 
$\mathit\Gamma_{{\rm i}\to {\rm f}}$,
the probability per unit time of the transition
${\rm i}\to {\rm f}$, in the limit of an infinitely slow switching on
and off of the atom-field interaction.
In our formalism, its expression can be easily obtained 
by taking the time derivative of ${\cal P}_{{\rm i}\to {\rm f}}(t)$ 
at $t=\,0$; in the case of (\ref{4.49}), one then finds
\begin{equation}
{\mathit\Gamma}_{{\rm i}\to {\rm f}}={\omega\over\pi}\,
{1\over {\rm e}^{\beta \omega}-1}\ .
\label{4.50}
\end{equation}
\hfill$\Box$}

\bigskip

\subsection{Two Atoms: Environment Induced Entanglement Generation}

We have thus far analyzed the dynamics of an atom immersed
in a bath of scalar massless fields at finite temperature, 
and found that, through
decoherence effects, the atom state is driven towards a purely
thermal equilibrium state, characterized by the same 
field temperature. When the subsystem in interaction with the field
environment consists of two, non-interacting two-level atoms
one thus expects similar mixing-enhancing phenomena to occur,
leading in particular to loss of the mutual quantum correlation 
(entanglement) that might have been present at the beginning.

However, even though not directly coupled, the external field
through which the two atoms move may provide an indirect 
interaction between them, and thus a means to entangle them.
Indeed, entanglement generation through the action of an external
heat bath has been shown to occur in certain circumstances;
it is therefore of physical interest to investigate the same issue
also in the present physical situation.

We shall therefore start by considering a system 
composed by two, equal two-level atoms,
that start interacting with the external fields at time $t=\,0$.
Being independent, without direct mutual interaction,
their internal dynamics can again be taken 
to be described by the generic Hamiltonian (\ref{4.1}).
Then, the total two-system Hamiltonian $H_S$ is now the sum of 
two terms:
\begin{equation}
H_S=H_S^{(1)}+H_S^{(2)}\ ,
\quad
H_S^{(1)}={\omega\over 2}\sum_{i=1}^3 n_i(\sigma_i\otimes{\bf 1})\ ,
\quad
H_S^{(2)}={\omega\over 2}\sum_{i=1}^3 n_i({\bf 1}\otimes \sigma_i)\ .
\label{4.51}
\end{equation}
Similarly, being immersed in the same set of free fields and
within the weak coupling hypothesis,
the atom-field interaction Hamiltonian can be most simply assumed
to be the generalization of that in (\ref{4.2}):
\begin{equation}
H'=\sum_{i=0}^3\Big[\big(\sigma_i\otimes{\bf 1}\big)\otimes\Phi_i(x)
+\big({\bf 1}\otimes\sigma_i\big)\otimes\Psi_i(x)\Big]\ .
\label{4.52}
\end{equation}
On the other hand, the Hamiltonian describing the environment dynamics
remains that of a collection of free, independent scalar fields $\Phi_i$
and $\Psi_i$.

\bigskip

\noindent
{\bf Remark 4.5}\quad
{
In writing the interaction terms in 
(\ref{4.52}), we have assumed the two atoms to interact
with the field bath through two different field operators, 
$\Phi$ and $\Psi$, respectively; as we shall see, this allows 
discussing entanglement generation in a more general setting.
For two equal atoms, the identification $\Phi=\Psi$ is however
physically justified, and we shall adopt it in most of the
following analysis. On the other hand, 
since the two atoms are assumed to be pointlike, with infinitesimal size,
there is no restriction in positioning them at the same point $x$,
and the coupling in $H'$ precisely reflects this simplified assumption.%
\footnote{The case of spacially separated atoms can also be similarly
treated, adopting the same open system techniques; for details,
see Ref.[\refcite{bf18}].}
$\Box$}
\bigskip

The derivation of the appropriate master equation describing the
dynamics of the two atoms proceeds 
as in the case of a single atom, discussed in the previous Section.
One starts from the Liouville-von Neumann equation (\ref{4.11}) generating
the time-evolution of the state $\rho_{\rm tot}(0)$
of the total system $\{ {\rm atoms}
+ {\rm external\ fields}\}$, and then traces over the fields degrees
of freedom, assuming a factorized initial state
$\rho_{\rm tot}(0)=\rho(0)\otimes \rho_\beta$.
In the {\it weak coupling limit}, the two-atom system state $\rho(t)$,
now a $4\times4$ density matrix,
is seen evolving in time according to a quantum dynamical
semigroup of completely positive maps, generated by a master equation
in Kossakowski-Lindblad form:
\begin{equation}
{\partial\rho(t)\over \partial t}= -i \big[{\cal H}_{\rm eff},\, \rho(t)\big]
 + \mathbb{D}[\rho(t)]\ .
\label{4.53}
\end{equation}

The unitary term depends on an effective Hamiltonian
which is the sum of $H_S$ in (\ref{4.51}) and suitable Lamb
contributions:
${\cal H}_{\rm eff}=H_{\rm eff}^{(1)}+H_{\rm eff}^{(2)}+H_{\rm eff}^{(12)}$.
The first two represent single system contributions;
they can be written as in (\ref{4.51}), but in general
with different terms for the two atoms:
\begin{equation}
H^{(1)}_{\rm eff}=\sum_{i=1}^3 H_i^{(1)} (\sigma_i\otimes {\bf 1})\ ,\quad
H^{(2)}_{\rm eff}=\sum_{i=1}^3 H_i^{(2)} ({\bf 1}\otimes \sigma_i)\ .
\label{4.54}
\end{equation}
The third piece is a field-generated
direct two-atom coupling term, that in general can be written as:
\begin{equation}
H_{\rm eff}^{(12)}=\sum_{i,j=1}^3 H^{(12)}_{ij}\, (\sigma_i\otimes\sigma_j)\ .
\label{4.55}
\end{equation}
As in the single atom case, a suitable temperature-independent renormalization
procedure needs to be implemented in order to make all these contributions
well defined.

The dissipative contribution $\mathbb{D}[\rho(t)]$ can be written as 
in~(\ref{lindblad}),
\begin{equation}
\mathbb{D}[\rho]=\sum_{\alpha,\beta=1}^6\,
C_{\alpha\beta}\ \Bigg[ 
{\cal F}_\beta\, \rho\ {\cal F}_\alpha\, -\, {1\over 2}
\Bigl\{{\cal F}_\alpha {\cal F}_\beta\,,\,\rho\Bigr\}\Bigg]\ ,
\label{4.56}
\end{equation}
with the help of the matrices 
${\cal F}_\alpha=\sigma_\alpha\otimes{\bf 1}$ for $\alpha=1,2,3$,
${\cal F}_\alpha={\bf 1}\otimes \sigma_{\alpha-3}$ for $\alpha=4,5,6$.
The Kossakowski matrix $C_{\alpha\beta}$ is now a positive $6\times 6$
matrix, that can be represented as
\begin{equation}
C=\pmatrix{
  {\cal A} & {\cal B} \cr
  {\cal B}^{\dagger} & {\cal C}}\ ,
\label{4.57}
\end{equation}
by means of the $3\times 3$ matrices ${\cal A}={\cal A}^\dagger$, 
${\cal C}={\cal C}^\dagger$ and ${\cal B}$.
In terms of this decomposition, $\mathbb{D}[\rho]$ 
can be rewritten in the following more explicit form:
\begin{eqnarray}
\nonumber
&&\hskip -.5cm \mathbb{D}[\rho]\!=\!\!
\sum_{i,j=1}^3\!\Bigg(\!
{\cal A}_{ij}\Bigg[(
\sigma_j\otimes{\bf 1})\,\rho\,(\sigma_i\otimes{\bf 1})
-\frac{1}{2}\Big\{(\sigma_i\sigma_j\otimes{\bf 1})\,,\,\rho\Big\}
\Bigg]\\
\nonumber
&&\hskip 1.1cm
+{\cal C}_{ij}\Bigg[(
{\bf 1}\otimes\sigma_j)\,\rho\,({\bf 1}\otimes\sigma_i)
           -\frac{1}{2}\Big\{({\bf 1}\otimes\sigma_i\sigma_j)\,,\,\rho\Big\}
\Bigg]\\
\nonumber
&&\hskip 1.1cm
+{\cal B}_{ij}\Bigg[(\sigma_j\otimes{\bf 1})\,\rho\,({\bf 1}\otimes\sigma_i)
           -\frac{1}{2}\Big\{(\sigma_j\otimes\sigma_i)\,,\,\rho\Big\}
\Bigg]\\
&&\hskip 1.1cm
+{\cal B}^\dagger_{ij}\Bigg[({\bf 1}\otimes\sigma_j)\,\rho\,(\sigma_i\otimes{\bf 1})
           -\frac{1}{2}\Big\{(\sigma_i\otimes\sigma_j)\,,\,\rho\Big\}
\Bigg]
\Bigg)\ .
\label{4.58}
\end{eqnarray}
The structure of the various contributions reveals their direct
physical meaning. Indeed, the first two contributions are dissipative 
terms that affect the first, respectively the second, atom in absence 
of the other. On the contrary, the last two pieces represent the way 
in which the noise generated by the external fields
may correlate the two, otherwise independent, atoms; this effect
is present only if the matrix $\cal B$ is different from zero.

\bigskip

\noindent
{\bf Remark 4.6}\quad 
{
As in the single atom case, the environment contributions
to ${\cal H}_{\rm eff}$ and the entries of the matrix $C_{\alpha\beta}$
in (\ref{4.57}) can be expressed 
in terms of the Hilbert, respectively the Fourier transforms of the thermal
Wightmann functions. In particular, the matrices $H_{ij}^{(12)}$ in
(\ref{4.55}) and ${\cal B}_{ij}$ in (\ref{4.58}) 
do not vanish only if the bath state $\rho_\beta$ 
correlates bath-operators coupled to different atoms, 
that is, if
the expectations ${\rm Tr}[\rho_\beta\,\Phi_i(t)\,\Psi_j(0)]$ are nonzero. 
Only in this case, entanglement has a chance to be created by the
action of the bath.
Indeed, if $H^{(12)}_{ij}=\,0$ and ${\cal B}_{ij}=\,0$, the two atoms
evolve independently and initially separable states may become more
mixed, but certainly not entangled.
$\Box$
}
\bigskip

In order to investigate whether the thermal bath made of
free fields can actually entangle two independent atoms,
one can use the partial transposition criterion, that was introduced
in Theorem 2.3. In fact,
we are dealing here with a couple of two-level systems,
and therefore Theorem 2.4 applies.
In more precise terms, the environment
is not able to create entanglement if and only if the operation of partial 
transposition preserves the positivity of the state $\rho(t)$ 
for all times.

From the operational point of view, one then prepares the two subsystems
at $t=\,0$ in a separable state: 
\begin{equation}
\label{4.59}
\rho(0)=\vert \varphi\rangle\langle \varphi\vert\otimes 
\vert \psi\rangle \langle \psi\vert\ ;
\end{equation}
without loss of generality, $\rho(0)$ can be taken to be pure: indeed,
if the environment is unable to create entanglement out of pure states, 
it will certainly not entangle their mixtures.
Then, one acts with the operation
of partial transposition (over the second
factor, for sake of definiteness) on both sides of (\ref{4.53}).
One thus obtains a master equation for the matrix $\tilde\rho(t)$, 
the partially transposed $\rho(t)$, that can be cast in the
following form:\cite{bf19}
\begin{equation}
\label{4.60}
\partial_t\tilde{\rho}(t)=-i\Big[\widetilde{\cal H}_{\rm eff}\,,\,
\tilde{\rho}(t)\Big]\,+\,
\widetilde{\mathbb{D}}[\tilde{\rho}(t)]\ ;
\end{equation}
here, $\widetilde{\cal H}_{\rm eff}$ is a new Hamiltonian to 
which both the unitary and the dissipative term in (\ref{4.53}) contribute:
\begin{eqnarray}
\nonumber
\widetilde{\cal H}_{\rm eff}=\sum_{i=1}^3 H_i^{(1)}&& {\hskip -.2cm} (\sigma_i\otimes {\bf 1})
+\sum_{ij=1}^3 H_i^{(2)} E_{ij} ({\bf 1}\otimes\sigma_j)\\
&&+\sum_{ij=1}^3 {\cal I}m\big(B\cdot E\big)_{ij} (\sigma_i\otimes\sigma_j)\ ,
\label{4.61}
\end{eqnarray}
where $E$ is the diagonal $3\times3$ matrix given by: $E={\rm diag}(-1,1,-1)$.
The additional piece $\widetilde{\mathbb{D}}[\,\cdot\,]$ 
is of the form~(\ref{4.56}), but
with a new matrix $C\to {\cal E}\cdot \widetilde{C}\cdot {\cal E}$,
where
\begin{eqnarray}
&&\widetilde{C}=\pmatrix{{\cal A}& {\cal R}e({\cal B})+iH^{(12)}\cr
 {\cal R}e({\cal B}^T)-iH^{(12)}{}^T & {\cal C}^T},\qquad
\label{4.62}\\
&&{\cal E}=\pmatrix{{\bf 1}_3&0\cr0&E}\ ,
\label{4.63}
\end{eqnarray}
and ${}^T$ denotes full matrix transposition, while
$H^{(12)}$ is the coefficient matrix in~(\ref{4.55}). 

Although $\tilde{\rho}(t)$ evolves according to a master equation formally
of Kossakowski-Lindblad form, the new coefficient matrix $\widetilde{C}$
need not be positive.
As a consequence, the time-evolution generated by (\ref{4.60}) may
result to be neither
completely positive, nor positive and need not preserve 
the positivity of the initial state $\tilde{\rho}(0)\equiv \rho(0)$.

\bigskip

\noindent
{\bf Remark 4.7}\quad
{ 
Notice that both the Hamiltonian and the dissipative terms of the original
master equation (\ref{4.53}) contribute to the piece 
$\widetilde{\mathbb{D}}[\,\cdot\,]$ in (\ref{4.60}), the
only one that can possibly produce negative eigenvalues.
In particular, this makes more transparent the physical mechanism according to
which a direct Hamiltonian coupling $H^{(12)}_{\rm eff}$ among the 
two systems can induce entanglement (as studied for instance in
Refs.[\refcite{zanardi1}-\refcite{zyckowski}]): on 
$\tilde{\rho}(t)$, $H^{(12)}_{\rm eff}$ ``acts'' as a generic
dissipative contribution, which in general does not preserve positivity.
$\Box$
}
\bigskip

In order to check the presence of negative eigenvalues in $\tilde{\rho}(t)$,
instead of examining the full equation (\ref{4.60}) it is convenient 
to study the quantity
\begin{equation}
{\cal Q}(t)=\langle\chi\vert\, \tilde{\rho}(t)\, \vert\chi\rangle\ ,
\label{4.64}
\end{equation}
where $\chi$ is any $4$-dimensional vector.\cite{bf19}
Assume that an initial separable state $\tilde{\rho}$ has indeed developed a 
negative eigenvalue crossing the zero value at time $t^*$.
Then, there exists a vector state $\vert\chi\rangle$ 
such that ${\cal Q}(t^*)=0$, ${\cal Q}(t)>0$ for $t< t^*$ and
${\cal Q}(t)<0$ for $t> t^*$. 
The sign of entanglement creation may thus be given by a negative
first derivative of ${\cal Q}(t)$ at $t=t^*$.
Moreover, by assumption, the state $\rho(t^*)$ is separable.
Without loss of generality, one can 
set $t^*=0$ and, as already remarked, restrict 
the attention to factorized pure initial states.

In other words, the two atoms, initially prepared in a 
state $\rho(0)=\tilde\rho(0)$ as in (\ref{4.59}), will become entangled
by the noisy dynamics induced by their independent interaction with the bath
if there exists a suitable vector $|\chi\rangle$, such that: \hfill

{1)} ${\cal Q}(0)=\,0$ and 

{2)} $\partial_t {\cal Q}(0)<0$.\hfill\break
\noindent
Note that the vector $\vert\chi\rangle$ needs to be chosen entangled,
since otherwise ${\cal Q}(t)$ is never negative.
\bigskip

\noindent
{\bf Remark 4.8} The criterion just stated is clearly sufficient
for entanglement creation. In fact, 
when $\partial_t{\cal Q}(0)>0$ for all choices 
of the initial state $\rho(0)$ and probe vector $|\chi\rangle$, 
entanglement can not be generated by the environment, since $\tilde\rho$
remains positive. However, the treatment of the case $\partial_t{\cal Q}(0)=\,0$
requires special care: in order to check entanglement creation,
higher order derivatives of $\cal Q$, possibly with a time dependent
$|\chi\rangle$, need to be examined. $\Box$

\bigskip

A more manageable test of entanglement creation, valid for any probe
vector $|\chi\rangle$, can be obtained by a suitable manipulation of
the expression $\partial_t{\cal Q}(0)$.
In the two-dimensional Hilbert spaces pertaining to the two atoms,
consider first the orthonormal basis 
$\{|\varphi\rangle,\ |\tilde\varphi\rangle\}$,
$\{|\psi\rangle,\ |\tilde\psi\rangle\}$, obtained by augmenting
with the two states $|\tilde\varphi\rangle$ and
$|\tilde\psi\rangle$ the ones that define
$\rho(0)$ in (\ref{4.59}). 
They can be both unitarily rotated to the standard basis
$\{|-\rangle,\ |+\rangle\}$ of eigenvectors of $\sigma_3$:
\begin{eqnarray}
&&|\varphi\rangle= U |-\rangle \qquad |\tilde\varphi\rangle= U |+\rangle\ ,
\nonumber\\
&&|\psi\rangle= V |-\rangle \qquad |\tilde\psi\rangle= V |+\rangle\ .
\label{4.65}
\end{eqnarray}
Similarly, the unitary transformations $U$ and $V$ induce orthogonal 
transformations
$\cal U$ and $\cal V$, respectively, on the Pauli matrices:
\begin{equation}
U^\dagger \sigma_i U=\sum_{j=1}^3 {\cal U}_{ij}\sigma_j\ ,\quad
V^\dagger \sigma_i V=\sum_{j=1}^3 {\cal V}_{ij}\sigma_j\ .
\label{4.66}
\end{equation}
Direct computation then shows that $\partial_t{\cal Q}(0)$
can be written as a quadratic form in the components
of the probe vector $|\chi\rangle$. 
As a consequence, vectors $|\chi\rangle$ exist making this form
negative, {\it i.e.} $\partial_t{\cal Q}(0)<0$,
if and only if its corresponding discriminant is negative;
explicitly:
\begin{equation}
\langle u | {\cal A} | u \rangle \, \langle v | {\cal C}^T | v \rangle <
\big|\langle u | \big({\cal R}e({\cal B}) +iH^{(12)}\big)| v \rangle \big|^2\ .
\label{4.67}
\end{equation}
The three-dimensional vectors $|u\rangle$ and $|v\rangle$ 
contain the information about the starting factorized state (\ref{4.59}):
their components can be in fact expressed as:
\begin{equation}
u_i=\sum_{j=1}^3 {\cal U}_{ij}\, \langle +| \sigma_j |-\rangle\ ,\quad
v_i=\sum_{j=1}^3 {\cal V}_{ij}\, \langle -| \sigma_j |+\rangle\ .
\label{4.68}
\end{equation}
Therefore, the external quantum fields will be able to entangle the two 
atoms evolving with the Markovian dynamics generated by (\ref{4.53}) 
and characterized by
the Kossakowski matrix (\ref{4.57}), if there exists an initial
state $|\varphi\rangle\langle \varphi|\otimes |\psi\rangle\langle \psi|$,
or equivalently orthogonal transformations $\cal U$ and $\cal V$,
for which the inequality (\ref{4.67}) is satisfied.

The test in (\ref{4.67}) is very general and can
be applied to all situations in which two independent subsystems
are immersed in a common bath. It can be satisfied only if
the coupling between the two subsystems induced by
the environment through the mixed correlations
${\rm Tr}[\rho_\beta\Phi_i\Psi_j]$, encoded in the
coefficients of $\cal B$ and $H^{(12)}$, are sufficiently strong
with respect to the remaining contributions.

As already remarked, in the specific case of two atoms in interaction
with the same set of external fields, it is physically reasonable
to take $\Psi_i=\Phi_i$, so that the just mentioned mixed, environment
generated couplings result maximal and equal to the
diagonal ones, {\it i.e.} ${\cal A}_{ij}={\cal B}_{ij}={\cal C}_{ij}$.%
\footnote{This particular choice for the Kossakowski matrix
(\ref{4.57}) is of relevance also in phenomenological
applications; for instance, it is adopted in the 
analysis of the phenomenon of 
resonance fluorescence.\cite{agarwal,puri}}
Furthermore, in such a case, recalling (\ref{4.15}),
all field correlations become proportional to the standard,
scalar field temperature Wightmann function
$G_\beta(x)$ introduced in (\ref{4.16});
as a consequence, the three matrices ${\cal A}$, 
${\cal B}$ and ${\cal C}$ 
in (\ref{4.57}) become all equal
to the Kossakowski matrix (\ref{4.33}) pertaining to a single atom:
\begin{equation}
{\cal A}_{ij}={\cal B}_{ij}={\cal C}_{ij}=A\, \delta_{ij}-iB\, 
\epsilon_{ijk}\, n_k + C\, n_i\, n_j\ .
\label{4.69}
\end{equation}
In such a situation, also the Hamiltonian pieces (\ref{4.54})
and (\ref{4.55}) simplify: $H^{(1)}_{\rm eff}$ and 
$H^{(2)}_{\rm eff}$ can be written exactly as the original
system Hamiltonian (\ref{4.51}), with the frequency $\omega$
replaced by the renormalized one $\tilde\omega$ given in 
(\ref{4.38}), while the direct two-atom coupling term
takes the form
\begin{eqnarray}
&&H_{\rm eff}^{(12)}=-{i\over2}\sum_{i,j=1}^3\Big\{ \big[{\cal K}_\beta(\omega)
+{\cal K}_\beta(-\omega)\big]\,\delta_{ij}
\nonumber\\
&&\hskip 3.5cm +\big[2{\cal K}_\beta(0)-{\cal K}_\beta(\omega)
-{\cal K}_\beta(-\omega)\big]\,n_i n_j\Big\}\
\sigma_i\otimes\sigma_j\ ,
\label{4.70}
\end{eqnarray}
with ${\cal K}_\beta(\zeta)$ as in (\ref{4.29}).
\bigskip

\noindent
{\bf Remark 4.9}\quad
{
Recall that this function can be
split as in (\ref{4.31}) into a vacuum and temperature dependent piece.
Since, as observed there, the temperature dependent contribution
to ${\cal K}_\beta(\zeta)$ is odd in $\zeta$, one deduces that
$H_{\rm eff}^{(12)}$ does not
actually involves $T=1/\beta$: it is the same Lamb term that would have been
generated in the case of a two-atom system in the vacuum. 
Being interested in temperature induced phenomena, we shall not consider
this vacuum generated term any further and concentrate the attention
on the effects produced by the dissipative contribution 
$\mathbb{D}[\rho]$ in (\ref{4.56}).~$\Box$
}
\bigskip

Let us then consider again the test for entanglement creation
given in (\ref{4.67}). Because of condition (\ref{4.69}), it
now involves just the
hermitian $3\times3$ matrix ${\cal A}_{ij}$. 
Its expression can be 
further simplified by choosing $u_i=v_i$ in (\ref{4.68}); recalling
the definitions (\ref{4.65}), (\ref{4.66}), this
in turns implies $|\psi\rangle=|\tilde\phi\rangle$ (in other terms,
if in the initial state $\rho(0)$ we choose $|\phi\rangle=|-\rangle$,
then $|\psi\rangle$ must be taken to be $|+\rangle$). In this case,
(\ref{4.67}) simply becomes:
\begin{equation}
\big|\langle u|{\cal I}m({\cal A})|u\rangle\big|^2>0\ .
\label{4.71}
\end{equation}
As long as ${\cal A}_{ij}$ is not real, {\it i.e.} the parameter
$B$ in (\ref{4.69}) is nonvanishing, this condition is satisfied
for every $|u\rangle$ outside the null eigenspace of
${\cal I}m({\cal A})$. Take for instance the initial state
$\rho(0)=|-\rangle\langle -|\otimes |+\rangle\langle +|$,
so that the three-dimensional vector $|u\rangle$
has components $u_i=\{1, -i, 0\}$; from (\ref{4.69})
one easily finds: 
$\big|\langle u|{\cal I}m({\cal A})|u\rangle\big|^2=(B n_3)^2$,
which is in general non vanishing.

We can thus conclude that entanglement 
between the two atoms is indeed generated through the weak coupling with
the external quantum fields. This happens at the beginning
of the time-evolution, as soon as $t>0$. Note however that the test
in (\ref{4.71}) is unable to determine the fate of this
quantum correlations, as time becomes large.
In order to discuss asymptotic entanglement, one has to
analyze directly the structure of the dynamics generated
by the master equation in (\ref{4.53}).

\subsection{Two Atom Reduced Dynamics: Entanglement Enhancement}

As in the case of the single atom evolution, it is convenient
to decompose the $4\times4$ density matrix $\rho(t)$
describing the state of the two atoms along the Pauli matrices:
\begin{equation}
\rho(t)={1\over4}\bigg[{\bf 1}\otimes{\bf 1}+
\sum_{i=1}^3 \rho_{0i}(t)\ {\bf 1}\otimes\sigma_i
+ \sum_{i=1}^3 \rho_{i0}(t)\ \sigma_i\otimes{\bf 1}
+ \sum_{i,j=1}^3 \rho_{ij}(t)\ \sigma_i\otimes\sigma_j\bigg]\ ,
\label{4.72}
\end{equation}
where the components $\rho_{0i}(t)$, $\rho_{i0}(t)$, $\rho_{ij}(t)$
are all real.
Substitution of this expansion in the master equation (\ref{4.53}) allows 
deriving
the corresponding evolution equations for the above components of $\rho(t)$.
As explained in Remark 4.9, we shall ignore the Hamiltonian piece
since it can not give rise to temperature dependent entanglement phenomena,
and concentrate on the study of the effects induced by the
dissipative part in (\ref{4.58}), with the elements of the
Kossakowski matrix as in (\ref{4.69}).

Let us first observe that when the three submatrices $\cal A$,
$\cal B$ and $\cal C$ in (\ref{4.57}) are all equal, 
the the form of the dissipative contribution in
(\ref{4.58}) simplifies so that the evolution equation
can be rewritten as
\begin{equation}
{\partial\rho(t)\over \partial t}=\mathbb{D}[\rho(t)]=
\sum_{i,j=1}^3
{\cal A}_{ij}\bigg[
\Sigma_j\,\rho(t)\,\Sigma_i
-\frac{1}{2}\Big\{\Sigma_i\Sigma_j\,,\,\rho(t)\Big\}
\bigg]\ ,
\label{4.72-1}
\end{equation}
in terms of the following symmetrized two-system operators
\begin{equation}
\Sigma_i=\sigma_i\otimes{\bf 1} + {\bf 1}\otimes\sigma_i\ ,
\quad i=1,2,3\ .
\label{4.72-2}
\end{equation}
One easily checks that these operators 
obey the same $su(2)$ Lie algebra of the Pauli matrices;
further, together with
\begin{equation}
S_{ij}=\sigma_i\otimes\sigma_j + \sigma_j\otimes\sigma_i\ ,
\quad i,j=1,2,3\ ,
\label{4.72-3}
\end{equation}
they form a closed algebra under matrix multiplication.
For later reference, we give below its explicit expression:
it can be easily obtained using 
$\sigma_i\sigma_j=\delta_{ij}+i\sum_{k=1}^3\varepsilon_{ijk}\, \sigma_k$:
\begin{eqnarray}
\nonumber
&&\Sigma_i\, \Sigma_j=2\,\delta_{ij}\, {\bf 1}\otimes{\bf 1}
+i\sum_{k=1}^3\varepsilon_{ijk}\,\Sigma_k+S_{ij}\ ,\\
\nonumber
&&S_{ij}\, \Sigma_k=\delta_{ik}\, \Sigma_j + \delta_{jk}\, \Sigma_i
+i\sum_{l=1}^3\varepsilon_{ikl}\, S_{lj} + i\sum_{l=1}^3\varepsilon_{jkl}\, S_{il}\ ,\\
\nonumber
&&\Sigma_k\, S_{ij}=\delta_{ik}\, \Sigma_j + \delta_{jk}\, \Sigma_i
-i\sum_{l=1}^3\varepsilon_{ikl}\, S_{lj} - i\sum_{l=1}^3\varepsilon_{jkl}\, S_{il}\ ,\\
\nonumber
&&S_{ij}\, S_{kl}= 2\big(\delta_{ik}\,\delta_{jl}+\delta_{il}\,\delta_{jk}\big)
{\bf 1}\otimes{\bf 1}
+i\sum_{r=1}^3\Big(\delta_{ik}\varepsilon_{jlr}+\delta_{jk}\varepsilon_{ilr}
+\delta_{il}\varepsilon_{jkr}+\delta_{jl}\varepsilon_{ikr}\Big)\, \Sigma_r\\
\nonumber
&&\hskip 1.5cm -\Big(2\,\delta_{ij}\,\delta_{kl}-\delta_{ik}\,\delta_{jl}
-\delta_{il}\,\delta_{jk}\Big)\, S
+2\Big(\delta_{ij}\, S_{kl} + \delta_{kl}\, S_{ij}\Big)\\
&&\hskip 1.5cm -\delta_{ik}\, S_{jl} -\delta_{il}\, S_{jk}-\delta_{jk}\, S_{il}
-\delta_{jl}\, S_{ik}\ ,
\label{4.72-4}
\end{eqnarray}
where $S\equiv\sum_{r=i}^3 S_{ii}$.

Inserting the explicit expression for ${\cal A}_{ij}$ given in (\ref{4.69})
into (\ref{4.72-1}) and using the decomposition (\ref{4.72}), a 
straightforward but lengthy calculation allows to derive the following evolution
equations for the components of $\rho(t)$:
\begin{eqnarray}
\nonumber
&&{\partial\rho_{0i}(t)\over \partial t}=-2\sum_{k=1}^3\Big\{
\Big[\big(2A+C\big)\,\delta_{ik}
-C\, n_i n_k\Big]\, \rho_{0k}(t)
-Bn_k\, \rho_{ik}(t)\Big\}
-2B(2+\tau)\,n_i\ ,\\
\nonumber
&&{\partial\rho_{i0}(t)\over \partial t}=-2\sum_{k=1}^3\Big\{
\Big[\big(2A+C\big)\,\delta_{ik}
-C\, n_i n_k\Big]\, \rho_{k0}(t)
-Bn_k\, \rho_{ki}(t)\Big\}
-2B(2+\tau)\,n_i\ ,\\
\nonumber
&&{\partial\rho_{ij}(t)\over \partial t}=-4\Big[\big(2A+C\big)\,\rho_{ij}(t)
+\big(A+C\big)\,\rho_{ji}(t)- \Big(\big(A+C\big)\,\delta_{ij}
-C n_i n_j\Big)\tau\Big]\\
\nonumber
&&\hskip 1.6cm -4B\big[n_i\,\rho_{0j}(t)+n_j\,\rho_{i0}(t)\big]
-2B\big[n_i\,\rho_{j0}(t)+n_j\,\rho_{0i}(t)\big]\\
\nonumber
&&\hskip 1.6cm +2\sum_{k=1}^3\Big\{ B\, \delta_{ij}\,
n_k\big[\rho_{k0}(t)
+\rho_{0k}(t)\big]
+C n_i n_k\big[\rho_{kj}(t)+2\rho_{jk}(t)\big]\\
&&\hskip 1.6cm + C n_j n_k\big[\rho_{ik}(t)+2\rho_{ki}(t)\big]\Big\}
-4C\, \delta_{ij}\sum_{k,l=1}^3 n_k n_l\, \rho_{kl}(t)\ .
\label{4.73}
\end{eqnarray}
In these formulae, the quantity $\tau=\sum_{i=1}^3\rho_{ii}$ 
represents the trace of $\rho_{ij}$; 
it is a constant of motion, as easily seen by taking the trace
of both sides of the last equation above. Despite this,
the value of $\tau$ can not be chosen arbitrarily; the requirement
of positivity of the initial density matrix $\rho(0)$ readily
implies: \hbox{$-3\leq\tau\leq 1$}.

The system of first order differential equations in (\ref{4.73})
naturally splits
into two independent sets, involving the symmetric,
$\rho_{(0i)}=\rho_{0i}+\rho_{i0}$, $\rho_{(ij)}=\rho_{ij}+\rho_{ji}$,
and antisymmetric,
$\rho_{[0i]}=\rho_{0i}-\rho_{i0}$, $\rho_{[ij]}=\rho_{ij}-\rho_{ji}$,
variables. 
By examining the structure 
of the two sets of differential
equations, one can conclude that the antisymmetric variables
involve exponentially decaying factors, so that they vanish for
large times. Then, using the definitions (\ref{4.72-2}) and (\ref{4.72-3}),
the study of the equilibrium states $\hat\rho$ of the evolution equation
(\ref{4.72-1}) can be limited to density matrices of the form
\begin{equation}
\hat\rho={1\over4}\bigg[{\bf 1}\otimes{\bf 1}+
\sum_{i=1}^3 \hat\rho_{i}\, \Sigma_i
+ \sum_{i,j=1}^3 \hat\rho_{ij}\,  S_{ij}\bigg]\ ,
\label{4.73-1}
\end{equation}
with $\hat\rho_{ij}=\hat\rho_{ji}$.

The approach to equilibrium of semigroups whose generator
is of the generic Kossakowski-Lindblad form has been studied
in general and some rigorous mathematical results are
available.\cite{frigerio,spohn} We shall present such results
by adapting them to the case of the evolution
generated by the equation (\ref{4.72-1}).

First of all, one notices that in the case
of a finite dimensional Hilbert space,
there always exists at least one stationary state $\hat\rho_0$:
this can be understood by recalling that in
finite dimensions the ergodic average of the action of 
a completely positive one-parameter 
semigroup on any initial state always exists;
the result is clearly a stationary state.

Let us now introduce the operators 
${\cal V}_i=\sum_{j=1}^3{\cal A}^{1/2}_{ij}\, \Sigma_j$
(recall that ${\cal A}$ is non-negative), so that 
the r.h.s. of (\ref{4.72-1})
can be rewritten in diagonal form:
\begin{equation}
\mathbb{D}[\rho]\!=\!\!
\sum_{i,j=1}^3
\bigg[
{\cal V}_j\,\rho\,{\cal V}_i^\dagger
-\frac{1}{2}\Big\{{\cal V}_i^\dagger {\cal V}_j\,,\,\rho\Big\}
\bigg]\ .
\label{4.73-2}
\end{equation}
When the set ${\cal M}$ formed by all operators that commute with the 
linear span of $\{{\cal V}_i,\ {\cal V}_i^\dagger,\ i=1,2,3\}$
contains only the identity, one can show that 
the stationary state $\hat\rho_0$ 
is unique, and of maximal rank. 
On the other hand, when there are several stationary states,
they can be generated in a canonical way from a $\hat\rho_0$ with
maximal rank using the elements of the set $\cal M$.

In the case of the evolution equation (\ref{4.72-1}), 
$\cal M$ contains the operator
$S\equiv\sum_{i=1}^3 S_{ii}$, besides the identity;
indeed, with the help of the algebraic
relations in (\ref{4.72-4}), one immediately
finds: $[S,\ \Sigma_i]=\,0$.
Out of these two elements of $\cal M$, one can now construct
two mutually orthogonal projection operators:%
\footnote{One easily checks that $P$ is the projection operator
$Q^{(2)}_-$ on one of the maximally entangled Bell states introduced
in Example 2.3.}
\begin{equation}
P={1\over4}\bigg[{\bf 1}\otimes{\bf 1}-\frac{S}{2}\bigg]\ ,\qquad
Q=1-P\ .
\label{4.73-3}
\end{equation}
Then, one can show that any given initial state $\rho(0)$ 
will be mapped by the evolution (\ref{4.72-1}) into the following
equilibrium state:
\begin{equation}
\rho(0)\to\hat\rho=\frac{P\, \hat\rho_0\, P}{{\rm Tr}\big[P\, \hat\rho_0\, P\big]}\
{\rm Tr}\big[P\, \rho(0)\big]
+\frac{Q\, \hat\rho_0\, Q}{{\rm Tr}\big[Q\, \hat\rho_0\, Q\big]}\
{\rm Tr}\big[Q\, \rho(0)\big]\ .
\label{4.73-4}
\end{equation}
That this state is indeed stationary can be easily proven 
by recalling that $P$ and $Q$ commute with $\Sigma_i$, $i=1,2,3$,
and thus with the ${\cal V}_i$ as well;
therefore, $\mathbb{D}[\hat\rho\,]=\,0$, for any $\rho(0)$,
as a consequence of $\mathbb{D}[\hat\rho_0]=\,0$.

The problem of finding all invariant states of the dynamics (\ref{4.72-1})
is then reduced to that of identifying a stationary state
$\hat\rho_0$ with all eigenvalues nonzero.
Although in principle this amounts to solving a linear algebraic equation,
in practice it can be rather difficult for
general master equations of the form (\ref{4.12}).
Nevertheless, in the case at hand, the problem can be explicitly solved,
yielding:
\begin{equation}
\hat\rho_0={1\over2}\Big({\bf 1}_2-R\,\vec n\cdot\vec\sigma\Big)\ \otimes\
{1\over2}\Big({\bf 1}_2-R\,\vec n\cdot\vec\sigma\Big)\ ,
\label{4.73-5}
\end{equation}
where $R=B/A$ is the temperature dependent ratio already introduced 
in~(\ref{4.44}): note that $0\leq R\leq1$, where the two boundary 
values correspond to the infinite and zero
temperature limits, respectively.

Inserting this result in the expression (\ref{4.73-4}) allows deriving
the expression of the set of all equilibrium states of the
dynamics (\ref{4.72-1}); as expected, they take the symmetric form
of (\ref{4.73-1}), with the nonvanishing components
given by:
\begin{eqnarray}
&&\hat\rho_i=-{R\over 3+R^2}\big(\tau+3\big)\ n_i\ ,
\nonumber\\
&&\hat\rho_{ij}={1\over 2(3+R^2)}\Big[ \big(\tau-R^2\big)\ \delta_{ij}
+R^2\big(\tau+3\big)\ n_i\, n_j\Big]\ .
\label{4.74}
\end{eqnarray}
These stationary density matrices depend on the initial condition $\rho(0)$
only through the value of the parameter $\tau$, that as already mentioned
is a constant of motion for the dynamics in (\ref{4.72-1}).

As we have explicitly shown before, an environment made of the quantum
free fields
is in general able to initially entangle two independent subsystems weakly
interacting with it. Nevertheless, by examining the dynamics of a subsystem 
made of a single atom, we have also seen that such an environment
produce dissipation and noise, leading to effects that generically
counteract entanglement production.
It is therefore remarkable to find the asymptotic state $\hat\rho$
in~(\ref{4.74}) to be still entangled.

To explicitly show this, one can as before act with the operation of
partial transposition on $\hat\rho$ to see whether negative eigenvalues
are present. Alternatively, one can resort to concurrence as discussed
in Section 2; indeed, the expressions in~(\ref{4.74}) are sufficiently
simple to allow a direct evaluation. The use of concurrence has another
advantage; it not only signals the presence of entanglement, it also
provides a measure of it: its value ranges from zero, for separable
states, to one, for fully entangled states, like the Bell states.

As explained in Theorem 2.5, in order to determine the concurrence 
of any $4\times4$ density matrix $\rho$
representing the state of two atoms, one computes the eigenvalues
of the auxiliary matrix 
$\rho\, (\sigma_2\otimes\sigma_2)\, \rho^*\, (\sigma_2\otimes\sigma_2)$,
which turn out to be non-negative; their
square roots $\lambda_\mu$, $\mu=1,2,3,4$, 
can be ordered decreasingly in value: 
$\lambda_1\geq\lambda_2\geq\lambda_3\geq\lambda_4$. The concurrence of 
$\rho$ is then defined to be:
${\cal C}[\rho]={\rm max}\{\lambda_1-\lambda_2-\lambda_3-\lambda_4, 0\}$.

In the case of the asymptotic state $\hat\rho$ in~(\ref{4.74}), the
above mentioned procedure gives:
\begin{equation}
{\cal C}[\hat\rho]={\rm max}\Bigg\{{\big(3-R^2\big)\over2\big(3+R^2\big)}\,
\bigg[ {5R^2-3\over 3-R^2} -\tau\bigg],\ 0\Bigg\}\ .
\label{4.75}
\end{equation}
This expression is indeed nonvanishing, provided we start with
an initial state $\rho(0)$ for which
\begin{equation}
\tau< {5R^2-3\over 3-R^2}\ .
\label{4.76}
\end{equation}
The concurrence is therefore a linearly decreasing function of $\tau$,
starting from its maximum  ${\cal C}=1$ for $\tau=-3$ 
and reaching zero at $\tau=(5R^2-3)/(3-R^2)$;
notice that this ratio is an admissible value for $\tau$, since
it is always within the interval $[-1, 1]$ for the
allowed values of $R$. 

This result is remarkable, since it implies that the dynamics
in~(\ref{4.73}) not only can initially generate entanglement: 
it can continue to enhance it even in the asymptotic long time
regime.
\bigskip

\eject

\noindent
{\bf Example 4.4}\quad
As initial state, consider the direct product of two pure states:
\begin{equation}
\rho(0)=\rho_{\vec x}\otimes\rho_{\vec y}\ , \qquad
\rho_{\vec x}={1\over2}\Big({\bf 1}_2 + \vec x\cdot\vec\sigma\Big)\ ,
\quad
\rho_{\vec y}={1\over2}\Big({\bf 1}_2 + \vec y\cdot\vec\sigma\Big)\ ,
\label{4.77}
\end{equation}
where $\vec x$ and $\vec y$ are two unit vectors.
In this case, one easily finds that $\tau=\vec x\cdot\vec y$,
so that, recalling~(\ref{4.75}) the asymptotic entanglement
is maximized when $\vec x$ and $\vec y$ are collinear
and pointing in opposite directions.
In this case on explicitly finds:
\begin{equation}
{\cal C}[\hat\rho]={2R^2\over 3+R^2}\ ,
\label{4.78}
\end{equation}
provided $R\neq0$. Notice that ${\cal C}[\hat\rho]$ 
reaches its maximum value of $1/2$ when $R=1$, {\it i.e.} at zero 
temperature, while it vanishes when the temperature
becomes infinitely large, {\it i.e.} $R=\,0$. This has to
be expected, since in this case the decoherence effects
of the bath become dominant. $\Box$
\bigskip 

Furthermore, one can easily check that the phenomenon 
of entanglement production takes place also when
the initial state $\rho(0)$ already has a non-vanishing concurrence.
Let us consider the following initial state,
\begin{equation}
\rho(0)={\varepsilon\over4} {\bf 1}\otimes{\bf 1}+(1-\varepsilon)P\ ,
\label{4.79}
\end{equation}
which interpolates between the completely mixed separable state (see
Example 2.3) and the projection $P$ in (\ref{4.73-3}).
Provided $\varepsilon< 2/3$, this state is entangled, with
${\cal C}[\rho(0)]=1-3\varepsilon/2$.  As 
this initial state evolves into its corresponding asymptote
$\hat\rho$, the difference
in concurrence turns out to be
\begin{equation}
{\cal C}[\hat\rho]-{\cal C}[\rho(0)]={3R^2\varepsilon\over 3+R^2}\ ,
\label{4.81}
\end{equation}
which is indeed non vanishing. 
By direct inspection, one sees that the state
$P$ above is a fixed point of the dynamics generated
by (\ref{4.73}) and therefore it coincides with its corresponding asymptotic
state $\hat\rho$. This is in agreement with the already observed fact
that maximal concurrence require $\tau=-3$.
\bigskip

\noindent
{\bf Remark 4.10}\quad 
{
The formalism used to analyze the behaviour of
stationary atoms in a thermal field bath can be adapted to study
the dynamics of a uniformly accelerating point-like detector
through a vacuum scalar field. Indeed, the detector can be modeled
as a two-level system, in weak interaction with the external,
relativistic quantum field, which plays the role of environment.

In the comoving frame, the detector is seen evolving with a master
equation of the form (\ref{4.12}) as if it were immersed 
in a thermal bath, with temperature proportional to its proper acceleration.
This phenomenon is known in the literature as the Unruh effect.
The discussion of the previous sections suggests that for
an accelerating subsystem composed of two, non-interacting two-level atoms,
quantum correlations should be generated between the two subsystems
as a result of the effective open
system dynamics. A detailed treatment, similar to the one outlined 
in the last Section,\cite{bf20} precisely confirms that
in general the asymptotic density matrix describing the equilibrium
state of the two atoms turns out to be entangled. $\Box$
}
\bigskip

In summary, we have seen that
the entanglement generating properties of 
a bath of quantum fields at finite temperature 
can be ascertained through the study, with different techniques,
of two separate time regimes in the evolution
of two subsystems immersed in it. Entanglement production
in the short time region is regulated by the properties of the
generator of their subdynamics, while 
asymptotic entanglement is measured by the concurrence
of the final equilibrium state.

The same techniques can also be used to study
entanglement production in more general environments,
for which the two subsystems time-evolution
is still generated by master equations of type (\ref{4.72-1}),
but with a generic matrix ${\cal A}_{ij}$, not necessarily
of the form (\ref{4.69}).\cite{bf21} These extended dynamics
are useful in phenomenological applications, in particular
in quantum optics:\cite{puri,ficek}
the fact that also in these generalized situations the
asymptotic entanglement results non-vanishing
may give important feedback for the actual realization
of elementary quantum computational devices. 

\bigskip










%

\end{document}